%% file: mw181.tex
\ifx\mnmacrosloaded\undefined \input mn\fi
\newif\ifAMStwofonts
\AMStwofontstrue

\ifCUPmtplainloaded \else
  \NewTextAlphabet{textbfit} {cmbxti10} {}
  \NewTextAlphabet{textbfss} {cmssbx10} {}
  \NewMathAlphabet{mathbfit} {cmbxti10} {} 
  \NewMathAlphabet{mathbfss} {cmssbx10} {} 
  \ifAMStwofonts
    \NewSymbolFont{upmath} {eurm10}
    \NewSymbolFont{AMSa} {msam10}
    \NewMathSymbol{\upi}     {0}{upmath}{19}
    \NewMathSymbol{\umu}     {0}{upmath}{16}
    \NewMathSymbol{\upartial}{0}{upmath}{40}
    \NewMathSymbol{\leqslant}{3}{AMSa}{36}
    \NewMathSymbol{\geqslant}{3}{AMSa}{3E}

    \let\leq=\leqslant 
    \let\geq=\geqslant 
  \else
    \def\umu{\mu}
    \def\upi{\pi}
    \def\upartial{\partial}
  \fi
\fi


\pageoffset{-2.5pc}{0pc}

\loadboldmathnames



\pagerange{000--000}    
\pubyear{0000}
\volume{000}

\begintopmatter  

\title{Collapse of Primordial Clouds}

\author{Sandra  R. Oliveira, Oswaldo D. Miranda, Jos\'e C. N. de Araujo
        \hfill\break and Reuven Opher}

\affiliation{Instituto Astron\^omico e Geof\'{\i}sico -- Universidade de
S\~ao Paulo \hfill\break
Av. Miguel St\'efano 4200, S\~ao Paulo, 04301-904, SP, Brazil}

\shortauthor{S.R. Oliveira et al.}
\shorttitle{Collapse of Primordial Clouds}

\abstract{

We present here studies of collapse of purely baryonic Population III 
objects with masses ranging from $10M_\odot$ to $10^6M_\odot$. 
A spherical Lagrangian hydrodynamic code has been written to study the
formation and evolution of the primordial clouds, from the beginning of the
recombination era ($z_{rec} \sim 1500$) until the redshift when the collapse
occurs. All the relevant processes are included in the calculations, as well
as, the expansion of the Universe.
As initial condition we take different values for the Hubble constant and for
the baryonic density parameter (considering however a purely baryonic
Universe), as well as different density perturbation spectra,
in order to see their influence  on the behavior of the Population III
objects evolution. We find, for example, that the first mass that
collapses is $8.5\times10^4M_\odot$ for $h=1$, $\Omega=0.1$ and
$\delta_i={\delta\rho / \rho}=(M / M_o)^{-1/3}(1+z_{rec})^{-1}$
with the mass scale $M_o=10^{15}M_\odot$. For $M_o=4\times10^{17}M_\odot$
we obtain $4.4\times10^{4}M_\odot$ for the first mass that collapses.
The cooling-heating and photon drag processes have a key role in the collapse 
of the clouds and in their thermal history. Our results show, for example,
that when we disregard the Compton cooling-heating, the collapse of the
objects with masses $>8.5\times10^4M_\odot$ occurs
earlier. On the other hand, disregarding the photon drag process, the
collapse occurs at a higher redshift.} 


\maketitle

\section{Introduction}

The formation of galaxies, or pre-galactic objects, in the early Universe is
one of the most important fields of study in cosmology. 
It is not completely known, however, how the various structures observed have 
been formed. It is not completely clear how the density perturbations, 
that are the seeds for the formation of the structures of the Universe, were 
produced and how they evolved to produce the structures that we observe.

Some authors argued that for scales below $0.1Mpc$ the purely gravitational
simulations are not completely appropriate, due to the fact that
non-gravitational physical processes are preponderant (Cen 1992). Thus, the
more convenient manner of studying these scales is to use a hydrodynamic
code and to include all the physical processes that are important during and
after the recombination era, starting the calculations from the beginning
of the recombination era. As it will be shown in the present paper, we
make such a study.

Studies including the cooling-heating processes of the gas, in particular 
of pre-galactic composition, have been considered since the  1960's (e.g.,
Matsuda et al 1969, Hutchins et al 1976, Calberg 1981, Palla et al 1983
among others). These articles consider pressure free gas clouds collapsing
and analyzed the effects of the physical processes. 
Recently, some authors considered the pressure of the gas among other 
physical processes in the gas dynamics (de Araujo 1990, de Araujo \& 
Opher 1988, 1989, 1994, Haiman et al 1995, Thoul \& Weinberg 1995 among 
others).

For example, de Araujo \& Opher (1988) studied the collapse of primordial
clouds using an isothermal density perturbation spectrum.
They included the internal pressure, the influence of the cosmic background 
radiation and several cooling-heating processes; among them are: photon 
cooling, photon drag, photoionization due to the CBR, collisional ionization
and a set of equations for the creation and destruction of $H_2$ molecules,
as well as the expansion of the Universe. All these processes are present
during and after the recombination era, therefore, they are essential to 
be taken into account.

In particular, de Araujo \& Opher (1989) used a complete set of hydrodynamic
equations and assumed that the density contrast profile is uniform (``top
hat'' like) throughout the calculation. Thus, the resulting system of
equations is dependent only on time. As a consequence, the velocity profile
is linear in the spherical radial coordinate. This approximation could be
good but a detailed analysis shows that the physical parameters can change
their values for different points inside the cloud, as we are showing in the
present work. Also, it is not realistic to consider that the cloud, as a
whole, will collapse; a possible partial collapse is more realistic.

In the present article we used the same set of hydrodynamic equations of
de Araujo \& Opher (1988). We rewrote it to use a spherical Lagrangian 
hydrodynamic code to follow the variables of the fluid. We analyzed the
internal structure of the cloud during the collapse, using the same
spectrum of density perturbations used by de Araujo \& Opher (1988), 
to verify the differences in the behavior of the physical parameters 
when we include the internal morphology. 

The spherical Lagrangian hydrodynamic code (written by one of us, ODM) 
divides the cloud in several concentric shells. This number of shells
is the optimal combination for the variable time step and the number
of shells that gives the shortest processing time. Our results show
that 450 shells are enough to model satisfactorily all the clouds here
studied. To verify if the number of shells is an optimal choice we ran
a couple of models with 4500 shells, and we see that the results are the same
as for the models with 450 shells. We also reproduced the results of de
Araujo \& Opher (1994) with good agreement, for the models that have small
pressure gradients. The shocks that appear are treated with the inclusion of
the artificial viscosity of Richmeyer and Newmann (Richmeyer \& Morton 1967). 

In another article, to appear elsewhere (Miranda 1998), all the tests and
details of the spherical Lagrangian hydrodynamic code used in the present
article are given.

We begin the calculations at $z_{rec}=1500$ ($T_{rec}=4000K$), the
beginning of the recombination era, and we follow the evolution of the
cloud until its collapse. The initial density perturbation can be given in
the form of a power law spectrum as considered, for example, by Gott \& Rees 
(1975) 

$$
\delta_i={\delta\rho\over\rho}=\bigl({M\over M_o}\bigl)^
{-{1/3}}(1+z_{rec})^{-1},\eqno\stepeq
$$
with $M$ the mass of the cloud, $M_o$ is the reference mass and $z_{rec}$
is the redshift of the recombination. We use $\Omega_o=0.1$ (baryonic
density parameter) and $H=100km\ s^{-1}Mpc^{-1}$ (the Hubble constant). 

It is worth noting that the above equation gives the rms density
fluctuations as a function of the mass scale. This equation sets the
initial density contrast in our top hat profile model.

In many studies, dealing with the evolution and collapse of primordial
structures (some of them referred to above), it is not specified when (at what
redshift) the calculations are started. It is also considered that the
proto-structure, under study, is already detached from the rest of the
Universe and, as a result, it is not taken into account the expansion of the
Universe. The merit of de Araujo and Opher's (1988, 1989, 1994) works was
to take into account all the necessary ingredients to study the evolution
of density perturbations but, by other hand, they unrealisticaly fixed the
density profile of the perturbations during all the time of the cloud
evolution.

In order to follow the evolution of density perturbations, it is necessary
to start the calculations at the beginning of the recombination era, as
mentioned above, including the expansion of the Universe and all the
relevant processes that take place during and after the recombination era.
Obviously, the results of the study will depend also on the density
parameter, on the Hubble constant and on the density perturbation spectrum
chosen as initial conditions. However, the three above
ingredients are not completely known and thus, in the present article, we
are considering different combinations of these three ingredients to
follow the influence caused in the evolution of the clouds under study.

In \S 2 we describe the basic equations, in \S 3 we discuss our results and 
finally in \S 4 we present our conclusions. 

\section{Basic Equations}

The hydrodynamic equations that describe the dynamics of the primordial
clouds are:

$$
{\partial \rho\over\partial t}+{1\over r^2}{\partial\over\partial r}
(r^2\rho v)=0,\eqno\stepeq
$$
the mass conservation equation written in spherical coordinates;
where $\rho$ is the density of the cloud, $r$ the radial coordinate and 
$v$ the velocity. In our code the continuity equation is calculated 
directly from the grid. The specific volumes of the Lagrangian shells
give the density of the cloud.

The equation of motion is given by:

$$
{D\vec v\over Dt}+{1\over\rho}\nabla P+\nabla\phi+{\sigma_{{}_T}bT_r^4x
\over m_pc}\biggl[\vec v-{\dot R(t)\over R(t)}\vec r\biggl]=0\eqno\stepeq
$$
where the velocity is $\vec v=\vec v_n+\vec v_{{}_H}$ with 
$\vec v_n$ the peculiar velocity of the cloud and $\vec v_{{}_H}=H\vec r$ 
the Hubble flow; the pressure of the gas is $P=kN\rho T_m(1+x)$, with
$k$ the Boltzmann constant, $N$ is the Avogadro's number, $T_m$ is the
temperature of the matter inside the cloud, and $x$ the degree of ionization.
The gravitational potential is $\phi$, $\sigma_{{}_T}$ the Thomson cross 
section, $b={4\sigma_{{}_{SB}} / c}$ (with $\sigma_{{}_{SB}}$ the 
Stefan-Boltzmann constant), $T_r$ is the temperature of the radiation, 
$m_p$ is the proton mass, $c$ the velocity of light, $H(t)=\dot R(t)/R(t)$
with $R(t)$ the scale factor and $\dot R(t)$ its time derivative, 
and $ D/Dt \equiv \partial_t + v\partial_r $ is the total derivative.

The energy equation is written as:

$$
{DE\over Dt}={P\over\rho^2}{D\rho\over Dt}-L\eqno\stepeq
$$
where $E$ is the thermal energy per gram and $L$ the cooling function. 

The cooling-heating processes are included in the cooling function $L$ which
is given by the summation of four mechanisms:

$$
L=L_{{}_R}+L_{{}_C}+L_{{}_{H_2}}+L_\alpha\eqno\stepeq
$$

The heat conduction is not taken into account due to the fact that the 
clouds, that we are studying, never reach temperatures in which heat 
conduction is important. Bremsstrahlung was also not included because the 
temperatures of the matter inside the clouds never reach values above
of $10^4$K. 

The cooling from recombination, $L_{{}_R}$ (see e.g. Schwarz et al 1972),
is given by: 

$$
\eqalignno{%
L_{{}_R} & =-kNT_m{Dx\over Dt} & \cr
         & =-kNT_m{C\biggl\{\beta e^{-(B_1-B_2)/kT_r}(1-x)
         -{a\rho x^2\over m_p}\biggl\}+I}, & \stepeq\cr}
$$
where the constants $B_1$, $B_2$, $C$, $\beta$, $\Lambda_{2s,1s}$ are
defined below. The function $I$ (see Eq. 17) is the collisional ionization 
(see, e.g., Defouw 1970). Also, the term $L_{{}_R}$ takes into account 
the photoionization and the ionization due to collisions.

The Compton cooling-heating $L_{{}_C}$ is given by (see, e.g., Peebles 1968):

$$
L_{{}_C}={4kN\sigma_{{}_T}bT_r^4x\over m_ec}(T_m-T_r).\eqno\stepeq
$$

At high redshifts the Compton cooling-heating provides an important 
mechanism of energy exchange between the CBR and the electrons.

For the cooling by molecular hydrogen, $L_{{}_{H_2}}$, we follow
Lepp \& Shull (1983), who give the following equations: 

$$
L_{{}_{H_2}}={\Lambda\over\rho}\ \ \ \ \ \ (erg\ g^{-1}s^{-1})\eqno\stepeq
$$

The electronic ground state of these molecules may be excited to rotational 
and vibrational levels as a result of collisions with atomic hydrogen,
emiting infrared photons and hence cooling the cloud. We use the equations
given by Lepp \& Shull (1983) where they represent a good fit for
$100K\leq T_m\leq10^5K$. 

The equation to determine $\Lambda$ is given by:

$$
\Lambda=n_{{}_{H_2}}\biggl[{L_{RH}\over1+{L_{RH}\over
L_{RL}}}+{L_{VH}\over1+{L_{VH}\over L_{VL}}}\biggl]\eqno\stepeq
$$
with:

\beginlist
\item (i) $L_{{}_{VH}}=1.1\times10^{-18}e^{-(6744/T_m)}$; 
\bigskip
\item (ii) $L_{{}_{VL}}=8.18\times10^{-25}n_{{}_H}T^{1/2}_m\;e^{-(1000/
T_m)}$ \par
\noindent (for $T_m\geq1635K$) \par
or \par
$L_{{}_{VL}}=11.45\times10^{-26}n_{{}_H}e^{[(T_m/125)-(T_m/577)^2]}$ \par
\noindent (for $T_m<1635K$);
\bigskip
\item (iii) $L_{{}_{RH}}=3.9\times10^{-19}e^{-(6118/T_m)}$ \par
\noindent (for $T_m\geq1087K$) \par
or \par
$L_{{}_{RH}}=10^{(-19.24+0.47y-1.247y^2)}$ \par
\noindent (for $T_m<1087K$);
\bigskip
\item (iv) $L_{{}_{RL}}=(n_{{}_{H_2}}^{0.77}+1.2n_{{}_H}^{0.77})\times1.38
\times10^{-22} e^{-(9234/T_m)}$ \par
\noindent (for $T_m\geq4031K$) \par
or \par
$L_{{}_{RL}}=(n_{{}_{H_2}}^{0.77}+1.2n_{{}_H}^{0.77})\times
10^{(-22.9-0.553y-1.148y^2)}$ \par
\noindent (for $T_m<4031K$).
\endlist

\bigskip

\par In the above equations $n_{{}_{H_2}}$ is the numerical density of
$H_2$, $n_{{}_H}$ is the numerical density of $H$ and  $y=\log(T_r/10^4)$. 

The reactions which produce $H_2$, used in our calculations, are shown 
below: 

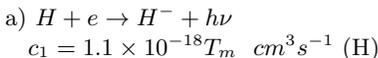
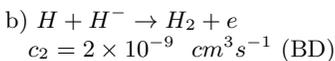
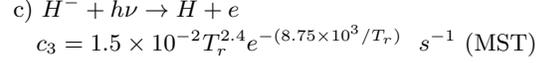
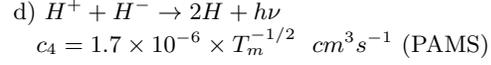
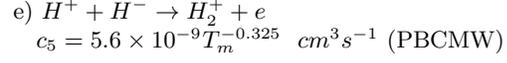
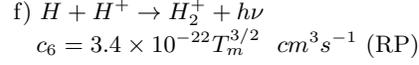
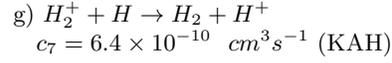
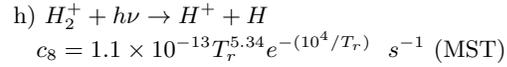
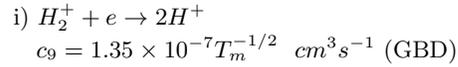
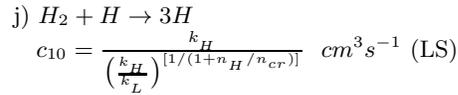
\beginlist
\item a)
$H+e\rightarrow H^-+h\nu$ \par
$c_1=1.1\times10^{-18}T_m \ \ cm^3s^{-1}$ (H) 
\bigskip
\item b)
$H+H^-\rightarrow H_2+e$ \par
$c_2=2\times10^{-9}\ \ cm^3s^{-1}$ (BD)
\bigskip
\item c)
$H^-+h\nu\rightarrow H+e$ \par
$c_3=1.5\times10^{-2}T_r^{2.4}e^{-(8.75\times10^3/T_r)}\ \ s^{-1}$ (MST)
\bigskip
\item d)
$H^++H^-\rightarrow 2H+h\nu$ \par
$c_4=1.7\times10^{-6}\times T_m^{-1/2}\ \ cm^3s^{-1}$ (PAMS)
\bigskip
\item e)
$H^++H^-\rightarrow H_2^++e$ \par
$c_5=5.6\times10^{-9}T_m^{-0.325}\ \ cm^3s^{-1}$ (PBCMW)
\bigskip
\item f)
$H+H^+\rightarrow H_2^++h\nu$ \par
$c_6=3.4\times10^{-22}T_m^{3/2}\ \ cm^3s^{-1}$ (RP)
\bigskip
\item g)
$H_2^++H\rightarrow H_2+H^+$ \par
$c_7=6.4\times10^{-10}\ \ cm^3s^{-1}$ (KAH)
\bigskip
\item h)
$H_2^++h\nu\rightarrow H^++H$ \par
$c_8=1.1\times10^{-13}T_r^{5.34}e^{-(10^4/T_r)}\ \ s^{-1}$ (MST)
\bigskip
\item i)
$H_2^++e\rightarrow 2H^+$ \par
$c_9=1.35\times10^{-7}T_m^{-{1/2}}\ \ cm^3s^{-1}$ (GBD)
\bigskip
\item j)
$H_2+H\rightarrow 3H$ \par
$c_{10}={k_{{}_H}\over{\bigl({k_{{}_H}\over k_{{}_L}}\bigl)^
{[1/(1+n_{{}_H}/ n_{{}_{cr}})] }}} \ \ cm^3s^{-1}$ (LS)
\endlist

\bigskip

\noindent where 
$c_i$ are the rates of production (or destruction) of the ions or molecules, 
\bigskip
$n_{{}_{cr}}=10^{(4-0.416s-0.327s^2)} \ \ cm^3$, $s=\log(T_m/10^4)$,
\bigskip
$k_{{}_H}=3.52\times10^{-9}e^{-(43900/T_m)} \ \ cm^3s^{-1}$,
\bigskip
$k_{{}_L}=6.11\times10^{-14}e^{-(29300/T_m)}\ \  cm^3s^{-1}\ \ 
{\rm if}\ T_m\geq7390K$ \par
\noindent or \par
$k_{{}_L}=2.67\times10^{-15}e^{-({6750/T_m})^2}\ \ 
cm^3s^{-1} \ \ {\rm if}\ T_m<7390K.$
\bigskip

The references used for the above rates are:
(H) --  Hirasawa 1969, 
(BD) -- Browne \& Dalgarno 1969, 
(MST) -- Matsuda, Sato \& Takeda 1969, 
(PAMS) -- Peterson, Abertz, Moseley \& Sheridan 1971, 
(PBCMW) -- Poulaert, Brouillard, Claeys, McGowant \& Wassenhove 1978, 
(RP) -- Ramaker \& Peek 1976, 
(KAH) -- Karpas, Anicich \& Huntress 1979, 
(GBD) -- Giusti--Suzor, Bardsley \& Derkits 1983, 
(LS) -- Lepp \& Shull 1983.
\bigskip

Due to the fact that the clouds here studied are made of pure hydrogen,
the above reactions are the most relevant to the formation and destruction
of molecular hydrogen.
Even if we included He, or even Li, in our studies, they would not 
contribute significantly to the cooling of the cloud, due to the temperatures 
and densities involved in our calculations.

The rates and production of ions and molecules are calculated by: 

\beginlist
\item a) Molecular formation rate via $H^-$
$$
x_{{}_{H^-}}={n_{{}_{H_-}}\over n_{{}_H}}={x(1-x)nc_1\over n[(1-x)
c_2+x(c_4+c_5)]+c_3}\eqno\stepeq
$$

$$
{dx_{{}_{H_2}}\over dt}=(c_2x_{{}_{H^-}}-c_{10}x_{{}_{H_2}})n(1-x)
\eqno\stepeq
$$

\item b) Molecular formation rate via $H_2^+$

$$
x_{{}_{H_2^+}}={n_{{}_{H_2^+}}\over n_{{}_H}}={nx[x_{{}_{H^-}}c_5-
(1+x)c_6]\over n[(1-x)c_7+xc_9]+c_8}\eqno\stepeq
$$

$$
{dx_{{}_{H_2}}\over dt}=n(1-x)[c_7x_{{}_{H_2^+}}-c_{10}x_{{}_{H_2}}]
\eqno\stepeq
$$
\endlist

The Lyman-$\alpha$ cooling is (e.g., Calberg 1981) given by:

$$
L_\alpha=1.25\times10^{-11}C_{12}{A_{2\gamma}\over
A_{2\gamma}+C_{21}}\ \ \ \ \ (erg/s)\eqno\stepeq
$$
where $A_{2\gamma}=8.272\ {\rm s}^{-1}$ is the two photon emission rate, 
$C_{21}=1.2\times10^{-6}T_m^{-{1/2}}xn_{{}_H}\ {\rm s}^{-1}$ 
the collisional de-exitation rate and 
$C_{12}=2C_{21}e^{-(1.19\times 10^5/T_m)}$.\par

For the ionization degree we follow Peebles (1968) modified to
take into account the collisional ionization. The ionization rate is
then written as follows:

$$
{Dx\over Dt}=C\biggl\{\beta e^{-(B_1-B_2)/ kT_r}(1-x)-
{a\rho x^2\over m_p}\biggl\}+I\eqno\stepeq
$$
where
$B_1$, is the bound energy of the ground state and
$B_2$ is the bound energy of the first excited state and:

$$
C={\Lambda_{2s,1s}\over\Lambda_{2s,1s}+\beta},\ \ \ \ \ 
\beta={(2\pi m_ekT_r)^{3/2}\over h^3}e^{-(B_2/kT_r)}a\eqno\stepeq
$$
with $\Lambda_{2s,1s}=8.227\ s^{-1}$ and $a=2.84\times10^{-11}T_m^{-1/2}
\ cm^3s^{-1}$, the recombination rate.\par

The collisional ionization rate $I$ (see e.g., Defouw 1970) is given by:

$$
I=1.23\times10^{-5}xN(1+x)\rho{k\over B_1}T_m^{1/2}e^{-(B_1/kT_m)}
\eqno\stepeq
$$

\section{Boundary Conditions}

The clouds evolve within a medium (the Universe) of density $\rho_u$
and thus, we take into account the influence of the external medium
on the evolution of the density perturbation. The expansion of the Universe
is also considered because the physical parameters of the primordial clouds
such as the density, the temperature and the degree of ionization depends
on the Universe expansion rate. It is worth stressing that in many studies of
structure formation the influence of the external medium on the structure
under study was not taken into consideration. 

In order to calculate the temperature of the matter of the Universe we
consider, in the cooling function, only the contribution due to the Compton
cooling-heating and the cooling due to the recombination processes. The other
physical processes considered for the cloud are not relevant for the
evolution of the matter temperature of the Universe.

The energy equation is then written as

$$
{dE_u\over dt}={P\over\rho^2_u}{d\rho_u\over dt}-L_{{}_R}-L_{{}_C}.
\eqno\stepeq
$$

For the degree of ionization of the Universe, we use directly the results
of the study done by Peebles (1968), therefore without taking into account
the collisional ionization, since it is not important for the matter
evolution of the Universe. Thus, the degree of ionization is given by:

$$
{dx_u\over\ dt}=C\biggl\{\beta e^{-(B_1-B_2)/kT_r}(1-x_u)
-{a\rho_ux_u^2\over m_p}\biggl\},\eqno\stepeq$$
$$C={\Lambda_{2s,1s}\over\Lambda_{2s,1s}+\beta},\ \ \ \ \  
\beta={(2\pi m_ekT_r)^{3/2}\over h^3}e^{-(B_2/kT_r)}a\eqno\stepeq
$$
with $\Lambda_{2s,1s}=8.227\ s^{-1}$ and $a=2.84\times10^{-11}T_{mu}^{-1/2}
\ cm^3s^{-1}.$

\section{Calculations and Discussions}

 The formation of objects in the Universe may be produced directly either
from primordial density perturbations or through fragmentation or still
agregation  of objects that are also produced from primordial density 
perturbations.

 We study here, in particular, the formation of the Population III 
objects ranging from $10M_\odot$ to $10^6M_\odot$. In the
present work we study the formation of these objects considering that
they could be produced directly from a primordial density perturbation
spectrum. 

 In many studies dealing with structure formation, it is considered
that the proto-object is already detached from the rest of the Universe
and as a result does not undergo the influence of the expansion of the 
Universe. These calculations do not take into account how a density
perturbation becomes a coupled object, and the studies begin at a redshift 
where a couple of the physical processes are not important any more.

 The structures that we observe today certainly depend on
the kind of the density perturbations produced primordially, and also
on the physical processes present during and after the recombination
era that are very important in development of these density perturbations.

 What we do here is to study one step before the kind of study
considered elsewhere, we mean, to see how a density perturbation
could become a proto-object. In so doing, we start our calculations
from the recombination era, and take into account a series of physical
processes present during and after the recombination era. The evolution
of a given density perturbation depends also on, 
the kind of density perturbation spectrum and on the kind of cosmology 
considered (i.e., values of the Hubble constant and density parameters for
baryonic and non-baryonic dark matter).

To see how the formation of proto-objects depend on the particular
spectrum used we consider different kinds of spectra. Also, we consider
different cosmologies, restricting, however, to those cosmologies
without non-baryonic dark matter. In another article (Oliveira et al
1998, hereafter paper II) we study structure formation taking into 
account the presence of non-baryonic dark matter.

Still, to see how the physical processes contribute to the evolution
of a given density perturbation, particularly those effects in general
not considered by other authors, we performed some calculations 
with and without such effects (below we see in details which physical
processes we are referring to).

Concerning the presentation of our results, we show, in particular, three
points in the cloud: the centre (that is the first shell), the middle shell,
and the last shell (that is the edge of the cloud) as a function of time.
In order to analyze the radial behavior, we show five different times:  
$t_i$ --  time at the recombination era,
$t_2=(t_3-t_i)/2$,
$t_3=(t_f-t_i)/2$,
$t_4=(t_f-t_3)/2$ and 
$t_f$ -- the time at the collapse of the core.

Also, as mentioned before, all the calculations were started at the
beginning of the recombination era ($z_{rec}\sim1500$). Due to the
expansion of the Universe, even the clouds that are already Jeans unstable
and eventually collapse, undergo, initially, an expansion phase before
collapsing. It can be shown (see, e.g., Peebles 1993 and Coles \& Lucchin
1995) that for a cloud which stops expanding, its mean density relative
to the background is given by:

$$
\biggl({\bar\rho_{cloud}\over\rho}\biggr)=\biggl({3\pi\over4}\biggr)^2
\simeq5.6\eqno\stepeq
$$

The above figure corresponds to: 

$$
\biggl({\delta\bar\rho\over\rho}\biggr)\simeq4.6\eqno\stepeq
$$

It is worth stressing that the above number is a lower limit since
it depends only on the expansion rate of the Universe (in this particular
case an Einstein -- de Sitter Universe) and on the gravity of the cloud. For
the cases in which other physical processes are important the clouds
stop expanding for

$$
\biggl({\delta\bar\rho\over\rho}\biggr)>4.6,\eqno\stepeq
$$
far greater than the linear regime. In this way there
could exist clouds, in the nonlinear regime, that neither collapse nor
stop expanding (see also paper II).

In the work by de Araujo \& Opher (1988,1989) it is defined the collapse 
redshift when the cloud radius has 1/100 of the turn around radius (when the 
velocity $v=0$, i.e., when the cloud stop expanding). In our calculation we 
consider that the cloud collapses when ${\delta\rho/\rho}\geq10^4$. It is 
worth stressing that what we call a collapse it is, in fact, a condition for 
which  the collapse of a proto-object become irreversible.

Let us have a look at the models for a Universe with $\Omega = 0.1$ and  
$h=1.0$ (the Hubble constant in unit of 100 ${\rm km\ s^{-1}Mpc^{-1}}$),
and let us consider a density perturbation given by Eq.(1) with
$M_o=10^{15}M_\odot$.

Let us first have a look at the evolution of the density contrast (Fig.1)
that gives us information about the collapse of the clouds.
Generally speaking, we see that there is no collapse for $M < 10^4 M_\odot$.

As expected, for $10M_\odot$ the density contrast (as seen in 
Fig.1) falls rapidly in all shells and almost completely disappears when 
$z\sim1250$. For $M=10^2M_\odot$ the same happens when $z\sim1300$. 
For lower redshifts the density contrast oscillates for both masses.
These results are expected, because at the time of the recombination 
era, these clouds were not Jeans unstable. Still, they tend to 
oscillate, but due to photon-drag the perturbations are strongly dissipated.

For the masses $10^3M_\odot$ and $10^4M_\odot$ the fate is a little bit 
different. In these cases the perturbations oscillate for different shells,
as can be seen, for instance, in the intermediate and internal shells of 
$M=10^3M_\odot$. For the external shells the perturbation decreases almost 
without oscillation. For the mass $10^4M_\odot$ the behavior is similar to 
that of $M=10^3M_\odot$. Again, due to the fact that these clouds are not 
Jeans unstable at the time of the recombination era, the perturbations are 
strongly dissipated by the photon-drag. It is also worth noting that the
cooling mechanisms are not efficient enough to cool down the clouds and
turn them Jeans unstable.

The masses $10^5M_\odot$ and $10^6M_\odot$ suffer
partial collapse, i.e., the core collapses while the external shells
oscillate. The internal shells of, for example, $M=10^5M_\odot$, initially 
oscillate but, due to the cooling mechanisms, this part of the cloud becomes 
Jeans unstable and then collapses. 

It is worth stressing that we start all the calculations with a ``top hat'' 
density contrast profile. For the masses $10M_\odot$ and $10^{2}M_\odot$,
the density contrast remains almost uniform as a function of the radial
coordinate. For the masses $10^3M_\odot$ and $10^4M_\odot$ small
oscillations occur along the radial coordinate. For the masses $10^5M_\odot$
and $10^6M_\odot$, the density contrast profile begins approximately uniform,
but it does not remain so as time goes on. This behavior shows that the 
``top hat" profile is not maintained throughout the cloud evolution. A ``top
hat'' profile is maintained only when forces proportional to the spherical
coordinate appears. As soon as the pressure gradients begin to be important
in the cloud evolution the ``top hat'' density profile is destroyed.

Alternatively in Fig.2 we illustrate the time evolution of the density
of the cloud. We see, clearly the decrease in the density as time goes
on for clouds with masses smaller than $10^4M_\odot$ and the partial
collapse underwent by the clouds with $10^5M_\odot$ and $10^6M_\odot$.
We also show the morphological behavior of the density profile
(see Fig.3) in order to see how it is modified as a function of time.

Concerning the velocity profile, it begins linear but it changes with 
time. It is worth stressing that a linear velocity profile is consistent 
with a ``top hat'' density profile and vice-versa. The masses $10M_\odot$
to $10^4M_\odot$, for example, oscillate accompanying the similar behavior
of the density contrast. For the masses $10^5M_\odot$ and $10^6M_\odot$ the
velocity decreases when the density contrast increases, as expected. We
conclude that the linear profile of the velocity, that, e.g., de Araujo \&
Opher (1989) assumed, appears to be a good approximation for the masses
$10M_\odot$ to $10^4M_\odot$. But for the masses $10^5M_\odot$ and
$10^6M_\odot$ the approximation is not valid. The internal pressure and the
physical processes work to destroy the velocity linear profile, as a result
the density profile is not a ``top hat'' like any more.

In the study of the collapse, the thermal and chemical history of the 
clouds are worth studying  due to the fact that the evolution of the 
clouds in what concern, for example, their fragmentation depends on how 
the temperature of the clouds evolve which, by other hand, depend on
their chemical ingredients. For primordial clouds, in particular, the 
thermal history will also depend on the Compton cooling (heating) 
processes, that maintains the temperature of the cloud close to 
the temperature the CBR. This cooling processes is particularly relevant
for high values of z (i.e., $z > 300$).

As we explained above the heat conduction and the Bremsstrahlung is not taken 
into account. It is worth stressing that in our study we do not take, as
initial condition, a virialized object as considered by some authors. Our
calculations start at the recombination era when the temperature of the
matter inside the perturbation has the same value of that of the radiation.
For clouds that collapse at high redshifts, in general, the principal
mechanism that acts in the evolution and collapse of the perturbations is
the Compton heating-cooling (see results presented in Tables 5-8). Thus,
during the expansion phase of the perturbations, when the Compton 
heating-cooling is efficient, the temperature of the matter inside the
perturbation is almost the same temperature of the radiation.

At the redshift collapse of the clouds, the temperature of the matter
increases by a factor 4 or 5 in relation to its correspondent value at
the turn-around. Thus the temperature of the clouds that collapse, 
due to shocks and the collapse itself, can rise to $\sim 10^4K$ depending
on the power spectrum and on the mass of the cloud (see, e.g. Table 2).
Also, depending on the power spectrum and on the mass of the cloud, the
shocks and the collapse itself are not able to rise the temperature of the
cloud to $\sim 10^4K$ (see, e.g., Table 3 and paper II).

Certainly, these results are dependent on the particular way in which we
defined the collapse of the clouds. That is, we consider that an object
collapsed when the density contrast is greater than $10^4$. The results
also depend on the thermal history of the clouds (that depends on the
physical processes here considered). As a result no strong shocks occur
that could eventually heat the clouds to temperatures $>10^4$K. If we
follow the calculations to higher values of density than the established by
our collapse criterion, the temperature will rise above the values here
obtained. It is possible that to density contrasts higher than $10^5$
strong shocks occur and they would probably heat the gas to values 
of the order of the virial temperature, as a result other physical 
processes, as e.g. Bremsstrahlung, would be important to the evolution of 
the clouds.

Let us examine the temperature, pressure and molecular 
density formation of the clouds studied. The temperature, pressure and 
molecular density fall with time for the masses ranging from 
$10M_\odot$ to $10^4M_\odot$, as seen in Fig.4. These are, in fact,
expected results since the density perturbations are disappearing 
as time goes on. As a result, the cloud temperature, for example, 
tends to be equal to the temperature of the matter of the Universe.

For clouds with masses ranging from $10^5M_\odot$ to $10^6M_\odot$, the
behavior is very different. The pressure, for example, grows, since the
cloud begins to collapse. The temperature initially decreases, but as time
goes on it increases, but not significantly, which is so due to the fact 
that the cooling mechanisms become important and cool the cloud. During
the collapse process there is an efficient formation of molecular
hydrogen. The molecular hydrogen, and also the Compton cooling
process are the most important cooling mechanisms for primordial
clouds. It is worth mentioning that the Compton cooling (heating) process
is important for high redshift (i.e. $z > 300$) as above mentioned, 
while the molecular hydrogen cooling is important for $z < 300$, which
is so due to the fact that the CBR photoionize 
the $H^-$ and $H^+_2$ ions that intermediate the formation of molecular
hydrogen.

An analysis of the molecular density versus temperature of the cloud is 
shown in Fig.5. For the cloud of $10^6M_\odot$, for example, the molecular 
density increases when the temperature is in the interval $100K<T<500K$,
which occurs during the cloud collapse. The molecular hydrogen is very
important in the cooling processes of the cloud, as already mentioned.

For the masses $10M_\odot$ to $10^4M_\odot$ the pressure profile remains 
constant with time along the radial coordinate; only tiny variations are seen 
for the mass $10^4M_\odot$. For other masses, the profile changes rapidly with 
time along the radial coordinate. The production of the molecules, and the 
consequent cooling of the cloud, is always more important in the central part 
of the cloud. 

In Fig.6 we show how the radii of the shells of the clouds
change with time. For the masses $10^5M_\odot$ and $10^6M_\odot$, for 
example, we clearly see how the different shells of the clouds evolve
with time. It is also possible to see that these clouds undergo
a partial collapse, the outer most shells do not stop expanding.

In Table 1 we present, in particular, in its second and third column, 
a comparison between our results and the results of de Araujo (1990) for 
$M_o=10^{15}M_\odot$. There we observe, e.g., that the mass $10^6M_\odot$
has collapse redshift similar to that obtained by de Araujo (1990). The
mass $10^4M_\odot$, however, collapses at $z_c=5.4$ in the models of de
Araujo (1990) but here there is no collapse for such a mass. The mass
$10^5M_\odot$ collapses at $z_c=260$ according to de Araujo (1990), but we
obtained a collapse at $z_c=22$. For clouds with masses larger than
$10^6 M_\odot$ the collapse redshift is almost the same, for these cases
the pressure is not very important in the evolution of the cloud, this
explain why the results are the same (see also paper II). For 
those cases in which the internal pressure is important on the evolution of 
the cloud the results are too different. In conclusion, it is not a good
assumption to maintain a ``top hat'' density profile for clouds
with masses less than $10^6 M_\odot$, as de Araujo (1990) considered
throughout their calculations.

\begintable{1}
\caption{{\bf Table 1.} The redshifts of collapse.} 
\halign{#\hfil & \qquad \hfil#\hfil\qquad &
         \hfil#\hfil\qquad & \hfil#\hfil\qquad &
         \hfil#\hfil\qquad & \hfil#\hfil \cr 

$M/M_\odot$ & $z_c$ & $z_c^A$ & $z_c^*$ & $\delta_{ta}^A$
& $\delta_{ta}$ \cr
\noalign{\vskip 10pt}
$10^4$           & NC  & 5.4 & NC  & 8.0 & --  \cr
$4.4\times 10^4$ & --  & --  & 40  & --  & --  \cr
$8.5\times 10^4$ & 12  & --  & 498 & --  & 6.3 \cr
$10^5$           & 22  & 260 & 477 & 6.7 & 5.7 \cr
$10^6$           & 197 & 180 & 643 & 5.1 & 5.0 \cr
}
\tabletext{The redshifts of collapse of the present paper, $z_c$, 
and de Araujo's studies (de Araujo 1990), $z_c^A$, for $M_o=10^{15}M_\odot$;
and redshifts of collapse of the present paper for
$M_o=4\times10^{17}M_\odot$, $z_c^*$. The density contrast at the turn
around for de Araujo's studies, $\delta_{ta}^A$, and for the present paper,
$\delta_{ta}$, for $M_o=10^{15}M_\odot$. NC stands for ``no collapse".} 
\endtable

As mentioned earlier the clouds stop expanding satisfying Eq.(23),
therefore well above the linear regime. Again we see in Table 1 in particular
for a cloud of $10^6 M_\odot$ that the present results agree with those by de 
Araujo (1990). For clouds in which the pressure is important the turn around 
occurs for values above the value given by Eq.(23). In the paper II we
discuss this issue for clouds with masses larger than $10^6 M_\odot$, where
we also include the presence of non-baryonic dark matter.

Another interesting comparison has to do with the minimum mass that
can collapse in our model. We have found that the minimum mass is
$\simeq 8.5\times 10^4 M_\odot$ (the collapse redshift is 12), 
that is almost one order of magnitude larger than that found by de 
Araujo (1990). This shows again that to fix the density profile is 
not a good assumption. 

We obtained also a mass limit that does not collapse but remain bounded, 
namely, $3.8\times10^4M_\odot$. This bounded cloud reaches a density 
contrast of $10^4$, but the collapse does not evolve to a denser stage, 
it oscillates until the present time. In Fig.7 we see the behavior of 
the radii of the shells as a function of time for this mass.

It is worth stressing that these results just discussed are for a Universe
model with $h=1$ and $\Omega=0.1$, with a reference mass
$M_o=10^{15}M_\odot$. 

Obviously the results presented here depend on the particular spectrum of 
perturbations used. The minimum mass, for example, that collapses
would be different if we used a different spectrum of density perturbations. 
In Table 1 we also include the collapse redshift for the spectrum normalized
with a reference mass $M_o=4\times10^{17}M_\odot$ (instead of 
$10^{15}M_\odot$). The density perturbations with $M_o=4\times10^{17}M_\odot$ 
are seven times greater than those with $M_o=10^{15}M_\odot$. 
We may see that the collapse redshifts are altered significantly, 
and also the minimum mass that collapses is now $4.4\times10^{4}M_\odot$ 
with a collapse redshift of $\sim 40$ (see Table 2). 

\begintable{2}
\caption{{\bf Table 2.} Values of the variables when we used $h=1.0$, 
$\Omega=0.1$ and $M_o=4\times10^{17}M_\odot$.}
\halign{#\hfil & \quad \hfil#\hfil\quad &
         \hfil#\hfil\quad & \hfil#\hfil\quad &
         \hfil#\hfil\quad & \hfil#\hfil \cr
$M/M_\odot$ & $z_{{}_{c}}$  & $n_{{}_{H_2}}$ ($cm^{-3}$)  & $T$ ($K$)
& $\rho$ ($g$ $cm^{-3}$) \cr
\noalign{\vskip 10pt} 
$10^4$ & NC & -- & -- & -- \cr
$4.4\times10^4$  & 40 & $3\times10^{-2}$ & $7\times10^3$ & $8\times10^{-19}$
\cr
$10^5$ & 48 & $3\times10^{-2}$ & $7\times10^3$ & $8\times10^{-19}$  \cr
$10^6$ & 624 & $2\times10^{-2}$ & $7\times10^3$ & $5\times10^{-17}$ \cr
\noalign{\vskip 10pt}
}
\endtable

Due to the fact that the collapses occur at a higher redshift for a large
$M_o$, there is also a modification in the importance of the physical
processes in the evolution of the clouds. The molecular hydrogen cooling,
for example, is less significant for collapses which occur at high redshift.

When we change the content of baryonic matter from $\Omega=0.1$ to 
$\Omega=0.2$ and maintain $h=1$ and $M_o=10^{15}M_\odot$ the minimum mass 
becomes  $5\times10^4M_\odot$; the collapse occuring at $z$=9 (see Table 3).
In Table 3 we show the collapse redshifts for clouds of different masses.

\begintable{3}
\caption{{\bf Table 3.} Values of the variables when we used $h=1.0$, 
$\Omega=0.2$ and $M_o=10^{15}M_\odot$.}
\halign{#\hfil & \quad \hfil#\hfil\quad &
         \hfil#\hfil\quad & \hfil#\hfil\quad &
         \hfil#\hfil\quad & \hfil#\hfil \cr

$M/M_\odot$ & $z_{{}_{c}}$ & $n_{{}_{H_2}}$ ($cm^{-3}$) & $T$ ($K$)
& $\rho$ ($g$ $cm^{-3}$) \cr
\noalign{\vskip 10pt} 
$5\times10^4$ & 9 & $300$ & $100$ & $3\times10^{-22}$ \cr
$8.5\times10^4$ & 90 & $100$ & $500$ & $3\times10^{-19}$ \cr
$10^5$ & 123 & $320$ & $500$ & $1\times 10^{-18}$ \cr
$10^6$ & 207 & $500$ & $10^3$ & $3\times10^{-18}$ \cr
\noalign{\vskip 10pt}
}
\endtable

For $h=0.5$, $\Omega=0.1$ and $M_o=10^{15}M_\odot$ we obtain that 
$2\times10^5M_\odot$ is the minimum mass that
collapses. The collapse occurs at $z$=0.78 (see Table 4). 

The above results show that the minimum mass does not depend strongly
neither on the kind of spectra used nor on the values of the Hubble
constant and the density parameter assumed. By other hand, the 
collapse redshifts depend strongly on the Universe model and 
on the spectra adopted.

\begintable{4}
\caption{{\bf Table 4.} Values of the variables when we used $h=0.5$,
$\Omega=0.1$ and $M_o=10^{15}M_\odot$.}
\halign{#\hfil & \quad \hfil#\hfil\quad &
         \hfil#\hfil\quad & \hfil#\hfil\quad &
         \hfil#\hfil\quad & \hfil#\hfil \cr

$M/M_\odot$ & $z_{{}_{c}}$ & $n_{{}_{H_2}}$ ($cm^{-3}$) & $T$ ($K$)
& $\rho$ ($g$ $cm^{-3}$) \cr
\noalign{\vskip 10pt} 
$8.5\times10^4$ & NC & $2\times10^{-12}$ & $10$ & $3\times10^{-31}$  \cr
$10^5$ & NC & $3\times10^{-13}$ & $10$ & $3\times10^{-31}$ \cr
$2\times10^5$ & 0.8 & $3\times10^{-6}$ & $50$ & $3\times10^{-25}$ \cr
$10^6$ & 158 & $63$ & $320$ & $2\times10^{-19}$ \cr
}
\endtable

As we have already mentioned, the present study includes a series of physical
processes not in general considered in the literature, namely, the 
Compton cooling (heating), the photon-drag due to the cosmic background 
radiation are examples.

We thus studied the influence of the cooling-heating processes and the  
photon-drag on the evolution of the clouds here considered. We chose, in 
particular, for this study, the clouds of $3.8\times10^4$, 
$8.5\times10^4$, $10^5$ and $10^6M_\odot$, the results are presented 
in Tables 5-8 respectively. 
To do this we studied individually the effects of all the four 
cooling-heating processes and the photon drag. The values in these tables 
have been taken for $z_{{}_{c}}$, that is, the redshift when the 
collapse occurs.

First let us analyze the cooling from recombination. For the mass 
$3.8\times10^4M_\odot$ we again do not have collapse when we disregard 
this cooling mechanism; for $8.5\times10^4M_\odot$ the collapse is delayed. 
The collapse redshift changes from $z\sim 12$ with all processes to 
$z\sim 6.8$ without the recombination cooling. For the masses 
$10^5M_\odot$ and $10^6M_\odot$ this cooling does not affect the collapse 
redshift.

The Compton cooling-heating works against the collapse for the masses from 
$3.8\times10^4M_\odot$ to $10^5M_\odot$. When a cloud (density perturbation)
begins to evolve, at the beginning of the recombination era, it first expands
and then stop expanding if it becomes Jeans unstable, as already mentioned.
During the expansion phase it would cool down adiabatically if we did not
have any cooling-heating processes. In this way, the cloud could stop
expanding and collapses earlier. Due to the Compton heating the cloud is
heated to temperatures near to that of the cosmic background radiation, as
a result the cloud stop expanding and collapses later.

Also, disregarding the Compton heating a cloud of $3.8\times10^4M_\odot$ 
collapses. This means that the minimum mass depends on this very processes.
For  masses with $M< 10^6M_\odot$ the collapse occurs earlier as we may 
see in Tables 5-8. 
For clouds with $M > 10^6M_\odot$, however, the collapse redshift is not
affected, again due to the fact that the pressure for this cloud
does not significantly affect the collapse, because these
clouds become Jeans unstable just after the recombination era.

\begintable*{5}
\caption{{\bf Table 5.} Values of the variables for $M=3.8\times10^4M_\odot$
with $h=1.0$, $\Omega=0.1$ and $M_o=10^{15}M_\odot$.}
\halign{#\hfil & \quad \hfil#\hfil\quad &
         \hfil#\hfil\quad & \hfil#\hfil\quad &
         \hfil#\hfil\quad & \hfil#\hfil\quad &
         \hfil#\hfil \cr

          & with all  & without    & without    & without
& without & without \cr
Variable  & processes & $L_{{}_R}$ & $L_{{}_C}$ & $L_{{}_{H_2}}$
& $L_\alpha$ & photon drag \cr
\noalign{\vskip 3pt\hrule\vskip 3pt} 
$z_{{}_{c}}$ & NC & NC & 156 & NC & NC & NC \cr
$n_{{}_{H_2}}$ ($cm^{-3}$) & $10^{-8}$ & $10^{-8}$ & $10^2$ & --
& $10^{-8}$ & $10^{-10}$ \cr
$T$ ($K$) & $ 50$ & $ 50$ & $ 10^2$ & $ 50$ & $ 50$ & $ 50$ \cr
$\rho$ ($g$ $cm^{-3}$) & $3\times10^{-27}$ & $3\times10^{-27}$
& $3\times10^{-19}$ & $2\times10^{-27}$ & $3\times10^{-27}$
& $1\times 10^{-30}$ \cr
\noalign{\vskip 3pt\hrule\vskip 10pt}
}
\endtable

\begintable*{6}
\caption{{\bf Table 6} Values of the variables for $M=8.5\times10^4M_\odot$
with $h=1.0$, $\Omega=0.1$ and $M_o=10^{15}M_\odot$.}
\halign{#\hfil & \quad \hfil#\hfil\quad &
         \hfil#\hfil\quad & \hfil#\hfil\quad &
         \hfil#\hfil\quad & \hfil#\hfil\quad & 
         \hfil#\hfil \cr
 
          & with all  & without    & without    & without
& without & without \cr
Variable  & processes & $L_{{}_R}$ & $L_{{}_C}$ & $L_{{}_{H_2}}$
& $L_\alpha$ & photon drag \cr
\noalign{\vskip 3pt\hrule\vskip 3pt}    
$z_{{}_{c}}$ & 12 & 6.8 & 252 & 2.7 & 6.8 & 2 \cr
$n_{{}_{H_2}}$ ($cm^{-3}$) & $10^{-2}$ & $10^{-2}$ & $3\times10^{-2}$
& -- & $10^{-2}$ & $3\times10^{-4}$ \cr
$T$ ($K$) & $10^2$ & $10^2$ & $7\times10^3$ & $4\times10^{2}$ & $10^2$
& $10^2$ \cr
$\rho$ ($g$ $cm^{-3}$) & $1\times 10^{-22}$ & $1\times 10^{-22}$
& $6\times10^{-20}$ & $1\times 10^{-23}$ & $1\times 10^{-22}$
& $3\times10^{-24}$ \cr
\noalign{\vskip 3pt\hrule\vskip 10pt}
}
\endtable

\begintable*{7}
\caption{{\bf Table 7.} Values of the variables for $M=10^5M_\odot$ 
with $h=1.0$, $\Omega=0.1$ and $M_o=10^{15}M_\odot$.}
\halign{#\hfil & \quad \hfil#\hfil\quad &
         \hfil#\hfil\quad & \hfil#\hfil\quad &
         \hfil#\hfil\quad & \hfil#\hfil\quad & 
         \hfil#\hfil \cr

          & with all  & without    & without    & without        & without
& without \cr
Variable  & processes & $L_{{}_R}$ & $L_{{}_C}$ & $L_{{}_{H_2}}$ & $L_\alpha$  
& photon drag \cr
\noalign{\vskip 3pt\hrule\vskip 3pt}  
$z_{{}_{c}}$ & 24 & 24 & 220 & 5 & 24 & 13 \cr
$n_{{}_{H_2}}$ ($cm^{-3}$) & $0.3$ & $0.5$ & $10^2$ & -- & $0.3$
& $6\times10^{-2}$ \cr
$T$ ($K$) & $200$ & $200$ & $2\times10^3$ & $200$ & $200$ & $10^2$ \cr
$\rho$ ($g$ $cm^{-3}$) & $3\times10^{-21}$ & $3\times10^{-21}$
& $2\times10^{-18}$ & $3\times10^{-21}$ & $3\times10^{-21}$
& $6\times10^{-22}$ \cr
\noalign{\vskip 3pt\hrule\vskip 10pt}
}
\endtable

\begintable*{8}
\caption{{\bf Table 8} Values of the variables for $M=10^6M_\odot$ 
with $h=1.0$, $\Omega=0.1$ and $M_o=10^{15}M_\odot$.}
\halign{#\hfil & \quad \hfil#\hfil\quad &
         \hfil#\hfil\quad & \hfil#\hfil\quad &
         \hfil#\hfil\quad & \hfil#\hfil\quad & 
         \hfil#\hfil \cr

          & with all  & without    & without    & without & without
& without \cr
Variable  & processes & $L_{{}_R}$ & $L_{{}_C}$ & $L_{{}_{H_2}}$ & $L_\alpha$  
& photon drag \cr
\noalign{\vskip 3pt\hrule\vskip 3pt}  
$z_{{}_{c}}$ & 194 & 194 & 196 & 157 & 194 & 246 \cr
$n_{{}_{H_2}}$ ($cm^{-3}$) & $800$ & $800$ & $800$ & -- & $10^2$ & $10^2$ \cr
$T$ ($K$) & $800$ & $800$ & $10^3$ & $8\times10^3$ & $10^3$ & $1300$ \cr
$\rho$ ($g$ $cm^{-3}$) & $2\times10^{-18}$ & $2\times10^{-18}$
& $2\times10^{-18}$ & $1\times 10^{-18}$ & $2\times10^{-18}$
& $3\times10^{-18}$ \cr
\noalign{\vskip 3pt\hrule\vskip 3pt}
}
\endtable

In  Fig.8 we present the time evolution of the temperature,
pressure and molecular hydrogen density, with all physical processes 
included and without the Compton cooling-heating processes.
It is seen that the evolution of the above physical quantities are very 
different, even for the case in which the redshift collapse is not
significantly altered, namely, for clouds of $10^6M_\odot$.
In conclusion, the thermal history of the clouds studied is strongly
dependent on the Compton heating process.

If we disregard the cooling by molecular hydrogen, the mass 
$3.8\times10^4M_\odot$ is not significantly affected, 
there is no collapse in this case, as already seen there is no collapse 
even considering the molecular hydrogen cooling.
The collapse for the other masses occur for lower values of redshift. 
Without such an important cooling process the pressure is greater and
as a result there is a delay in the collapse of the cloud.
In Fig.9 we can see the different behavior of the temperature, pressure 
and molecular density when this cooling is disregarded. 

For the Lyman-$\alpha$ cooling the behavior of the collapse is
similar to the cooling due to the recombination.

If we disregard the photon-drag process we observed that the collapse is 
delayed for masses $8.5\times10^4M_\odot$ and $10^5M_\odot$ and the collapse 
occurs earlier for the mass $10^6M_\odot$. 

The photon drag works like a brake for the expansion and for the collapse.
If we have masses around the Jeans mass the behavior of these masses near the 
epoch of the collapse is strongly influenced by the photon drag. 

When we disregard the photon drag for masses smaller than $10^6M_\odot$ 
the collapse is delayed, due to the fact that the cloud stop
expanding latter, beginning as a result to collapse latter. This
occurs, also because these clouds have $M \leq M_{J}$ (the Jeans mass).

If we have masses larger than $10^6M_\odot$ (Jeans unstable masses) 
when we disconsider the photon drag, the cloud stop 
expanding earlier, and due to the fact that the pressure is not so important
for these cases, the cloud collapses in a $t\sim t_{ff}$ (free-fall time). 
As a result the collapses occur earlier, without such an effect.

As a general conclusion we see that the cooling-heating processes are less 
effective for the mass $10^6M_\odot$, in particular, concerning
the epoch of the collapse, due to the fact that the  
pressure of the cloud in this case does not affect significantly its 
evolution. By other hand, the thermal history of the cloud, as well as,
the amount of molecular hydrogen produced depend significantly on the
physical processes here considered.

An important point concerning the present study is that we take into account 
the appropriate physics and begin the calculations in the appropriate era, 
i.e., at the recombination era. These facts strongly influence the evolution
of the clouds as we have shown above.

\section{Conclusions}

In the present work we studied the evolution of density perturbations
with masses within the range $10 - 10^6M_\odot$ taking into
account a series of physical processes which are present during and after the 
recombination era. In order to perform such a study a spherical Lagrangian 
hydrodynamical code was written, with which it was possible
to follow the spatial and time evolution of the density perturbations.

Our main conclusions are:

\beginlist
\item a) For clouds with $M<10^4M_\odot$, no collapse occurs. In this case,  
the density perturbations are strongly dissipated by the photon-drag.
Only a residual density perturbation remains; this conclusion
holds for the two density perturbations spectra and for the different
combinations of h and $\Omega$ adopted;

\item b) Clouds of $10^5$ and $10^6 M_\odot$ present a partial collapse;

\item c) The minimal mass which collapses does not change significantly 
when we alter $h,\ \Omega$ and $M_0$, namely,
$M_{min} \sim 10^4 - 10^5 M_\odot$;

\item d) When we disregard, for example, the Compton cooling-heating, the 
collapse for masses $>8.5\times10^4M_\odot$ occurs earlier;

\item e) The photon-drag is a very important process, the collapse redshift, 
for example, is changed significantly when we disregard it, but it 
works differently for higher or smaller masses. For the smaller masses 
the photon-drag dissipates their density perturbations and for the 
higher masses it causes a delay in the collapse of the clouds.
\endlist

\section*{Acknowledgements}

The calculations were performed on a HP--Apollo 9000 (purchased by the
Brazilian agency FAPESP) and on the CRAY computers EL98 and J90 (CCE/USP).   

We would like to thank Dr David Weinberg, the referee, for his helpful 
suggestions and comments that greatly improved the present version of our
paper. We would also like to thank the Brazilian agencies CAPES (SRO) and 
CNPq (ODM, JCNA and RO) for financial support.

\vskip 1truecm

\noindent{\bf FIGURE CAPTIONS}

\vskip 1truecm

  {\bf Fig.1 -- } The evolution of the density contrast ($\delta$) with
time. The solid line is the internal shell, the short dashed line is the
middle shell and the long dashed line is the external shell. Time is in years 
(the input parameters are $M_o=10^{15}M_\odot$, $h=1.0$, $\Omega=0.1$).

  {\bf Fig.2 -- } The evolution of the mass density $\rho$ ($g \ cm^{-3}$)
with time. The captions and input parameters for the curves are the same
as in Fig.1.

  {\bf Fig.3 -- } The mass density versus the radial coordinate (in $pc$)
of the cloud. The solid line is the initial time ($t_i$), the short dashed
line, long dashed line and very long dashed line are intermediate times
(respectively $t_2$, $t_3$ and $t_4$). The dashed-pointed line is 
the final time ($t_f$). The input parameters are the same as in Fig.1. 

  {\bf Fig.4 -- } The evolution of the temperature (T in $K$), pressure (P
in $dyn \ cm^{-2}$) and density of molecules ($n_{H_2}$ in $molecules
\ cm^{-3}$) with time. The captions and input parameters for the curves
are the same as in Fig.1.

  {\bf Fig.5 -- } The evolution of the density of molecules ($n_{H_2}$)
versus temperature. The captions and input parameters for the curves are the
same as in Fig.1.

  {\bf Fig.6 -- } The evolution of the radial coordinate of the clouds with
time. The captions and input parameters for the curves are the same as in
Fig.1. 

  {\bf Fig.7 -- } The evolution of the radial coordinate of the mass 
$M=3.8\times10^4M_\odot$ with time. The captions and input parameters for
the curves are the same as in Fig.1. 

{\bf Fig.8 -- }{The evolution of the temperature (T), pressure (P) and
density of molecules ($n_{H_2}$) with time for masses $M=3.8\times
10^4M_\odot$, $M=8.5\times10^4M_\odot$, $M=10^5M_\odot$ and $M=10^6M_\odot$
with all physical processes included and when we disregard the Compton
cooling-heating. The captions and input parameters for the curves are the
same as in Fig.1. 

  {\bf Fig.9 -- } The evolution of the temperature (T), pressure (P) and
density of molecules ($n_{H_2}$) for masses $M=3.8\times10^4M_\odot$,
$M=8.5\times10^4M_\odot$, $M=10^5M_\odot$ and $M=10^6M_\odot$ with all
physical processes included and when we disregard the $H_2$ cooling. The
captions and input parameters for the curves are the same as in Fig.1. 

\bye

%% file: mn.tex
%
%
%
%

\catcode `\@=11 

\def\@version{1.6}
\def\@verdate{18th September 1995}

%
%


\newif\ifprod@font

\ifx\@typeface\undefined
  \def\@typeface{Comp. Modern}\prod@fontfalse
\else
  \prod@fonttrue 
\fi

\def\newfam{\alloc@8\fam\chardef\sixt@@n} 

\ifprod@font
\font\fiverm=mtr10 at 5pt
\font\fivebf=mtbx10 at 5pt
\font\fiveit=mtti10 at 5pt
\font\fivesl=mtsl10 at 5pt
\font\fivett=cmtt8 at 5pt     \hyphenchar\fivett=-1
\font\fivecsc=mtcsc10 at 5pt
\font\fivesf=mtss10 at 5pt
\font\fivei=mtmi10 at 5pt      \skewchar\fivei='177
\font\fivesy=mtsy10 at 5pt     \skewchar\fivesy='60

\font\sixrm=mtr10 at 6pt
\font\sixbf=mtbx10 at 6pt
\font\sixit=mtti10 at 6pt
\font\sixsl=mtsl10 at 6pt
\font\sixtt=cmtt8 at 6pt      \hyphenchar\sixtt=-1
\font\sixcsc=mtcsc10 at 6pt
\font\sixsf=mtss10 at 6pt
\font\sixi=mtmi10 at 6pt       \skewchar\sixi='177
\font\sixsy=mtsy10 at 6pt      \skewchar\sixsy='60

\font\sevenrm=mtr10 at 7pt
\font\sevenbf=mtbx10 at 7pt
\font\sevenit=mtti10 at 7pt
\font\sevensl=mtsl10 at 7pt
\font\seventt=cmtt8 at 7pt     \hyphenchar\seventt=-1
\font\sevencsc=mtcsc10 at 7pt
\font\sevensf=mtss10 at 7pt
\font\seveni=mtmi10 at 7pt      \skewchar\seveni='177
\font\sevensy=mtsy10 at 7pt     \skewchar\sevensy='60

\font\eightrm=mtr10 at 8pt
\font\eightbf=mtbx10 at 8pt
\font\eightit=mtti10 at 8pt
\font\eighti=mtmi10 at 8pt      \skewchar\eighti='177
\font\eightsy=mtsy10 at 8pt     \skewchar\eightsy='60
\font\eightsl=mtsl10 at 8pt
\font\eighttt=cmtt8             \hyphenchar\eighttt=-1
\font\eightcsc=mtcsc10 at 8pt
\font\eightsf=mtss10 at 8pt

\font\ninerm=mtr10 at 9pt
\font\ninebf=mtbx10 at 9pt
\font\nineit=mtti10 at 9pt
\font\ninei=mtmi10 at 9pt      \skewchar\ninei='177
\font\ninesy=mtsy10 at 9pt     \skewchar\ninesy='60
\font\ninesl=mtsl10 at 9pt
\font\ninett=cmtt9             \hyphenchar\ninett=-1
\font\ninecsc=mtcsc10 at 9pt
\font\ninesf=mtss10 at 9pt

\font\tenrm=mtr10
\font\tenbf=mtbx10
\font\tenit=mtti10
\font\teni=mtmi10		\skewchar\teni='177
\font\tensy=mtsy10		\skewchar\tensy='60
\font\tenex=cmex10
\font\tensl=mtsl10
\font\tentt=cmtt10		\hyphenchar\tentt=-1
\font\tencsc=mtcsc10
\font\tensf=mtss10

\font\elevenrm=mtr10 at 11pt
\font\elevenbf=mtbx10 at 11pt
\font\elevenit=mtti10 at 11pt
\font\eleveni=mtmi10 at 11pt      \skewchar\eleveni='177
\font\elevensy=mtsy10 at 11pt     \skewchar\elevensy='60
\font\elevensl=mtsl10 at 11pt
\font\eleventt=cmtt10 at 11pt     \hyphenchar\eleventt=-1
\font\elevencsc=mtcsc10 at 11pt
\font\elevensf=mtss10 at 11pt

\font\twelverm=mtr10 at 12pt
\font\twelvebf=mtbx10 at 12pt
\font\twelveit=mtti10 at 12pt
\font\twelvesl=mtsl10 at 12pt
\font\twelvett=cmtt12             \hyphenchar\twelvett=-1
\font\twelvecsc=mtcsc10 at 12pt
\font\twelvesf=mtss10 at 12pt
\font\twelvei=mtmi10 at 12pt      \skewchar\twelvei='177
\font\twelvesy=mtsy10 at 12pt     \skewchar\twelvesy='60

\font\fourteenrm=mtr10 at 14pt
\font\fourteenbf=mtbx10 at 14pt
\font\fourteenit=mtti10 at 14pt
\font\fourteeni=mtmi10 at 14pt      \skewchar\fourteeni='177
\font\fourteensy=mtsy10 at 14pt     \skewchar\fourteensy='60
\font\fourteensl=mtsl10 at 14pt
\font\fourteentt=cmtt12 at 14pt     \hyphenchar\fourteentt=-1
\font\fourteencsc=mtcsc10 at 14pt
\font\fourteensf=mtss10 at 14pt

\font\seventeenrm=mtr10 at 17pt
\font\seventeenbf=mtbx10 at 17pt
\font\seventeenit=mtti10 at 17pt
\font\seventeeni=mtmi10 at 17pt      \skewchar\seventeeni='177
\font\seventeensy=mtsy10 at 17pt     \skewchar\seventeensy='60
\font\seventeensl=mtsl10 at 17pt
\font\seventeentt=cmtt12 at 17pt     \hyphenchar\seventeentt=-1
\font\seventeencsc=mtcsc10 at 17pt
\font\seventeensf=mtss10 at 17pt
\else
\font\fiverm=cmr5
\font\fivei=cmmi5             \skewchar\fivei='177
\font\fivesy=cmsy5            \skewchar\fivesy='60
\font\fivebf=cmbx5

\font\sixrm=cmr6
\font\sixi=cmmi6             \skewchar\sixi='177
\font\sixsy=cmsy6            \skewchar\sixsy='60
\font\sixbf=cmbx6

\font\sevenrm=cmr7
\font\sevenit=cmti7
\font\seveni=cmmi7             \skewchar\seveni='177
\font\sevensy=cmsy7            \skewchar\sevensy='60
\font\sevenbf=cmbx7

\font\eightrm=cmr8
\font\eightbf=cmbx8
\font\eightit=cmti8
\font\eighti=cmmi8			\skewchar\eighti='177
\font\eightsy=cmsy8			\skewchar\eightsy='60
\font\eightsl=cmsl8
\font\eighttt=cmtt8			\hyphenchar\eighttt=-1
\font\eightcsc=cmcsc10 at 8pt
\font\eightsf=cmss8

\font\ninerm=cmr9
\font\ninebf=cmbx9
\font\nineit=cmti9
\font\ninei=cmmi9			\skewchar\ninei='177
\font\ninesy=cmsy9			\skewchar\ninesy='60
\font\ninesl=cmsl9
\font\ninett=cmtt9			\hyphenchar\ninett=-1
\font\ninecsc=cmcsc10 at 9pt
\font\ninesf=cmss9

\font\tenrm=cmr10
\font\tenbf=cmbx10
\font\tenit=cmti10
\font\teni=cmmi10		\skewchar\teni='177
\font\tensy=cmsy10		\skewchar\tensy='60
\font\tenex=cmex10
\font\tensl=cmsl10
\font\tentt=cmtt10		\hyphenchar\tentt=-1
\font\tencsc=cmcsc10
\font\tensf=cmss10

\font\elevenrm=cmr10 scaled \magstephalf
\font\elevenbf=cmbx10 scaled \magstephalf
\font\elevenit=cmti10 scaled \magstephalf
\font\eleveni=cmmi10 scaled \magstephalf	\skewchar\eleveni='177
\font\elevensy=cmsy10 scaled \magstephalf	\skewchar\elevensy='60
\font\elevensl=cmsl10 scaled \magstephalf
\font\eleventt=cmtt10 scaled \magstephalf	\hyphenchar\eleventt=-1
\font\elevencsc=cmcsc10 scaled \magstephalf
\font\elevensf=cmss10 scaled \magstephalf

\font\twelverm=cmr10 scaled \magstep1
\font\twelvebf=cmbx10 scaled \magstep1
\font\twelvei=cmmi10 scaled \magstep1      \skewchar\twelvei='177
\font\twelvesy=cmsy10 scaled \magstep1     \skewchar\twelvesy='60

\font\fourteenrm=cmr10 scaled \magstep2
\font\fourteenbf=cmbx10 scaled \magstep2
\font\fourteenit=cmti10 scaled \magstep2
\font\fourteeni=cmmi10 scaled \magstep2		\skewchar\fourteeni='177
\font\fourteensy=cmsy10 scaled \magstep2	\skewchar\fourteensy='60
\font\fourteensl=cmsl10 scaled \magstep2
\font\fourteentt=cmtt10 scaled \magstep2	\hyphenchar\fourteentt=-1
\font\fourteencsc=cmcsc10 scaled \magstep2
\font\fourteensf=cmss10 scaled \magstep2

\font\seventeenrm=cmr10 scaled \magstep3
\font\seventeenbf=cmbx10 scaled \magstep3
\font\seventeenit=cmti10 scaled \magstep3
\font\seventeeni=cmmi10 scaled \magstep3	\skewchar\seventeeni='177
\font\seventeensy=cmsy10 scaled \magstep3	\skewchar\seventeensy='60
\font\seventeensl=cmsl10 scaled \magstep3
\font\seventeentt=cmtt10 scaled \magstep3	\hyphenchar\seventeentt=-1
\font\seventeencsc=cmcsc10 scaled \magstep3
\font\seventeensf=cmss10 scaled \magstep3
\fi

\def\hexnumber#1{\ifcase#1 0\or1\or2\or3\or4\or5\or6\or7\or8\or9\or
  A\or B\or C\or D\or E\or F\fi}

\def\makestrut{%
  \setbox\strutbox=\hbox{%
    \vrule height.7\baselineskip depth.3\baselineskip width \z@}%
}

\def\baselinestretch{1}
\newskip\tmp@bls

\def\b@ls#1{
  \tmp@bls=#1\relax
  \baselineskip=#1\relax\makestrut
  \normalbaselineskip=\baselinestretch\tmp@bls
  \normalbaselines
}

\def\nostb@ls#1{
  \normalbaselineskip=#1\relax
  \normalbaselines
  \makestrut
}

%

\newfam\scfam  
\newfam\sffam  

\def\mit{\fam\@ne}
\def\cal{\fam\tw@}
\def\em{\ifdim\fontdimen1\font>\z@ \rm\else\it\fi}

\textfont3=\tenex
\scriptfont3=\tenex
\scriptscriptfont3=\tenex

\setbox0=\hbox{\tenex B} \p@renwd=\wd0 

\def\eightpoint{
  \def\rm{\fam0\eightrm}%
  \textfont0=\eightrm \scriptfont0=\sixrm \scriptscriptfont0=\fiverm%
  \textfont1=\eighti  \scriptfont1=\sixi  \scriptscriptfont1=\fivei%
  \textfont2=\eightsy \scriptfont2=\sixsy \scriptscriptfont2=\fivesy%
  \textfont\itfam=\eightit\def\it{\fam\itfam\eightit}%
  \ifprod@font
    \scriptfont\itfam=\sixit
      \scriptscriptfont\itfam=\fiveit
  \else
    \scriptfont\itfam=\eightit
      \scriptscriptfont\itfam=\eightit
  \fi
  \textfont\bffam=\eightbf%
    \scriptfont\bffam=\sixbf%
      \scriptscriptfont\bffam=\fivebf%
  \def\bf{\fam\bffam\eightbf}%
  \textfont\slfam=\eightsl\def\sl{\fam\slfam\eightsl}%
  \ifprod@font
    \scriptfont\slfam=\sixsl
      \scriptscriptfont\slfam=\fivesl
  \else
    \scriptfont\slfam=\eightsl
      \scriptscriptfont\slfam=\eightsl
  \fi
  \textfont\ttfam=\eighttt\def\tt{\fam\ttfam\eighttt}%
  \ifprod@font
    \scriptfont\ttfam=\sixtt
      \scriptscriptfont\ttfam=\fivett
  \else
    \scriptfont\ttfam=\eighttt
      \scriptscriptfont\ttfam=\eighttt
  \fi
  \textfont\scfam=\eightcsc\def\sc{\fam\scfam\eightcsc}%
  \ifprod@font
    \scriptfont\scfam=\sixcsc
      \scriptscriptfont\scfam=\fivecsc
  \else
    \scriptfont\scfam=\eightcsc
      \scriptscriptfont\scfam=\eightcsc
  \fi
  \textfont\sffam=\eightsf\def\sf{\fam\sffam\eightsf}%
  \ifprod@font
    \scriptfont\sffam=\sixsf
      \scriptscriptfont\sffam=\fivesf
  \else
    \scriptfont\sffam=\eightsf
      \scriptscriptfont\sffam=\eightsf
  \fi
  \def\oldstyle{\fam\@ne\eighti}%
  \b@ls{10pt}\rm\@viiipt%
}
\def\@viiipt{}

\def\ninepoint{
  \def\rm{\fam0\ninerm}%
  \textfont0=\ninerm \scriptfont0=\sixrm \scriptscriptfont0=\fiverm%
  \textfont1=\ninei  \scriptfont1=\sixi  \scriptscriptfont1=\fivei%
  \textfont2=\ninesy \scriptfont2=\sixsy \scriptscriptfont2=\fivesy%
  \textfont\itfam=\nineit\def\it{\fam\itfam\nineit}%
  \ifprod@font
    \scriptfont\itfam=\sixit
      \scriptscriptfont\itfam=\fiveit
  \else
    \scriptfont\itfam=\nineit
      \scriptscriptfont\itfam=\nineit
  \fi
  \textfont\bffam=\ninebf%
    \scriptfont\bffam=\sixbf%
      \scriptscriptfont\bffam=\fivebf%
  \def\bf{\fam\bffam\ninebf}%
  \textfont\slfam=\ninesl\def\sl{\fam\slfam\ninesl}%
  \ifprod@font
    \scriptfont\slfam=\sixsl
      \scriptscriptfont\slfam=\fivesl
  \else
    \scriptfont\slfam=\ninesl
      \scriptscriptfont\slfam=\ninesl
  \fi
  \textfont\ttfam=\ninett\def\tt{\fam\ttfam\ninett}%
  \ifprod@font
    \scriptfont\ttfam=\sixtt
      \scriptscriptfont\ttfam=\fivett
  \else
    \scriptfont\ttfam=\ninett
      \scriptscriptfont\ttfam=\ninett
  \fi
  \textfont\scfam=\ninecsc\def\sc{\fam\scfam\ninecsc}%
  \ifprod@font
    \scriptfont\scfam=\sixcsc
      \scriptscriptfont\scfam=\fivecsc
  \else
    \scriptfont\scfam=\ninecsc
      \scriptscriptfont\scfam=\ninecsc
  \fi
  \textfont\sffam=\ninesf\def\sf{\fam\sffam\ninesf}%
  \ifprod@font
    \scriptfont\sffam=\sixsf
      \scriptscriptfont\sffam=\fivesf
  \else
    \scriptfont\sffam=\ninesf
      \scriptscriptfont\sffam=\ninesf
  \fi
  \def\oldstyle{\fam\@ne\ninei}%
  \b@ls{\TextLeading plus \Feathering}\rm\@ixpt%
}
\def\@ixpt{}

\def\tenpoint{
  \def\rm{\fam0\tenrm}%
  \textfont0=\tenrm \scriptfont0=\sevenrm \scriptscriptfont0=\fiverm%
  \textfont1=\teni  \scriptfont1=\seveni  \scriptscriptfont1=\fivei%
  \textfont2=\tensy \scriptfont2=\sevensy \scriptscriptfont2=\fivesy%
  \textfont\itfam=\tenit\def\it{\fam\itfam\tenit}%
  \ifprod@font
    \scriptfont\itfam=\sevenit
      \scriptscriptfont\itfam=\fiveit
  \else
    \scriptfont\itfam=\tenit
      \scriptscriptfont\itfam=\tenit
  \fi
  \textfont\bffam=\tenbf%
    \scriptfont\bffam=\sevenbf%
      \scriptscriptfont\bffam=\fivebf%
  \def\bf{\fam\bffam\tenbf}%
  \textfont\slfam=\tensl\def\sl{\fam\slfam\tensl}%
  \ifprod@font
    \scriptfont\slfam=\sevensl
      \scriptscriptfont\slfam=\fivesl
  \else
    \scriptfont\slfam=\tensl
      \scriptscriptfont\slfam=\tensl
  \fi
  \textfont\ttfam=\tentt\def\tt{\fam\ttfam\tentt}%
  \ifprod@font
    \scriptfont\ttfam=\seventt
      \scriptscriptfont\ttfam=\fivett
  \else
    \scriptfont\ttfam=\tentt
      \scriptscriptfont\ttfam=\tentt
  \fi
  \textfont\scfam=\tencsc\def\sc{\fam\scfam\tencsc}%
  \ifprod@font
    \scriptfont\scfam=\sevencsc
      \scriptscriptfont\scfam=\fivecsc
  \else
    \scriptfont\scfam=\tencsc
      \scriptscriptfont\scfam=\tencsc
  \fi
  \textfont\sffam=\tensf\def\sf{\fam\sffam\tensf}%
  \ifprod@font
    \scriptfont\sffam=\sevensf
      \scriptscriptfont\sffam=\fivesf
  \else
    \scriptfont\sffam=\tensf
      \scriptscriptfont\sffam=\tensf
  \fi
  \def\oldstyle{\fam\@ne\teni}%
  \b@ls{11pt}\rm\@xpt%
}
\def\@xpt{}

\def\elevenpoint{
  \def\rm{\fam0\elevenrm}%
  \textfont0=\elevenrm \scriptfont0=\eightrm \scriptscriptfont0=\sixrm%
  \textfont1=\eleveni  \scriptfont1=\eighti  \scriptscriptfont1=\sixi%
  \textfont2=\elevensy \scriptfont2=\eightsy \scriptscriptfont2=\sixsy%
  \textfont\itfam=\elevenit\def\it{\fam\itfam\elevenit}%
  \ifprod@font
    \scriptfont\itfam=\eightit
      \scriptscriptfont\itfam=\sixit
  \else
    \scriptfont\itfam=\elevenit
      \scriptscriptfont\itfam=\elevenit
  \fi
  \textfont\bffam=\elevenbf%
    \scriptfont\bffam=\eightbf%
      \scriptscriptfont\bffam=\sixbf%
  \def\bf{\fam\bffam\elevenbf}%
  \textfont\slfam=\elevensl\def\sl{\fam\slfam\elevensl}%
  \ifprod@font
    \scriptfont\slfam=\eightsl
      \scriptscriptfont\slfam=\sixsl
  \else
    \scriptfont\slfam=\elevensl
      \scriptscriptfont\slfam=\elevensl
  \fi
  \textfont\ttfam=\eleventt\def\tt{\fam\ttfam\eleventt}%
  \ifprod@font
    \scriptfont\ttfam=\eighttt
      \scriptscriptfont\ttfam=\sixtt
  \else
    \scriptfont\ttfam=\eleventt
      \scriptscriptfont\ttfam=\eleventt
  \fi
  \textfont\scfam=\elevencsc\def\sc{\fam\scfam\elevencsc}%
  \ifprod@font
    \scriptfont\scfam=\eightcsc
      \scriptscriptfont\scfam=\sixcsc
  \else
    \scriptfont\scfam=\elevencsc
      \scriptscriptfont\scfam=\elevencsc
  \fi
  \textfont\sffam=\elevensf\def\sf{\fam\sffam\elevensf}%
  \ifprod@font
    \scriptfont\sffam=\eightsf
      \scriptscriptfont\sffam=\sixsf
  \else
    \scriptfont\sffam=\elevensf
      \scriptscriptfont\sffam=\elevensf
  \fi
  \def\oldstyle{\fam\@ne\eleveni}%
  \b@ls{13pt}\rm\@xipt%
}
\def\@xipt{}

\def\fourteenpoint{
  \def\rm{\fam0\fourteenrm}%
  \textfont0\fourteenrm  \scriptfont0\tenrm  \scriptscriptfont0\sevenrm%
  \textfont1\fourteeni   \scriptfont1\teni   \scriptscriptfont1\seveni%
  \textfont2\fourteensy  \scriptfont2\tensy  \scriptscriptfont2\sevensy%
  \textfont\itfam=\fourteenit\def\it{\fam\itfam\fourteenit}%
  \ifprod@font
    \scriptfont\itfam=\tenit
      \scriptscriptfont\itfam=\sevenit
  \else
    \scriptfont\itfam=\fourteenit
      \scriptscriptfont\itfam=\fourteenit
  \fi
  \textfont\bffam=\fourteenbf%
    \scriptfont\bffam=\tenbf%
      \scriptscriptfont\bffam=\sevenbf%
  \def\bf{\fam\bffam\fourteenbf}%
  \textfont\slfam=\fourteensl\def\sl{\fam\slfam\fourteensl}%
  \ifprod@font
    \scriptfont\slfam=\tensl
      \scriptscriptfont\slfam=\sevensl
  \else
    \scriptfont\slfam=\fourteensl
      \scriptscriptfont\slfam=\fourteensl
  \fi
  \textfont\ttfam=\fourteentt\def\tt{\fam\ttfam\fourteentt}%
  \ifprod@font
    \scriptfont\ttfam=\tentt
      \scriptscriptfont\ttfam=\seventt
  \else
    \scriptfont\ttfam=\fourteentt
      \scriptscriptfont\ttfam=\fourteentt
  \fi
  \textfont\scfam=\fourteencsc\def\sc{\fam\scfam\fourteencsc}%
  \ifprod@font
    \scriptfont\scfam=\tencsc
      \scriptscriptfont\scfam=\sevencsc
  \else
    \scriptfont\scfam=\fourteencsc
      \scriptscriptfont\scfam=\fourteencsc
  \fi
  \textfont\sffam=\fourteensf\def\sf{\fam\sffam\fourteensf}%
  \ifprod@font
    \scriptfont\sffam=\tensf
      \scriptscriptfont\sffam=\sevensf
  \else
    \scriptfont\sffam=\fourteensf
      \scriptscriptfont\sffam=\fourteensf
  \fi
  \def\oldstyle{\fam\@ne\fourteeni}%
  \b@ls{17pt}\rm\@xivpt%
}
\def\@xivpt{}

\def\seventeenpoint{
  \def\rm{\fam0\seventeenrm}%
  \textfont0\seventeenrm  \scriptfont0\twelverm  \scriptscriptfont0\tenrm%
  \textfont1\seventeeni   \scriptfont1\twelvei   \scriptscriptfont1\teni%
  \textfont2\seventeensy  \scriptfont2\twelvesy  \scriptscriptfont2\tensy%
  \textfont\itfam=\seventeenit\def\it{\fam\itfam\seventeenit}%
  \ifprod@font
    \scriptfont\itfam=\twelveit
      \scriptscriptfont\itfam=\tenit
  \else
    \scriptfont\itfam=\seventeenit
      \scriptscriptfont\itfam=\seventeenit
  \fi
  \textfont\bffam=\seventeenbf%
    \scriptfont\bffam=\twelvebf%
      \scriptscriptfont\bffam=\tenbf%
  \def\bf{\fam\bffam\seventeenbf}%
  \textfont\slfam=\seventeensl\def\sl{\fam\slfam\seventeensl}%
  \ifprod@font
    \scriptfont\slfam=\twelvesl
      \scriptscriptfont\slfam=\tensl
  \else
    \scriptfont\slfam=\seventeensl
      \scriptscriptfont\slfam=\seventeensl
  \fi
  \textfont\ttfam=\seventeentt\def\tt{\fam\ttfam\seventeentt}%
  \ifprod@font
    \scriptfont\ttfam=\twelvett
      \scriptscriptfont\ttfam=\tentt
  \else
    \scriptfont\ttfam=\seventeentt
      \scriptscriptfont\ttfam=\seventeentt
  \fi
  \textfont\scfam=\seventeencsc\def\sc{\fam\scfam\seventeencsc}%
  \ifprod@font
    \scriptfont\scfam=\twelvecsc
      \scriptscriptfont\scfam=\tencsc
  \else
    \scriptfont\scfam=\seventeencsc
      \scriptscriptfont\scfam=\seventeencsc
  \fi
  \textfont\sffam=\seventeensf\def\sf{\fam\sffam\seventeensf}%
  \ifprod@font
    \scriptfont\sffam=\twelvesf
      \scriptscriptfont\sffam=\tensf
  \else
    \scriptfont\sffam=\seventeensf
      \scriptscriptfont\sffam=\seventeensf
  \fi
  \def\oldstyle{\fam\@ne\seventeeni}%
  \b@ls{20pt}\rm\@xviipt%
}
\def\@xviipt{}

\lineskip=1pt      \normallineskip=\lineskip
\lineskiplimit=\z@ \normallineskiplimit=\lineskiplimit


\def\loadboldmathnames{%
  \def\balpha{{\bmath{\alpha}}}%
  \def\bbeta{{\bmath{\beta}}}%
  \def\bgamma{{\bmath{\gamma}}}%
  \def\bdelta{{\bmath{\delta}}}%
  \def\bepsilon{{\bmath{\epsilon}}}%
  \def\bzeta{{\bmath{\zeta}}}%
  \def\boldeta{{\bmath{\eta}}}%
  \def\btheta{{\bmath{\theta}}}%
  \def\biota{{\bmath{\iota}}}%
  \def\bkappa{{\bmath{\kappa}}}%
  \def\blambda{{\bmath{\lambda}}}%
  \def\bmu{{\bmath{\mu}}}%
  \def\bnu{{\bmath{\nu}}}%
  \def\bxi{{\bmath{\xi}}}%
  \def\bpi{{\bmath{\pi}}}%
  \def\brho{{\bmath{\rho}}}%
  \def\bsigma{{\bmath{\sigma}}}%
  \def\btau{{\bmath{\tau}}}%
  \def\bupsilon{{\bmath{\upsilon}}}%
  \def\bphi{{\bmath{\phi}}}%
  \def\bchi{{\bmath{\chi}}}%
  \def\bpsi{{\bmath{\psi}}}%
  \def\bomega{{\bmath{\omega}}}%
  \def\bvarepsilon{{\bmath{\varepsilon}}}%
  \def\bvartheta{{\bmath{\vartheta}}}%
  \def\bvarpi{{\bmath{\varpi}}}%
  \def\bvarrho{{\bmath{\varrho}}}%
  \def\bvarsigma{{\bmath{\varsigma}}}%
  \def\bvarphi{{\bmath{\varphi}}}%
  \def\baleph{{\bmath{\aleph}}}%
  \def\bimath{{\bmath{\imath}}}%
  \def\bjmath{{\bmath{\jmath}}}%
  \def\bell{{\bmath{\ell}}}%
  \def\bwp{{\bmath{\wp}}}%
  \def\bRe{{\bmath{\Re}}}%
  \def\bIm{{\bmath{\Im}}}%
  \def\bpartial{{\bmath{\partial}}}%
  \def\binfty{{\bmath{\infty}}}%
  \def\bprime{{\bmath{\prime}}}%
  \def\bemptyset{{\bmath{\emptyset}}}%
  \def\bnabla{{\bmath{\nabla}}}%
  \def\btop{{\bmath{\top}}}%
  \def\bbot{{\bmath{\bot}}}%
  \def\btriangle{{\bmath{\triangle}}}%
  \def\bforall{{\bmath{\forall}}}%
  \def\bexists{{\bmath{\exists}}}%
  \def\bneg{{\bmath{\neg}}}%
  \def\bflat{{\bmath{\flat}}}%
  \def\bnatural{{\bmath{\natural}}}%
  \def\bsharp{{\bmath{\sharp}}}%
  \def\bclubsuit{{\bmath{\clubsuit}}}%
  \def\bdiamondsuit{{\bmath{\diamondsuit}}}%
  \def\bheartsuit{{\bmath{\heartsuit}}}%
  \def\bspadesuit{{\bmath{\spadesuit}}}%
  \def\bsmallint{{\bmath{\smallint}}}%
  \def\btriangleleft{{\bmath{\triangleleft}}}%
  \def\btriangleright{{\bmath{\triangleright}}}%
  \def\bbigtriangleup{{\bmath{\bigtriangleup}}}%
  \def\bbigtriangledown{{\bmath{\bigtriangledown}}}%
  \def\bwedge{{\bmath{\wedge}}}%
  \def\bvee{{\bmath{\vee}}}%
  \def\bcap{{\bmath{\cap}}}%
  \def\bcup{{\bmath{\cup}}}%
  \def\bddagger{{\bmath{\ddagger}}}%
  \def\bdagger{{\bmath{\dagger}}}%
  \def\bsqcap{{\bmath{\sqcap}}}%
  \def\bsqcup{{\bmath{\sqcup}}}%
  \def\buplus{{\bmath{\uplus}}}%
  \def\bamalg{{\bmath{\amalg}}}%
  \def\bdiamond{{\bmath{\diamond}}}%
  \def\bbullet{{\bmath{\bullet}}}%
  \def\bwr{{\bmath{\wr}}}%
  \def\bdiv{{\bmath{\div}}}%
  \def\bodot{{\bmath{\odot}}}%
  \def\boslash{{\bmath{\oslash}}}%
  \def\botimes{{\bmath{\otimes}}}%
  \def\bominus{{\bmath{\ominus}}}%
  \def\boplus{{\bmath{\oplus}}}%
  \def\bmp{{\bmath{\mp}}}%
  \def\bpm{{\bmath{\pm}}}%
  \def\bcirc{{\bmath{\circ}}}%
  \def\bbigcirc{{\bmath{\bigcirc}}}%
  \def\bsetminus{{\bmath{\setminus}}}%
  \def\bcdot{{\bmath{\cdot}}}%
  \def\bast{{\bmath{\ast}}}%
  \def\btimes{{\bmath{\times}}}%
  \def\bstar{{\bmath{\star}}}%
  \def\bpropto{{\bmath{\propto}}}%
  \def\bsqsubseteq{{\bmath{\sqsubseteq}}}%
  \def\bsqsupseteq{{\bmath{\sqsupseteq}}}%
  \def\bparallel{{\bmath{\parallel}}}%
  \def\bmid{{\bmath{\mid}}}%
  \def\bdashv{{\bmath{\dashv}}}%
  \def\bvdash{{\bmath{\vdash}}}%
  \def\bnearrow{{\bmath{\nearrow}}}%
  \def\bsearrow{{\bmath{\searrow}}}%
  \def\bnwarrow{{\bmath{\nwarrow}}}%
  \def\bswarrow{{\bmath{\swarrow}}}%
  \def\bLeftrightarrow{{\bmath{\Leftrightarrow}}}%
  \def\bLeftarrow{{\bmath{\Leftarrow}}}%
  \def\bRightarrow{{\bmath{\Rightarrow}}}%
  \def\bleq{{\bmath{\leq}}}%
  \def\bgeq{{\bmath{\geq}}}%
  \def\bsucc{{\bmath{\succ}}}%
  \def\bprec{{\bmath{\prec}}}%
  \def\bapprox{{\bmath{\approx}}}%
  \def\bsucceq{{\bmath{\succeq}}}%
  \def\bpreceq{{\bmath{\preceq}}}%
  \def\bsupset{{\bmath{\supset}}}%
  \def\bsubset{{\bmath{\subset}}}%
  \def\bsupseteq{{\bmath{\supseteq}}}%
  \def\bsubseteq{{\bmath{\subseteq}}}%
  \def\bin{{\bmath{\in}}}%
  \def\bni{{\bmath{\ni}}}%
  \def\bgg{{\bmath{\gg}}}%
  \def\bll{{\bmath{\ll}}}%
  \def\bnot{{\bmath{\not}}}%
  \def\bleftrightarrow{{\bmath{\leftrightarrow}}}%
  \def\bleftarrow{{\bmath{\leftarrow}}}%
  \def\brightarrow{{\bmath{\rightarrow}}}%
  \def\bmapstochar{{\bmath{\mapstochar}}}%
  \def\bsim{{\bmath{\sim}}}%
  \def\bsimeq{{\bmath{\simeq}}}%
  \def\bperp{{\bmath{\perp}}}%
  \def\bequiv{{\bmath{\equiv}}}%
  \def\basymp{{\bmath{\asymp}}}%
  \def\bsmile{{\bmath{\smile}}}%
  \def\bfrown{{\bmath{\frown}}}%
  \def\bleftharpoonup{{\bmath{\leftharpoonup}}}%
  \def\bleftharpoondown{{\bmath{\leftharpoondown}}}%
  \def\brightharpoonup{{\bmath{\rightharpoonup}}}%
  \def\brightharpoondown{{\bmath{\rightharpoondown}}}%
  \def\blhook{{\bmath{\lhook}}}%
  \def\brhook{{\bmath{\rhook}}}%
  \def\bldotp{{\bmath{\ldotp}}}%
  \def\bcdotp{{\bmath{\cdotp}}}%
}

\def\,{\relax\ifmmode \mskip\thinmuskip\else \thinspace\fi}
\let\protect=\relax

\long\def\@ifundefined#1#2#3{\expandafter\ifx\csname
  #1\endcsname\relax#2\else#3\fi}




\newtoks\math@groups \math@groups={}
\def\addtom@thgroup#1#2{#1\expandafter{\the#1#2}} 



\def\addtosizeh@ok#1#2#3#4{%
  \expandafter\def\csname @#1pt\endcsname{%
    \def\s@ze{#2}\def\ss@ze{#3}\def\sss@ze{#4}\the\math@groups%
  }%
}



\let\resetsizehook=\addtosizeh@ok


\ifprod@font
  \addtosizeh@ok{viii} {8} {6}  {5}
  \addtosizeh@ok{ix}   {9} {6}  {5}
  \addtosizeh@ok{x}    {10}{7}  {5}
  \addtosizeh@ok{xi}   {11}{8}  {6}
  \addtosizeh@ok{xiv}  {14}{10} {7}
  \addtosizeh@ok{xvii} {17}{12}{10}
\else
  \addtosizeh@ok{viii} {8}     {6}     {5}
  \addtosizeh@ok{ix}   {9}     {6}     {5}
  \addtosizeh@ok{x}    {10}    {7}     {5}
  \addtosizeh@ok{xi}   {10.95} {8}     {6}
  \addtosizeh@ok{xiv}  {14.4}  {10}    {7}
  \addtosizeh@ok{xvii} {17.28} {12}    {10}
\fi

\def\get@font#1#2#3{%
  \edef\fonts@ze{\romannumeral#3}
  \edef\fontn@me{\fonts@ze#1}
  \@ifundefined{\fontn@me}%
    {
     \global\expandafter\font\csname \fontn@me\endcsname=#2 at #3pt}%
    {}%
}

\def\ass@tfont#1#2{%
  \xdef\fam@name{\csname #1\endcsname}%
  \xdef\font@name{\csname #2\endcsname}%
  \let\textfont@name\font@name
  \textfont\fam@name\textfont@name
}

\def\ass@sfont#1#2{%
  \xdef\fam@name{\csname #1\endcsname}%
  \xdef\font@name{\csname #2\endcsname}%
  \let\textfont@name\font@name
  \scriptfont\fam@name\textfont@name
}

\def\ass@ssfont#1#2{%
  \xdef\fam@name{\csname #1\endcsname}%
  \xdef\font@name{\csname #2\endcsname}%
  \let\textfont@name\font@name
  \scriptscriptfont\fam@name\textfont@name
}


\def\NewSymbolFont#1#2{%
  \expandafter\ifx\csname sym#1fam\endcsname\relax 
    \expandafter\newfam\csname sym#1fam\endcsname
    \expandafter\edef\csname sym#1fam\endcsname{\the\allocationnumber}%
    \addtom@thgroup\math@groups{%
      \get@font{#1}{#2}{\s@ze}%
      \ass@tfont{sym#1fam}{\fontn@me}%
      \get@font{#1}{#2}{\ss@ze}%
      \ass@sfont{sym#1fam}{\fontn@me}%
      \get@font{#1}{#2}{\sss@ze}%
      \ass@ssfont{sym#1fam}{\fontn@me}%
    }%
  \else
    \errmessage{Family `#1' already defined}%
  \fi
}


\def\NewMathSymbol#1#2#3#4{%
  \edef\f@mly{\expandafter\hexnumber{\csname sym#3fam\endcsname}}%
  \mathchardef#1="#2\f@mly#4\relax
}


\newif\ifd@f

\def\NewMathDelimiter#1#2#3#4#5#6{%
  \d@ftrue
  \expandafter\ifx\csname sym#3fam\endcsname\relax
    \d@ffalse \errmessage{Family `#3' is not defined}%
  \fi
  \expandafter\ifx\csname sym#5fam\endcsname\relax
    \d@ffalse \errmessage{Family `#5' is not defined}%
  \fi
  \ifd@f
    \edef\f@mly{\expandafter\hexnumber{\csname sym#3fam\endcsname}}%
    \edef\f@mlytw@{\expandafter\hexnumber{\csname sym#5fam\endcsname}}%
    \xdef#1{\delimiter"#2\f@mly #4\f@mlytw@ #6\relax}%
  \fi
}


\def\setboxz@h{\setbox\z@\hbox}
\def\wdz@{\wd\z@}
\def\boxz@{\box\z@}
\def\setbox@ne{\setbox\@ne}
\def\wd@ne{\wd\@ne}

\def\math@atom#1#2{%
   \binrel@{#1}\binrel@@{#2}}
\def\binrel@#1{\setboxz@h{\thinmuskip0mu
  \medmuskip\m@ne mu\thickmuskip\@ne mu$#1\m@th$}%
 \setbox@ne\hbox{\thinmuskip0mu\medmuskip\m@ne mu\thickmuskip
  \@ne mu${}#1{}\m@th$}%
 \setbox\tw@\hbox{\hskip\wd@ne\hskip-\wdz@}}
\def\binrel@@#1{\ifdim\wd2<\z@\mathbin{#1}\else\ifdim\wd\tw@>\z@
 \mathrel{#1}\else{#1}\fi\fi}

\def\m@thit{1}

\def\set@skchar#1{\global\expandafter\skewchar
  \csname\fontn@me\endcsname=#1\relax}

\def\NewMathAlphabet#1#2#3{%
  \def\tst{#3}%
  \ifx\tst\empty\else 
    \expandafter\gdef\csname #1@sc\endcsname{}
  \fi
  \expandafter\def\csname #1\endcsname{
    \protect\csname @#1\endcsname}%
  \expandafter\def\csname @#1\endcsname##1{
    {%
    \begingroup
      \get@font{#1}{#2}{\s@ze}%
      \@ifundefined{#1@sc}{}{\set@skchar{#3}}%
      \ass@tfont{m@thit}{\fontn@me}%
      \get@font{#1}{#2}{\ss@ze}%
      \@ifundefined{#1@sc}{}{\set@skchar{#3}}%
      \ass@sfont{m@thit}{\fontn@me}%
      \get@font{#1}{#2}{\sss@ze}%
      \@ifundefined{#1@sc}{}{\set@skchar{#3}}%
      \ass@ssfont{m@thit}{\fontn@me}%
      \math@atom{##1}{%
      \mathchoice%
        {\hbox{$\m@th\displaystyle##1$}}%
        {\hbox{$\m@th\textstyle##1$}}%
        {\hbox{$\m@th\scriptstyle##1$}}%
        {\hbox{$\m@th\scriptscriptstyle##1$}}}%
    \endgroup
    }%
  }%
}


\newif\iffirstta  \firsttatrue

\def\set@hchar#1{\global\expandafter\hyphenchar
  \csname\fontn@me\endcsname=#1\relax}

\def\NewTextAlphabet#1#2#3{%
  \iffirstta
    \global\firsttafalse
    \newfam\scratchfam
    \edef\scrt@fam{\the\allocationnumber}%
  \fi
  \def\tst{#3}%
  \ifx\tst\empty\else 
    \expandafter\gdef\csname #1@hc\endcsname{}
  \fi
  \expandafter\def\csname #1\endcsname{
    \protect\csname t@#1\endcsname}%
  \long\expandafter\def\csname t@#1\endcsname##1{
    \ifmmode
      \typeout{Warning: do not use \expandafter\string\csname #1\endcsname
        \space in math mode}\fi%
    {%
      \get@font{#1}{#2}{\s@ze}\let\t@xtfnt=\fontn@me\relax
      \@ifundefined{#1@hc}{}{\set@hchar{#3}}%
      \ass@tfont{scrt@fam}{\fontn@me}%
      \get@font{#1}{#2}{\ss@ze}%
      \@ifundefined{#1@hc}{}{\set@hchar{#3}}%
      \ass@sfont{scrt@fam}{\fontn@me}%
      \get@font{#1}{#2}{\sss@ze}%
      \@ifundefined{#1@hc}{}{\set@hchar{#3}}%
      \ass@ssfont{scrt@fam}{\fontn@me}%
      \fam\scratchfam\csname\t@xtfnt\endcsname
    ##1%
    }%
  }%
  \expandafter\def\csname #1shape
    \endcsname{\protect\csname @#1shape\endcsname}%
  \expandafter\def\csname @#1shape\endcsname{
    \ifmmode
      \typeout{Warning: do not use \expandafter\string\csname
        #1shape\endcsname \space in math mode}\fi
      \get@font{#1}{#2}{\s@ze}\let\t@xtfnt=\fontn@me\relax
      \@ifundefined{#1@hc}{}{\set@hchar{#3}}%
      \ass@tfont{scrt@fam}{\fontn@me}%
      \get@font{#1}{#2}{\ss@ze}%
      \@ifundefined{#1@hc}{}{\set@hchar{#3}}%
      \ass@sfont{scrt@fam}{\fontn@me}%
      \get@font{#1}{#2}{\sss@ze}%
      \@ifundefined{#1@hc}{}{\set@hchar{#3}}%
      \ass@ssfont{scrt@fam}{\fontn@me}%
      \fam\scratchfam\csname\t@xtfnt\endcsname
  }%
}


\ifprod@font
  \def\math@itfnt{mtmib10}
  \def\math@syfnt{mtbsy10}
\else
  \def\math@itfnt{cmmib10}
  \def\math@syfnt{cmbsy10}
\fi

\def\m@thsy{2}

\def\bmath{\protect\@bmath}
\def\@bmath#1{%
  {%
  \begingroup
    \get@font{mthit}{\math@itfnt}{\s@ze}\set@skchar{'177}%
    \ass@tfont{m@thit}{\fontn@me}%
    \get@font{mthit}{\math@itfnt}{\ss@ze}\set@skchar{'177}%
    \ass@sfont{m@thit}{\fontn@me}%
    \get@font{mthit}{\math@itfnt}{\sss@ze}\set@skchar{'177}%
    \ass@ssfont{m@thit}{\fontn@me}%
    \get@font{mthsy}{\math@syfnt}{\s@ze}\set@skchar{'60}%
    \ass@tfont{m@thsy}{\fontn@me}%
    \get@font{mthsy}{\math@syfnt}{\ss@ze}\set@skchar{'60}%
    \ass@sfont{m@thsy}{\fontn@me}%
    \get@font{mthsy}{\math@syfnt}{\sss@ze}\set@skchar{'60}%
    \ass@ssfont{m@thsy}{\fontn@me}%
    \math@atom{#1}{%
    \mathchoice%
      {\hbox{$\m@th\displaystyle#1$}}%
      {\hbox{$\m@th\textstyle#1$}}%
      {\hbox{$\m@th\scriptstyle#1$}}%
      {\hbox{$\m@th\scriptscriptstyle#1$}}}%
  \endgroup
  }%
}



\def\diameter{{\ifmmode\mathchoice
{\ooalign{\hfil\hbox{$\displaystyle/$}\hfil\crcr
{\hbox{$\displaystyle\mathchar"20D$}}}}
{\ooalign{\hfil\hbox{$\textstyle/$}\hfil\crcr
{\hbox{$\textstyle\mathchar"20D$}}}}
{\ooalign{\hfil\hbox{$\scriptstyle/$}\hfil\crcr
{\hbox{$\scriptstyle\mathchar"20D$}}}}
{\ooalign{\hfil\hbox{$\scriptscriptstyle/$}\hfil\crcr
{\hbox{$\scriptscriptstyle\mathchar"20D$}}}}
\else{\ooalign{\hfil/\hfil\crcr\mathhexbox20D}}%
\fi}}

\def\sq{\ifmmode\squareforqed\else{\unskip\nobreak\hfil
\penalty50\hskip1em\null\nobreak\hfil\squareforqed
\parfillskip=0pt\finalhyphendemerits=0\endgraf}\fi}
\def\squareforqed{\hbox{\rlap{$\sqcap$}$\sqcup$}}


\def\bbbc{{\mathchoice {\setbox0=\hbox{$\displaystyle\rm C$}\hbox{\hbox
to0pt{\kern0.4\wd0\vrule height0.9\ht0\hss}\box0}}
{\setbox0=\hbox{$\textstyle\rm C$}\hbox{\hbox
to0pt{\kern0.4\wd0\vrule height0.9\ht0\hss}\box0}}
{\setbox0=\hbox{$\scriptstyle\rm C$}\hbox{\hbox
to0pt{\kern0.4\wd0\vrule height0.9\ht0\hss}\box0}}
{\setbox0=\hbox{$\scriptscriptstyle\rm C$}\hbox{\hbox
to0pt{\kern0.4\wd0\vrule height0.9\ht0\hss}\box0}}}}
\def\bbbq{{\mathchoice {\setbox0=\hbox{$\displaystyle\rm
Q$}\hbox{\raise
0.15\ht0\hbox to0pt{\kern0.4\wd0\vrule height0.8\ht0\hss}\box0}}
{\setbox0=\hbox{$\textstyle\rm Q$}\hbox{\raise
0.15\ht0\hbox to0pt{\kern0.4\wd0\vrule height0.8\ht0\hss}\box0}}
{\setbox0=\hbox{$\scriptstyle\rm Q$}\hbox{\raise
0.15\ht0\hbox to0pt{\kern0.4\wd0\vrule height0.7\ht0\hss}\box0}}
{\setbox0=\hbox{$\scriptscriptstyle\rm Q$}\hbox{\raise
0.15\ht0\hbox to0pt{\kern0.4\wd0\vrule height0.7\ht0\hss}\box0}}}}
\def\bbbt{{\mathchoice {\setbox0=\hbox{$\displaystyle\rm
T$}\hbox{\hbox to0pt{\kern0.3\wd0\vrule height0.9\ht0\hss}\box0}}
{\setbox0=\hbox{$\textstyle\rm T$}\hbox{\hbox
to0pt{\kern0.3\wd0\vrule height0.9\ht0\hss}\box0}}
{\setbox0=\hbox{$\scriptstyle\rm T$}\hbox{\hbox
to0pt{\kern0.3\wd0\vrule height0.9\ht0\hss}\box0}}
{\setbox0=\hbox{$\scriptscriptstyle\rm T$}\hbox{\hbox
to0pt{\kern0.3\wd0\vrule height0.9\ht0\hss}\box0}}}}
\def\bbbs{{\mathchoice
{\setbox0=\hbox{$\displaystyle     \rm S$}\hbox{\raise0.5\ht0\hbox
to0pt{\kern0.35\wd0\vrule height0.45\ht0\hss}\hbox
to0pt{\kern0.55\wd0\vrule height0.5\ht0\hss}\box0}}
{\setbox0=\hbox{$\textstyle        \rm S$}\hbox{\raise0.5\ht0\hbox
to0pt{\kern0.35\wd0\vrule height0.45\ht0\hss}\hbox
to0pt{\kern0.55\wd0\vrule height0.5\ht0\hss}\box0}}
{\setbox0=\hbox{$\scriptstyle      \rm S$}\hbox{\raise0.5\ht0\hbox
to0pt{\kern0.35\wd0\vrule height0.45\ht0\hss}\raise0.05\ht0\hbox
to0pt{\kern0.5\wd0\vrule height0.45\ht0\hss}\box0}}
{\setbox0=\hbox{$\scriptscriptstyle\rm S$}\hbox{\raise0.5\ht0\hbox
to0pt{\kern0.4\wd0\vrule height0.45\ht0\hss}\raise0.05\ht0\hbox
to0pt{\kern0.55\wd0\vrule height0.45\ht0\hss}\box0}}}}
\def\bbbz{{\mathchoice {\hbox{$\sf\textstyle Z\kern-0.4em Z$}}
{\hbox{$\sf\textstyle Z\kern-0.4em Z$}}
{\hbox{$\sf\scriptstyle Z\kern-0.3em Z$}}
{\hbox{$\sf\scriptscriptstyle Z\kern-0.2em Z$}}}}


\def\Nulle{0} 
\def\Afe{1}   
\def\Hae{2}   
\def\Hbe{3}   
\def\Hce{4}   
\def\Hde{5}   


\newcount\LastMac       \LastMac=\Nulle

\newskip\half      \half=5.5pt plus 1.5pt minus 2.25pt
\newskip\one       \one=11pt plus 3pt minus 5.5pt
\newskip\onehalf   \onehalf=16.5pt plus 5.5pt minus 8.25pt
\newskip\two       \two=22pt plus 5.5pt minus 11pt

\def\Half{\addvspace{\half}}
\def\One{\addvspace{\one}}
\def\OneHalf{\addvspace{\onehalf}}
\def\Two{\addvspace{\two}}

\def\Raggedright{
  \rightskip=\z@ plus \hsize\relax
}

\def\Fullout{
  \rightskip=\z@\relax
}

\def\Hang#1#2{
  \hangindent=#1%
  \hangafter=#2\relax
}


\newif\ifsp@page
\def\pagestyle#1{\csname ps@#1\endcsname}
\def\thispagestyle#1{\global\sp@pagetrue\gdef\sp@type{#1}}

\def\ps@titlepage{%
  \def\@oddhead{\eightpoint\noindent \the\CatchLine
    \ifprod@font\else\qquad Printed\ \today\qquad
      (MN plain \TeX\ macros\ v\@version)\fi \hfil}%
  \let\@evenhead=\@oddhead
  \def\@oddfoot{\eightpoint\copyright\ \@pubyear\ RAS\hfil}%
  \def\@evenfoot{\hfil\eightpoint\noindent\copyright\ \@pubyear\ RAS}%
}

\def\ps@headings{%
  \def\@oddhead{\elevenpoint\it\noindent
    \hfill\the\RightHeader\hskip1.5em\rm\folio}%
  \def\@evenhead{\elevenpoint\noindent
    \folio\hskip1.5em\it\the\LeftHeader\hfill}%
  \def\@oddfoot{\eightpoint\noindent\copyright\ \@pubyear\ RAS,
    MNRAS {\bf \@volume}, \@pagerange\hfil}%
  \def\@evenfoot{\hfil\eightpoint\copyright\ \@pubyear\ RAS,
    MNRAS {\bf \@volume}, \@pagerange}%
}

\def\ps@plate{%
  \def\@oddhead{\eightpoint\noindent\plt@cap\hfil}%
  \def\@evenhead{\eightpoint\noindent\plt@cap\hfil}%
  \def\@oddfoot{\eightpoint\noindent\copyright\ \@pubyear\ RAS,
    MNRAS {\bf \@volume}, \@pagerange\hfil}%
  \def\@evenfoot{\hfil\eightpoint\copyright\ \@pubyear\ RAS,
    MNRAS {\bf \@volume}, \@pagerange}%
}



\def\title#1{
  \bgroup
    \vbox to 8pt{\vss}%
    \seventeenpoint
    \Raggedright
    \noindent \strut{\bf #1}\par
  \egroup
}

\def\author#1{
  \bgroup
    \ifnum\LastMac=\Afe \OneHalf\else \vskip 21pt\fi
    \fourteenpoint
    \Raggedright
    \noindent \strut #1\par
    \vskip 3pt%
  \egroup
}

\def\affiliation#1{
  \bgroup
    \vskip -4pt%
    \eightpoint
    \Raggedright
    \noindent \strut {\it #1}\par
  \egroup
  \LastMac=\Afe\relax
}

\def\acceptedline#1{
  \bgroup
    \Two
    \eightpoint
    \Raggedright
    \noindent \strut #1\par
  \egroup
}

\long\def\abstract#1{%
  \bgroup
    \vskip 20pt%
    \leftskip 11pc\rightskip\z@
    \noindent{\ninebf ABSTRACT}\par
    \tenpoint
    \Fullout
    \noindent #1\par
  \egroup
}

\long\def\keywords#1{
  \bgroup
    \Half
    \leftskip 11pc\rightskip\z@
    \tenpoint
    \Fullout
    \noindent\hbox{\bf Key words:}\ #1\par
  \egroup
}


\def\maketitle{%
  \EndOpening
  \ifsinglecol \else \MakePage\fi
}


\def\pageoffset#1#2{\hoffset=#1\relax\voffset=#2\relax}


\def\@nameuse#1{\csname #1\endcsname}
\def\arabic#1{\@arabic{\@nameuse{#1}}}
\def\alph#1{\@alph{\@nameuse{#1}}}
\def\Alph#1{\@Alph{\@nameuse{#1}}}
\def\@arabic#1{\number #1}
\def\@Alph#1{\ifcase#1\or A\or B\or C\or D\else\@Ialph{#1}\fi}
\def\@Ialph#1{\ifcase#1\or \or \or \or \or E\or F\or G\or H\or I\or J\or
   K\or L\or M\or N\or O\or P\or Q\or R\or S\or T\or U\or V\or W\or X\or
   Y\or Z\else\errmessage{Counter out of range}\fi}
\def\@alph#1{\ifcase#1\or a\or b\or c\or d\else\@ialph{#1}\fi}
\def\@ialph#1{\ifcase#1\or \or \or \or \or e\or f\or g\or h\or i\or j\or
   k\or l\or m\or n\or o\or p\or q\or r\or s\or t\or u\or v\or w\or x\or y\or
   z\else\errmessage{Counter out of range}\fi}


\newcount\Eqnno
\newcount\SubEqnno

\def\theeq{\arabic{Eqnno}}
\def\thesubeq{\alph{SubEqnno}}

\def\stepeq{\relax
  \global\SubEqnno \z@
  \global\advance\Eqnno \@ne\relax
  {\rm (\theeq)}%
}

\def\startsubeq{\relax
  \global\SubEqnno \z@
  \global\advance\Eqnno \@ne\relax
  \stepsubeq
}

\def\stepsubeq{\relax
  \global\advance\SubEqnno \@ne\relax
  {\rm (\theeq\thesubeq)}%
}


\newcount\Sec        
\newcount\SecSec
\newcount\SecSecSec

\def\thesection{\arabic{Sec}}
\def\thesubsection{\thesection.\arabic{SecSec}}
\def\thesubsubsection{\thesubsection.\arabic{SecSecSec}}

\Sec=\z@

\def\:{\let\@sptoken= } \:  
\def\:{\@xifnch} \expandafter\def\: {\futurelet\@tempc\@ifnch}

\def\@ifnextchar#1#2#3{%
  \let\@tempMACe #1%
  \def\@tempMACa{#2}%
  \def\@tempMACb{#3}%
  \futurelet \@tempMACc\@ifnch%
}

\def\@ifnch{%
\ifx \@tempMACc \@sptoken%
  \let\@tempMACd\@xifnch%
\else%
  \ifx \@tempMACc \@tempMACe%
    \let\@tempMACd\@tempMACa%
  \else%
    \let\@tempMACd\@tempMACb%
  \fi%
\fi%
\@tempMACd%
}

\def\@ifstar#1#2{\@ifnextchar *{\def\@tempMACa*{#1}\@tempMACa}{#2}}

\newskip\@tempskipb

\def\addvspace#1{%
  \ifvmode\else \endgraf\fi%
  \ifdim\lastskip=\z@%
    \vskip #1\relax%
  \else%
    \@tempskipb#1\relax\@xaddvskip%
  \fi%
}

\def\@xaddvskip{%
  \ifdim\lastskip<\@tempskipb%
    \vskip-\lastskip%
    \vskip\@tempskipb\relax%
  \else%
    \ifdim\@tempskipb<\z@%
      \ifdim\lastskip<\z@ \else%
        \advance\@tempskipb\lastskip%
        \vskip-\lastskip\vskip\@tempskipb%
      \fi%
    \fi%
  \fi%
}

\newskip\@tmpSKIP

\def\addpen#1{%
  \ifvmode
    \if@nobreak
    \else
      \ifdim\lastskip=\z@
        \penalty#1\relax
      \else
        \@tmpSKIP=\lastskip
        \vskip -\lastskip
        \penalty#1\vskip\@tmpSKIP
      \fi
    \fi
  \fi
}

\newcount\@clubpen   \@clubpen=\clubpenalty
\newif\if@nobreak    \@nobreakfalse

\def\@noafterindent{%
  \global\@nobreaktrue
  \everypar{\if@nobreak
              \global\@nobreakfalse
              \clubpenalty \@M
              {\setbox\z@\lastbox}%
              \LastMac=\Nulle\relax%
            \else
              \clubpenalty \@clubpen
              \everypar{}%
            \fi}%
}

\newcount\gds@cbrk   \gds@cbrk=-300

\def\@nohdbrk{\interlinepenalty \@M\relax}

\let\@par=\par
\def\@restorepar{\def\par{\@par}}

\newif\if@endpe   \@endpefalse
 
\def\@doendpe{\@endpetrue \@nobreakfalse \LastMac=\Nulle\relax%
     \def\par{\@restorepar\everypar{}\par\@endpefalse}%
              \everypar{\setbox\z@\lastbox\everypar{}\@endpefalse}%
}

\def\section{\@ifstar{\@ssection}{\@section}}

\def\@section#1{
  \if@nobreak
    \everypar{}%
    \ifnum\LastMac=\Hae \addvspace{\half}\fi
  \else
    \addpen{\gds@cbrk}%
    \addvspace{\two}%
  \fi
  \bgroup
    \ninepoint\bf
    \Raggedright
    \global\advance\Sec \@ne
    \ifappendix
      \global\Eqnno=\z@ \global\SubEqnno=\z@\relax
      \def\ch@ck{#1}%
      \ifx\ch@ck\empty \def\c@lon{}\else\def\c@lon{:}\fi
      \noindent\@nohdbrk APPENDIX\ \thesection\c@lon\hskip 0.5em%
        \uppercase{#1}\par
    \else
      \noindent\@nohdbrk\thesection\hskip 1pc \uppercase{#1}\par
    \fi
    \global\SecSec=\z@
  \egroup
  \nobreak
  \vskip\half
  \nobreak
  \@noafterindent
  \LastMac=\Hae\relax
}

\def\@ssection#1{
  \if@nobreak
    \everypar{}%
    \ifnum\LastMac=\Hae \addvspace{\half}\fi
  \else
    \addpen{\gds@cbrk}%
    \addvspace{\two}%
  \fi
  \bgroup
    \ninepoint\bf
    \Raggedright
    \noindent\@nohdbrk\uppercase{#1}\par
  \egroup
  \nobreak
  \vskip\half
  \nobreak
  \@noafterindent
  \LastMac=\Hae\relax
}

\def\subsection{\@ifstar{\@ssubsection}{\@subsection}}

\def\@subsection#1{
  \if@nobreak
    \everypar{}%
    \ifnum\LastMac=\Hae \addvspace{1pt plus 1pt minus .5pt}\fi
  \else
    \addpen{\gds@cbrk}%
    \addvspace{\onehalf}%
  \fi
  \bgroup
    \ninepoint\bf
    \Raggedright
    \global\advance\SecSec \@ne
    \noindent\@nohdbrk\thesubsection \hskip 1pc\relax #1\par
    \global\SecSecSec=\z@
  \egroup
  \nobreak
  \vskip\half
  \nobreak
  \@noafterindent
  \LastMac=\Hbe\relax
}

\def\@ssubsection#1{
  \if@nobreak
    \everypar{}%
    \ifnum\LastMac=\Hae \addvspace{1pt plus 1pt minus .5pt}\fi
  \else
    \addpen{\gds@cbrk}%
    \addvspace{\onehalf}%
  \fi
  \bgroup
    \ninepoint\bf
    \Raggedright
    \noindent\@nohdbrk #1\par
  \egroup
  \nobreak
  \vskip\half
  \nobreak
  \@noafterindent
  \LastMac=\Hbe\relax
}

\def\subsubsection{\@ifstar{\@ssubsubsection}{\@subsubsection}}

\def\@subsubsection#1{
  \if@nobreak
    \everypar{}%
    \ifnum\LastMac=\Hbe \addvspace{1pt plus 1pt minus .5pt}\fi
  \else
    \addpen{\gds@cbrk}%
    \addvspace{\onehalf}%
  \fi
  \bgroup
    \ninepoint\it
    \Raggedright
    \global\advance\SecSecSec \@ne
    \noindent\@nohdbrk\thesubsubsection \hskip 1pc\relax #1\par
  \egroup
  \nobreak
  \vskip\half
  \nobreak
  \@noafterindent
  \LastMac=\Hce\relax
}

\def\@ssubsubsection#1{
  \if@nobreak
    \everypar{}%
    \ifnum\LastMac=\Hbe \addvspace{1pt plus 1pt minus .5pt}\fi
  \else
    \addpen{\gds@cbrk}%
    \addvspace{\onehalf}%
  \fi
  \bgroup
    \ninepoint\it
    \Raggedright
    \noindent\@nohdbrk #1\par
  \egroup
  \nobreak
  \vskip\half
  \nobreak
  \@noafterindent
  \LastMac=\Hce\relax
}

\def\paragraph#1{
  \if@nobreak
    \everypar{}%
  \else
    \addpen{\gds@cbrk}%
    \addvspace{\one}%
  \fi%
  \bgroup%
    \ninepoint\it
    \noindent #1\ \nobreak%
  \egroup
  \LastMac=\Hde\relax
  \ignorespaces
}


\newif\ifappendix

\def\appendix{%
  \global\appendixtrue
  \def\thesection{\Alph{Sec}}%
  \def\thesubsection{\thesection\arabic{SecSec}}%
  \def\theeq{\thesection\arabic{Eqnno}}%
  \Sec=\z@ \SecSec=\z@ \SecSecSec=\z@ \Eqnno=\z@ \SubEqnno=\z@\relax
}




\def\beginlist{%
  \par\if@nobreak \else\addvspace{\half}\fi%
  \bgroup%
    \ninepoint
    \let\item=\list@item%
}

\def\list@item{%
  \par\noindent\hskip 1em\relax%
  \ignorespaces%
}

\def\endlist{\par\egroup\addvspace{\half}\@doendpe}


\def\beginrefs{%
  \par
  \bgroup
    \eightpoint
    \Fullout
    \let\bibitem=\bib@item
}

\def\bib@item{%
  \par\parindent=1.5em\Hang{1.5em}{1}%
  \everypar={\Hang{1.5em}{1}\ignorespaces}%
  \noindent\ignorespaces
}

\def\endrefs{\par\egroup\@doendpe}


\newtoks\CatchLine

\def\@journal{Mon.\ Not.\ R.\ Astron.\ Soc.\ }  
\def\@pubyear{1994}        
\def\@pagerange{000--000}  
\def\@volume{000}          
\def\@microfiche{}         %

\def\pubyear#1{\gdef\@pubyear{#1}\@makecatchline}
\def\pagerange#1{\gdef\@pagerange{#1}\@makecatchline}
\def\volume#1{\gdef\@volume{#1}\@makecatchline}
\def\microfiche#1{\gdef\@microfiche{and Microfiche\ #1}\@makecatchline}

\def\@makecatchline{%
  \global\CatchLine{%
    {\rm \@journal {\bf \@volume},\ \@pagerange\ (\@pubyear)\ \@microfiche}}%
}

\@makecatchline 

\newtoks\LeftHeader
\def\shortauthor#1{
  \global\LeftHeader{#1}%
}

\newtoks\RightHeader
\def\shorttitle#1{
  \global\RightHeader{#1}%
}

\def\PageHead{
  \begingroup
    \ifsp@page
      \csname ps@\sp@type\endcsname
    \fi
    \ifodd\pageno
      \let\the@head=\@oddhead
    \else
      \let\the@head=\@evenhead
    \fi
    \vbox to \z@{\vskip-22.5\p@%
      \hbox to \PageWidth{\vbox to8.5\p@{}%
        \the@head
      }%
    \vss}%
  \endgroup
  \nointerlineskip
}

\gdef\PageFoot{%
  \nointerlineskip%
  \begingroup
  \ifsp@page
    \csname ps@\sp@type\endcsname
    \global\sp@pagefalse
  \fi
  \vbox to 22pt{\vfil%
    \hbox to \PageWidth{%
      \eightpoint\strut\noindent
      \ifodd\pageno
        \@oddfoot
      \else
        \@evenfoot
      \fi
    }%
  }%
  \endgroup
}

\def\today{%
  \number\day\space
  \ifcase\month\or January\or February\or March\or April\or May\or June\or
    July\or August\or September\or October\or November\or December\fi
  \space\number\year%
}

\def\authorcomment#1{%
  \gdef\PageFoot{%
    \nointerlineskip%
    \vbox to 20pt{\vfil%
      \hbox to \PageWidth{\elevenpoint\noindent \hfil #1 \hfil}}%
  }%
}


\newif\ifplate@page
\newbox\plt@box

\def\beginplatepage{%
  \let\plate=\plate@head
  \let\caption=\fig@caption
  \global\setbox\plt@box=\vbox\bgroup
  \TEMPDIMEN=\PageWidth 
  \hsize=\PageWidth\relax
}

\def\endplatepage{\par\egroup\global\plate@pagetrue}
\def\plate@head#1{\gdef\plt@cap{#1}}


\def\letters{%
  \gdef\folio{\ifnum\pageno<\z@ L\romannumeral-\pageno
    \else L\number\pageno \fi}%
}


\newdimen\mathindent

\global\mathindent=\z@
\global\everydisplay{\global\@dspwd=\displaywidth\displaysetup}


\def\@displaylines#1{
  {}$\displ@y\hbox{\vbox{\halign{$\@lign\hfil\displaystyle##\hfil$\crcr
  #1\crcr}}}${}%
}

\def\@eqalign#1{\null\vcenter{\openup\jot\m@th
  \ialign{\strut\hfil$\displaystyle{##}$&$\displaystyle{{}##}$\hfil
      \crcr#1\crcr}}%
}

\def\@eqalignno#1{
  \global\advance\@dspwd by -\mathindent%
  {}$\displ@y\hbox{\vbox{\halign to\@dspwd%
  {\hfil$\@lign\displaystyle{##}$\tabskip\z@skip
  &$\@lign\displaystyle{{}##}$\hfil\tabskip\centering
  &\llap{$\@lign##$}\tabskip\z@skip\crcr
  #1\crcr}}}${}%
}


\global\let\displaylines=\@displaylines
\global\let\eqalign=\@eqalign
\global\let\eqalignno=\@eqalignno
\global\let\leqalignno=\@eqalignno

\newdimen\@dspwd   \@dspwd=\z@
\newif\if@eqno
\newif\if@leqno
\newtoks\@eqn
\newtoks\@eq

\def\displaysetup#1$${\displaytest#1\eqno\eqno\displaytest}

\def\displaytest#1\eqno#2\eqno#3\displaytest{%
 \if!#3!\ldisplaytest#1\leqno\leqno\ldisplaytest
 \else\@eqnotrue\@leqnofalse\@eqn={#2}\@eq={#1}\fi
 \generaldisplay$$}

\def\ldisplaytest#1\leqno#2\leqno#3\ldisplaytest{%
\@eq={#1}%
 \if!#3!\@eqnofalse\else\@eqnotrue\@leqnotrue
  \@eqn={#2}\fi}

\def\generaldisplay{%
  \if@eqno
    \if@leqno
      \hbox to \displaywidth{\noindent
        \rlap{$\displaystyle\the\@eqn$}%
        \hskip\mathindent$\displaystyle\the\@eq$\hfil}%
    \else
      \hbox to \displaywidth{\noindent
        \hskip\mathindent
        $\displaystyle\the\@eq$\hfil$\displaystyle\the\@eqn$}%
    \fi
  \else
    \hbox to \displaywidth{\noindent
      \hskip\mathindent$\displaystyle\the\@eq$\hfil}%
  \fi
}


\def\@notice{%
  \par\Two%
  \noindent{\b@ls{11pt}\ninerm This paper has been produced using the
    Royal Astronomical Society/Blackwell Science \TeX\ macros.\par}%
}

\outer\def\bye{\@notice\par\vfill\supereject\end}


\def\start@mess{%
  Monthly notices of the RAS journal style (\@typeface)\space
    v\@version,\space \@verdate.%
}

\everyjob{\Warn{\start@mess}}



\newif\if@debug \@debugfalse  

\def\Print#1{\if@debug\immediate\write16{#1}\else \fi}
\def\Warn#1{\immediate\write16{#1}}
\def\wlog#1{}

\newcount\Iteration 

\def\Single{0} \def\Double{1}                 
\def\Figure{0} \def\Table{1}                  

\def\InStack{0}  
\def\InZoneA{1}
\def\InZoneB{2}
\def\InZoneC{3}

\newcount\TEMPCOUNT 
\newdimen\TEMPDIMEN 
\newbox\TEMPBOX     
\newbox\VOIDBOX     

\newcount\LengthOfStack 
\newcount\MaxItems      
\newcount\StackPointer
\newcount\Point         
\newcount\NextFigure    
\newcount\NextTable     
\newcount\NextItem      

\newcount\StatusStack   
\newcount\NumStack      
\newcount\TypeStack     
\newcount\SpanStack     
\newcount\BoxStack      

\newcount\ItemSTATUS    
\newcount\ItemNUMBER    
\newcount\ItemTYPE      
\newcount\ItemSPAN      
\newbox\ItemBOX         
\newdimen\ItemSIZE      

\newdimen\PageHeight    
\newdimen\TextLeading   
\newdimen\Feathering    
\newcount\LinesPerPage  
\newdimen\ColumnWidth   
\newdimen\ColumnGap     
\newdimen\PageWidth     
\newdimen\BodgeHeight   
\newcount\Leading       

\newdimen\ZoneBSize  
\newdimen\TextSize   
\newbox\ZoneABOX     
\newbox\ZoneBBOX     
\newbox\ZoneCBOX     

\newif\ifFirstSingleItem
\newif\ifFirstZoneA
\newif\ifMakePageInComplete
\newif\ifMoreFigures \MoreFiguresfalse 
\newif\ifMoreTables  \MoreTablesfalse  

\newif\ifFigInZoneB 
\newif\ifFigInZoneC 
\newif\ifTabInZoneB 
\newif\ifTabInZoneC

\newif\ifZoneAFullPage

\newbox\MidBOX    
\newbox\LeftBOX
\newbox\RightBOX
\newbox\PageBOX   

\newif\ifLeftCOL  
\LeftCOLtrue

\newdimen\ZoneBAdjust

\newcount\ItemFits
\def\Yes{1}
\def\No{2}


\MaxItems=15
\NextFigure=\z@        
\NextTable=\@ne

\BodgeHeight=6pt
\TextLeading=11pt    
\Leading=11
\Feathering=\z@      
\LinesPerPage=61     
\topskip=\TextLeading
\ColumnWidth=20pc    
\ColumnGap=2pc       

\newskip\ItemSepamount  
\ItemSepamount=\TextLeading plus \TextLeading minus 4pt

\parskip=\z@ plus .1pt
\parindent=18pt
\widowpenalty=\z@
\clubpenalty=10000
\tolerance=1500
\hbadness=1500
\abovedisplayskip=6pt plus 2pt minus 1pt
\belowdisplayskip=6pt plus 2pt minus 1pt
\abovedisplayshortskip=6pt plus 2pt minus 1pt
\belowdisplayshortskip=6pt plus 2pt minus 1pt

\frenchspacing

\ninepoint 

\PageHeight=682pt
\PageWidth=2\ColumnWidth
\advance\PageWidth by \ColumnGap

\pagestyle{headings}




\newcount\DUMMY \StatusStack=\allocationnumber
\newcount\DUMMY \newcount\DUMMY \newcount\DUMMY 
\newcount\DUMMY \newcount\DUMMY \newcount\DUMMY 
\newcount\DUMMY \newcount\DUMMY \newcount\DUMMY
\newcount\DUMMY \newcount\DUMMY \newcount\DUMMY 
\newcount\DUMMY \newcount\DUMMY \newcount\DUMMY

\newcount\DUMMY \NumStack=\allocationnumber
\newcount\DUMMY \newcount\DUMMY \newcount\DUMMY 
\newcount\DUMMY \newcount\DUMMY \newcount\DUMMY 
\newcount\DUMMY \newcount\DUMMY \newcount\DUMMY 
\newcount\DUMMY \newcount\DUMMY \newcount\DUMMY 
\newcount\DUMMY \newcount\DUMMY \newcount\DUMMY

\newcount\DUMMY \TypeStack=\allocationnumber
\newcount\DUMMY \newcount\DUMMY \newcount\DUMMY 
\newcount\DUMMY \newcount\DUMMY \newcount\DUMMY 
\newcount\DUMMY \newcount\DUMMY \newcount\DUMMY 
\newcount\DUMMY \newcount\DUMMY \newcount\DUMMY 
\newcount\DUMMY \newcount\DUMMY \newcount\DUMMY

\newcount\DUMMY \SpanStack=\allocationnumber
\newcount\DUMMY \newcount\DUMMY \newcount\DUMMY 
\newcount\DUMMY \newcount\DUMMY \newcount\DUMMY 
\newcount\DUMMY \newcount\DUMMY \newcount\DUMMY 
\newcount\DUMMY \newcount\DUMMY \newcount\DUMMY 
\newcount\DUMMY \newcount\DUMMY \newcount\DUMMY

\newbox\DUMMY   \BoxStack=\allocationnumber
\newbox\DUMMY   \newbox\DUMMY \newbox\DUMMY 
\newbox\DUMMY   \newbox\DUMMY \newbox\DUMMY 
\newbox\DUMMY   \newbox\DUMMY \newbox\DUMMY 
\newbox\DUMMY   \newbox\DUMMY \newbox\DUMMY 
\newbox\DUMMY   \newbox\DUMMY \newbox\DUMMY

\def\wlog{\immediate\write\m@ne}


\def\GetItemAll#1{%
 \GetItemSTATUS{#1}
 \GetItemNUMBER{#1}
 \GetItemTYPE{#1}
 \GetItemSPAN{#1}
 \GetItemBOX{#1}
}

\def\GetItemSTATUS#1{%
 \Point=\StatusStack
 \advance\Point by #1
 \global\ItemSTATUS=\count\Point
}

\def\GetItemNUMBER#1{%
 \Point=\NumStack
 \advance\Point by #1
 \global\ItemNUMBER=\count\Point
}

\def\GetItemTYPE#1{%
 \Point=\TypeStack
 \advance\Point by #1
 \global\ItemTYPE=\count\Point
}

\def\GetItemSPAN#1{%
 \Point\SpanStack
 \advance\Point by #1
 \global\ItemSPAN=\count\Point
}

\def\GetItemBOX#1{%
 \Point=\BoxStack
 \advance\Point by #1
 \global\setbox\ItemBOX=\vbox{\copy\Point}
 \global\ItemSIZE=\ht\ItemBOX
 \global\advance\ItemSIZE by \dp\ItemBOX
 \TEMPCOUNT=\ItemSIZE
 \divide\TEMPCOUNT by \Leading
 \divide\TEMPCOUNT by 65536
 \advance\TEMPCOUNT \@ne
 \ItemSIZE=\TEMPCOUNT pt
 \global\multiply\ItemSIZE by \Leading
}


\def\JoinStack{%
 \ifnum\LengthOfStack=\MaxItems 
  \Warn{WARNING: Stack is full...some items will be lost!}
 \else
  \Point=\StatusStack
  \advance\Point by \LengthOfStack
  \global\count\Point=\ItemSTATUS
  \Point=\NumStack
  \advance\Point by \LengthOfStack
  \global\count\Point=\ItemNUMBER
  \Point=\TypeStack
  \advance\Point by \LengthOfStack
  \global\count\Point=\ItemTYPE
  \Point\SpanStack
  \advance\Point by \LengthOfStack
  \global\count\Point=\ItemSPAN
  \Point=\BoxStack
  \advance\Point by \LengthOfStack
  \global\setbox\Point=\vbox{\copy\ItemBOX}
  \global\advance\LengthOfStack \@ne
  \ifnum\ItemTYPE=\Figure 
   \global\MoreFigurestrue
  \else
   \global\MoreTablestrue
  \fi
 \fi
}


\def\LeaveStack#1{%
 {\Iteration=#1
 \loop
 \ifnum\Iteration<\LengthOfStack
  \advance\Iteration \@ne
  \GetItemSTATUS{\Iteration}
   \advance\Point by \m@ne
   \global\count\Point=\ItemSTATUS
  \GetItemNUMBER{\Iteration}
   \advance\Point by \m@ne
   \global\count\Point=\ItemNUMBER
  \GetItemTYPE{\Iteration}
   \advance\Point by \m@ne
   \global\count\Point=\ItemTYPE
  \GetItemSPAN{\Iteration}
   \advance\Point by \m@ne
   \global\count\Point=\ItemSPAN
  \GetItemBOX{\Iteration}
   \advance\Point by \m@ne
   \global\setbox\Point=\vbox{\copy\ItemBOX}
 \repeat}
 \global\advance\LengthOfStack by \m@ne
}


\newif\ifStackNotClean

\def\CleanStack{%
 \StackNotCleantrue
 {\Iteration=\z@
  \loop
   \ifStackNotClean
    \GetItemSTATUS{\Iteration}
    \ifnum\ItemSTATUS=\InStack
     \advance\Iteration \@ne
     \else
      \LeaveStack{\Iteration}
    \fi
   \ifnum\LengthOfStack<\Iteration
    \StackNotCleanfalse
   \fi
 \repeat}
}


\def\FindItem#1#2{%
 \global\StackPointer=\m@ne 
 {\Iteration=\z@
  \loop
  \ifnum\Iteration<\LengthOfStack
   \GetItemSTATUS{\Iteration}
   \ifnum\ItemSTATUS=\InStack
    \GetItemTYPE{\Iteration}
    \ifnum\ItemTYPE=#1
     \GetItemNUMBER{\Iteration}
     \ifnum\ItemNUMBER=#2
      \global\StackPointer=\Iteration
      \Iteration=\LengthOfStack 
     \fi
    \fi
   \fi
  \advance\Iteration \@ne
 \repeat}
}


\def\FindNext{%
 \global\StackPointer=\m@ne 
 {\Iteration=\z@
  \loop
  \ifnum\Iteration<\LengthOfStack
   \GetItemSTATUS{\Iteration}
   \ifnum\ItemSTATUS=\InStack
    \GetItemTYPE{\Iteration}
   \ifnum\ItemTYPE=\Figure
    \ifMoreFigures
      \global\NextItem=\Figure
      \global\StackPointer=\Iteration
      \Iteration=\LengthOfStack 
    \fi
   \fi
   \ifnum\ItemTYPE=\Table
    \ifMoreTables
      \global\NextItem=\Table
      \global\StackPointer=\Iteration
      \Iteration=\LengthOfStack 
    \fi
   \fi
  \fi
  \advance\Iteration \@ne
 \repeat}
}


\def\ChangeStatus#1#2{%
 \Point=\StatusStack
 \advance\Point by #1
 \global\count\Point=#2
}



\def\Zone{\InZoneA}

\ZoneBAdjust=\z@

\def\MakePage{
 \global\ZoneBSize=\PageHeight
 \global\TextSize=\ZoneBSize
 \global\ZoneAFullPagefalse
 \global\topskip=\TextLeading
 \MakePageInCompletetrue
 \MoreFigurestrue
 \MoreTablestrue
 \FigInZoneBfalse
 \FigInZoneCfalse
 \TabInZoneBfalse
 \TabInZoneCfalse
 \global\FirstSingleItemtrue
 \global\FirstZoneAtrue
 \global\setbox\ZoneABOX=\box\VOIDBOX
 \global\setbox\ZoneBBOX=\box\VOIDBOX
 \global\setbox\ZoneCBOX=\box\VOIDBOX
 \loop
  \ifMakePageInComplete
 \FindNext
 \ifnum\StackPointer=\m@ne
  \NextItem=\m@ne
  \MoreFiguresfalse
  \MoreTablesfalse
 \fi
 \ifnum\NextItem=\Figure
   \FindItem{\Figure}{\NextFigure}
   \ifnum\StackPointer=\m@ne \global\MoreFiguresfalse
   \else
    \GetItemSPAN{\StackPointer}
    \ifnum\ItemSPAN=\Single \def\Zone{\InZoneB}\relax
     \ifFigInZoneC \global\MoreFiguresfalse\fi
    \else
     \def\Zone{\InZoneA}
     \ifFigInZoneB \def\Zone{\InZoneC}\fi
    \fi
   \fi
   \ifMoreFigures\Print{}\FigureItems\fi
 \fi
\ifnum\NextItem=\Table
   \FindItem{\Table}{\NextTable}
   \ifnum\StackPointer=\m@ne \global\MoreTablesfalse
   \else
    \GetItemSPAN{\StackPointer}
    \ifnum\ItemSPAN=\Single\relax
     \ifTabInZoneC \global\MoreTablesfalse\fi
    \else
     \def\Zone{\InZoneA}
     \ifTabInZoneB \def\Zone{\InZoneC}\fi
    \fi
   \fi
   \ifMoreTables\Print{}\TableItems\fi
 \fi
   \MakePageInCompletefalse 
   \ifMoreFigures\MakePageInCompletetrue\fi
   \ifMoreTables\MakePageInCompletetrue\fi
 \repeat
 \ifZoneAFullPage
  \global\TextSize=\z@
  \global\ZoneBSize=\z@
  \global\vsize=\z@\relax
  \global\topskip=\z@\relax
  \vbox to \z@{\vss}
  \eject
 \else
 \global\advance\ZoneBSize by -\ZoneBAdjust
 \global\vsize=\ZoneBSize
 \global\hsize=\ColumnWidth
 \global\ZoneBAdjust=\z@
 \ifdim\TextSize<23pt
 \Warn{}
 \Warn{* Making column fall short: TextSize=\the\TextSize *}
 \vskip-\lastskip\eject\fi
 \fi
}

\def\MakeRightCol{
 \global\TextSize=\ZoneBSize
 \MakePageInCompletetrue
 \MoreFigurestrue
 \MoreTablestrue
 \global\FirstSingleItemtrue
 \global\setbox\ZoneBBOX=\box\VOIDBOX
 \def\Zone{\InZoneB}
 \loop
  \ifMakePageInComplete
 \FindNext
 \ifnum\StackPointer=\m@ne
  \NextItem=\m@ne
  \MoreFiguresfalse
  \MoreTablesfalse
 \fi
 \ifnum\NextItem=\Figure
   \FindItem{\Figure}{\NextFigure}
   \ifnum\StackPointer=\m@ne \MoreFiguresfalse
   \else
    \GetItemSPAN{\StackPointer}
    \ifnum\ItemSPAN=\Double\relax
     \MoreFiguresfalse\fi
   \fi
   \ifMoreFigures\Print{}\FigureItems\fi
 \fi
 \ifnum\NextItem=\Table
   \FindItem{\Table}{\NextTable}
   \ifnum\StackPointer=\m@ne \MoreTablesfalse
   \else
    \GetItemSPAN{\StackPointer}
    \ifnum\ItemSPAN=\Double\relax
     \MoreTablesfalse\fi
   \fi
   \ifMoreTables\Print{}\TableItems\fi
 \fi
   \MakePageInCompletefalse 
   \ifMoreFigures\MakePageInCompletetrue\fi
   \ifMoreTables\MakePageInCompletetrue\fi
 \repeat
 \ifZoneAFullPage
  \global\TextSize=\z@
  \global\ZoneBSize=\z@
  \global\vsize=\z@\relax
  \global\topskip=\z@\relax
  \vbox to \z@{\vss}
  \eject
 \else
 \global\vsize=\ZoneBSize
 \global\hsize=\ColumnWidth
 \ifdim\TextSize<23pt
 \Warn{}
 \Warn{* Making column fall short: TextSize=\the\TextSize *}
 \vskip-\lastskip\eject\fi
\fi
}

\def\FigureItems{
 \Print{Considering...}
 \ShowItem{\StackPointer}
 \GetItemBOX{\StackPointer} 
 \GetItemSPAN{\StackPointer}
  \CheckFitInZone 
  \ifnum\ItemFits=\Yes
   \ifnum\ItemSPAN=\Single
     \ChangeStatus{\StackPointer}{\InZoneB} 
     \global\FigInZoneBtrue
     \ifFirstSingleItem
      \hbox{}\vskip-\BodgeHeight
     \global\advance\ItemSIZE by \TextLeading
     \fi
     \unvbox\ItemBOX\ItemSep
     \global\FirstSingleItemfalse
     \global\advance\TextSize by -\ItemSIZE
     \global\advance\TextSize by -\TextLeading
   \else
    \ifFirstZoneA
     \global\advance\ItemSIZE by \TextLeading
     \global\FirstZoneAfalse\fi
    \global\advance\TextSize by -\ItemSIZE
    \global\advance\TextSize by -\TextLeading
    \global\advance\ZoneBSize by -\ItemSIZE
    \global\advance\ZoneBSize by -\TextLeading
    \ifFigInZoneB\relax
     \else
     \ifdim\TextSize<3\TextLeading
     \global\ZoneAFullPagetrue
     \fi
    \fi
    \ChangeStatus{\StackPointer}{\Zone}
    \ifnum\Zone=\InZoneC \global\FigInZoneCtrue\fi
  \fi
   \Print{TextSize=\the\TextSize}
   \Print{ZoneBSize=\the\ZoneBSize}
  \global\advance\NextFigure \@ne
   \Print{This figure has been placed.}
  \else
   \Print{No space available for this figure...holding over.}
   \Print{}
   \global\MoreFiguresfalse
  \fi
}

\def\TableItems{
 \Print{Considering...}
 \ShowItem{\StackPointer}
 \GetItemBOX{\StackPointer} 
 \GetItemSPAN{\StackPointer}
  \CheckFitInZone 
  \ifnum\ItemFits=\Yes
   \ifnum\ItemSPAN=\Single
    \ChangeStatus{\StackPointer}{\InZoneB}
     \global\TabInZoneBtrue
     \ifFirstSingleItem
      \hbox{}\vskip-\BodgeHeight
     \global\advance\ItemSIZE by \TextLeading
     \fi
     \unvbox\ItemBOX\ItemSep
     \global\FirstSingleItemfalse
     \global\advance\TextSize by -\ItemSIZE
     \global\advance\TextSize by -\TextLeading
   \else
    \ifFirstZoneA
    \global\advance\ItemSIZE by \TextLeading
    \global\FirstZoneAfalse\fi
    \global\advance\TextSize by -\ItemSIZE
    \global\advance\TextSize by -\TextLeading
    \global\advance\ZoneBSize by -\ItemSIZE
    \global\advance\ZoneBSize by -\TextLeading
    \ifFigInZoneB\relax
     \else
     \ifdim\TextSize<3\TextLeading
     \global\ZoneAFullPagetrue
     \fi
    \fi
    \ChangeStatus{\StackPointer}{\Zone}
    \ifnum\Zone=\InZoneC \global\TabInZoneCtrue\fi
   \fi
  \global\advance\NextTable \@ne
   \Print{This table has been placed.}
  \else
  \Print{No space available for this table...holding over.}
   \Print{}
   \global\MoreTablesfalse
  \fi
}


\def\CheckFitInZone{%
{\advance\TextSize by -\ItemSIZE
 \advance\TextSize by -\TextLeading
 \ifFirstSingleItem
  \advance\TextSize by \TextLeading
 \fi
 \ifnum\Zone=\InZoneA\relax
  \else \advance\TextSize by -\ZoneBAdjust
 \fi
 \ifdim\TextSize<3\TextLeading \global\ItemFits=\No
 \else \global\ItemFits=\Yes\fi}
}

\def\BeginOpening{%
  \ninepoint
  \thispagestyle{titlepage}%
  \global\setbox\ItemBOX=\vbox\bgroup%
    \hsize=\PageWidth%
    \hrule height \z@
    \ifsinglecol\vskip 6pt\fi 
}

\let\begintopmatter=\BeginOpening  

\def\EndOpening{%
  \One
  \egroup
  \ifsinglecol
    \box\ItemBOX%
    \vskip\TextLeading plus 2\TextLeading
    \@noafterindent
  \else
    \ItemNUMBER=\z@%
    \ItemTYPE=\Figure
    \ItemSPAN=\Double
    \ItemSTATUS=\InStack
    \JoinStack
  \fi
}


\newif\if@here  \@herefalse

\def\no@float{\global\@heretrue}
\let\nofloat=\relax 

\def\beginfigure{%
  \@ifstar{\global\@dfloattrue \@bfigure}{\global\@dfloatfalse \@bfigure}%
}

\def\@bfigure#1{%
  \par
  \if@dfloat
    \ItemSPAN=\Double
    \TEMPDIMEN=\PageWidth
  \else
    \ItemSPAN=\Single
    \TEMPDIMEN=\ColumnWidth
  \fi
  \ifsinglecol
    \TEMPDIMEN=\PageWidth
  \else
    \ItemSTATUS=\InStack
    \ItemNUMBER=#1%
    \ItemTYPE=\Figure
  \fi
  \bgroup
    \hsize=\TEMPDIMEN
    \global\setbox\ItemBOX=\vbox\bgroup
      \eightpoint\nostb@ls{10pt}%
      \let\caption=\fig@caption
      \ifsinglecol \let\nofloat=\no@float\fi
}

\def\fig@caption#1{%
  \vskip 5.5pt plus 6pt%
  \bgroup 
    \eightpoint\nostb@ls{10pt}%
    \setbox\TEMPBOX=\hbox{#1}%
    \ifdim\wd\TEMPBOX>\TEMPDIMEN
      \noindent \unhbox\TEMPBOX\par
    \else
      \hbox to \hsize{\hfil\unhbox\TEMPBOX\hfil}%
    \fi
  \egroup
}

\def\endfigure{%
  \par\egroup 
  \egroup
  \ifsinglecol
    \if@here \midinsert\global\@herefalse\else \topinsert\fi
      \unvbox\ItemBOX
    \endinsert
  \else
    \JoinStack
    \Print{Processing source for figure \the\ItemNUMBER}%
  \fi
}


\newbox\tab@cap@box
\def\tab@caption#1{\global\setbox\tab@cap@box=\hbox{#1\par}}

\newtoks\tab@txt@toks
\long\def\tab@txt#1{\global\tab@txt@toks={#1}\global\table@txttrue}

\newif\iftable@txt  \table@txtfalse
\newif\if@dfloat    \@dfloatfalse

\def\begintable{%
  \@ifstar{\global\@dfloattrue \@btable}{\global\@dfloatfalse \@btable}%
}

\def\@btable#1{%
  \par
  \if@dfloat
    \ItemSPAN=\Double
    \TEMPDIMEN=\PageWidth
  \else
    \ItemSPAN=\Single
    \TEMPDIMEN=\ColumnWidth
  \fi
  \ifsinglecol
    \TEMPDIMEN=\PageWidth
  \else
    \ItemSTATUS=\InStack
    \ItemNUMBER=#1%
    \ItemTYPE=\Table
  \fi
  \bgroup
    \eightpoint\nostb@ls{10pt}%
    \global\setbox\ItemBOX=\vbox\bgroup
      \let\caption=\tab@caption
      \let\tabletext=\tab@txt
      \ifsinglecol \let\nofloat=\no@float\fi
}

\def\endtable{%
  \par\egroup 
  \egroup
  \setbox\TEMPBOX=\hbox to \TEMPDIMEN{%
    \eightpoint\nostb@ls{10pt}%
    \hss
    \vbox{%
      \hsize=\wd\ItemBOX
      \ifvoid\tab@cap@box
      \else
        \noindent\unhbox\tab@cap@box
        \vskip 5.5pt plus 6pt%
      \fi
      \box\ItemBOX
      \iftable@txt
        \vskip 10pt%
        \noindent\the\tab@txt@toks
        \global\table@txtfalse
      \fi
    }%
    \hss
  }%
  \ifsinglecol
    \if@here \midinsert\global\@herefalse\else \topinsert\fi
      \box\TEMPBOX
    \endinsert
  \else
    \global\setbox\ItemBOX=\box\TEMPBOX
    \JoinStack
    \Print{Processing source for table \the\ItemNUMBER}%
  \fi
}

\def\UnloadZoneA{%
\FirstZoneAtrue
 \Iteration=\z@
  \loop
   \ifnum\Iteration<\LengthOfStack
    \GetItemSTATUS{\Iteration}
    \ifnum\ItemSTATUS=\InZoneA
     \GetItemBOX{\Iteration}
     \ifFirstZoneA \vbox to \BodgeHeight{\vfil}%
     \FirstZoneAfalse\fi
     \unvbox\ItemBOX\ItemSep
     \LeaveStack{\Iteration}
     \else
     \advance\Iteration \@ne
   \fi
 \repeat
}

\def\UnloadZoneC{%
\Iteration=\z@
  \loop
   \ifnum\Iteration<\LengthOfStack
    \GetItemSTATUS{\Iteration}
    \ifnum\ItemSTATUS=\InZoneC
     \GetItemBOX{\Iteration}
     \ItemSep\unvbox\ItemBOX
     \LeaveStack{\Iteration}
     \else
     \advance\Iteration \@ne
   \fi
 \repeat
}


\def\ShowItem#1{
  {\GetItemAll{#1}
  \Print{\the#1:
  {TYPE=\ifnum\ItemTYPE=\Figure Figure\else Table\fi}
  {NUMBER=\the\ItemNUMBER}
  {SPAN=\ifnum\ItemSPAN=\Single Single\else Double\fi}
  {SIZE=\the\ItemSIZE}}}
}

\def\ShowStack{%
 \Print{}
 \Print{LengthOfStack = \the\LengthOfStack}
 \ifnum\LengthOfStack=\z@ \Print{Stack is empty}\fi
 \Iteration=\z@
 \loop
 \ifnum\Iteration<\LengthOfStack
  \ShowItem{\Iteration}
  \advance\Iteration \@ne
 \repeat
}

\def\B#1#2{%
\hbox{\vrule\kern-0.4pt\vbox to #2{%
\hrule width #1\vfill\hrule}\kern-0.4pt\vrule}
}


\newif\ifsinglecol   \singlecolfalse

\def\onecolumn{%
  \global\output={\singlecoloutput}%
  \global\hsize=\PageWidth
  \global\vsize=\PageHeight
  \global\ColumnWidth=\hsize
  \global\TextLeading=12pt
  \global\Leading=12
  \global\singlecoltrue
  \global\let\onecolumn=\relax
  \global\let\footnote=\sing@footnote
  \global\let\vfootnote=\sing@vfootnote
  \ninepoint 
  \message{(Single column)}%
}

\def\singlecoloutput{%
  \shipout\vbox{\PageHead\vbox to \PageHeight{\pagebody\vss}\PageFoot}%
  \advancepageno
  \ifplate@page
    \shipout\vbox{%
      \sp@pagetrue
      \def\sp@type{plate}%
      \global\plate@pagefalse
      \PageHead\vbox to \PageHeight{\unvbox\plt@box\vfil}\PageFoot%
    }%
    \message{[plate]}%
    \advancepageno
  \fi
  \ifnum\outputpenalty>-\@MM \else\dosupereject\fi%
}

\def\ItemSep{\vskip\ItemSepamount\relax}

\def\ItemSepbreak{\par\ifdim\lastskip<\ItemSepamount
  \removelastskip\penalty-200\ItemSep\fi%
}


\let\@@endinsert=\endinsert 

\def\endinsert{\egroup 
  \if@mid \dimen@\ht\z@ \advance\dimen@\dp\z@ \advance\dimen@12\p@
    \advance\dimen@\pagetotal \advance\dimen@-\pageshrink
    \ifdim\dimen@>\pagegoal\@midfalse\p@gefalse\fi\fi
  \if@mid \ItemSep\box\z@\ItemSepbreak
  \else\insert\topins{\penalty100 
    \splittopskip\z@skip
    \splitmaxdepth\maxdimen \floatingpenalty\z@
    \ifp@ge \dimen@\dp\z@
    \vbox to\vsize{\unvbox\z@\kern-\dimen@}
    \else \box\z@\nobreak\ItemSep\fi}\fi\endgroup%
}


\def\gobbleone#1{}
\def\gobbletwo#1#2{}
\let\footnote=\gobbletwo 
\let\vfootnote=\gobbleone

\def\sing@footnote#1{\let\@sf\empty 
  \ifhmode\edef\@sf{\spacefactor\the\spacefactor}\/\fi
  \hbox{$^{\hbox{\eightpoint #1}}$}\@sf\sing@vfootnote{#1}%
}

\def\sing@vfootnote#1{\insert\footins\bgroup\eightpoint\b@ls{9pt}%
  \interlinepenalty\interfootnotelinepenalty
  \splittopskip\ht\strutbox 
  \splitmaxdepth\dp\strutbox \floatingpenalty\@MM
  \leftskip\z@skip \rightskip\z@skip \spaceskip\z@skip \xspaceskip\z@skip
  \noindent $^{\scriptstyle\hbox{#1}}$\hskip 4pt%
    \footstrut\futurelet\next\fo@t%
}

\def\footnoterule{\kern-3\p@ \hrule height \z@ \kern 3\p@}

\skip\footins=19.5pt plus 12pt minus 1pt
\count\footins=1000
\dimen\footins=\maxdimen

\def\note#1#2{%
  \let\@sf=\empty \ifhmode\edef\@sf{\spacefactor\the\spacefactor}\/\fi
  #1\insert\footins\bgroup
    \eightpoint\b@ls{10pt}\rm
    \interlinepenalty\interfootnotelinepenalty
    \splitmaxdepth\dp\strutbox \floatingpenalty\@MM
    \leftskip\z@skip \rightskip\z@skip \spaceskip\z@skip \xspaceskip\z@skip
    \noindent\footstrut #1$\,$#2\strut\par
  \egroup
  \@sf\relax}


\def\landscape{%
  \global\TEMPDIMEN=\PageWidth
  \global\PageWidth=\PageHeight
  \global\PageHeight=\TEMPDIMEN
  \global\let\landscape=\relax
  \onecolumn
  \message{(landscape)}%
  \raggedbottom
}


\output{%
  \ifLeftCOL
    \global\setbox\LeftBOX=\vbox to \ZoneBSize{\box255\unvbox\ZoneBBOX
      \ifvoid\footins\else
        \vskip\skip\footins\unvbox\footins\fi
    }%
    \global\LeftCOLfalse
    \MakeRightCol
  \else
    \setbox\RightBOX=\vbox to \ZoneBSize{\box255\unvbox\ZoneBBOX
      \ifvoid\footins\else
        \vskip\skip\footins\unvbox\footins\fi
    }%
    \setbox\MidBOX=\hbox{\box\LeftBOX\hskip\ColumnGap\box\RightBOX}%
    \setbox\PageBOX=\vbox to \PageHeight{%
      \UnloadZoneA\box\MidBOX\UnloadZoneC}%
    \shipout\vbox{\PageHead\vbox to \PageHeight{\box\PageBOX\vss}\PageFoot}%
    \advancepageno
    \ifplate@page
      \shipout\vbox{%
        \sp@pagetrue
        \def\sp@type{plate}%
        \global\plate@pagefalse
        \PageHead\vbox to \PageHeight{\unvbox\plt@box\vfil}\PageFoot%
      }%
      \message{[plate]}%
      \advancepageno
    \fi
    \global\LeftCOLtrue
    \CleanStack
    \MakePage
  \fi
}


\Warn{\start@mess}

\newif\ifCUPmtplainloaded 
\ifprod@font
  \global\CUPmtplainloadedtrue
\fi

\def\mnmacrosloaded{} 

\catcode `\@=12 

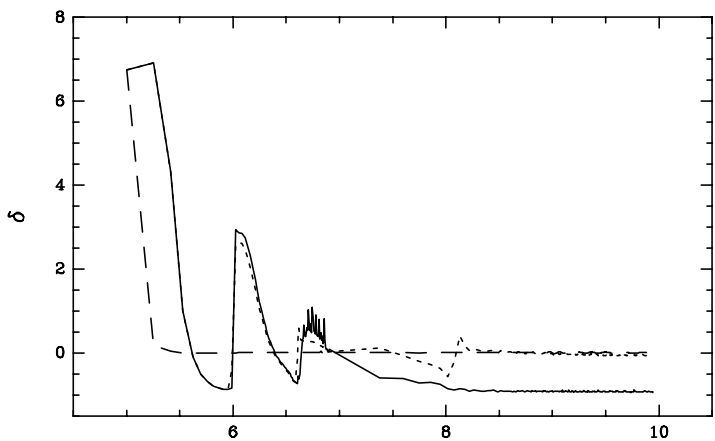
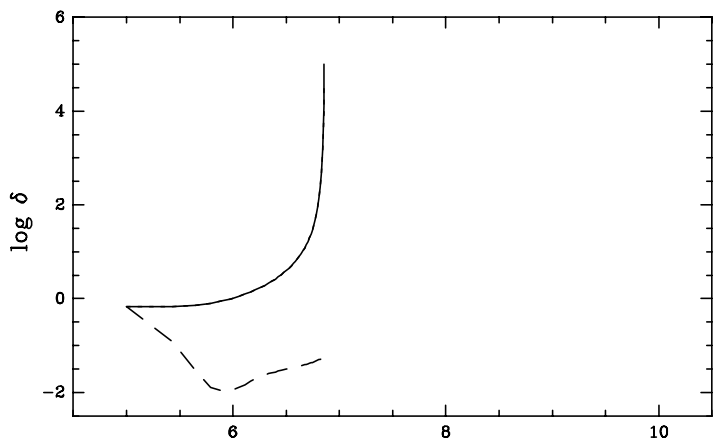
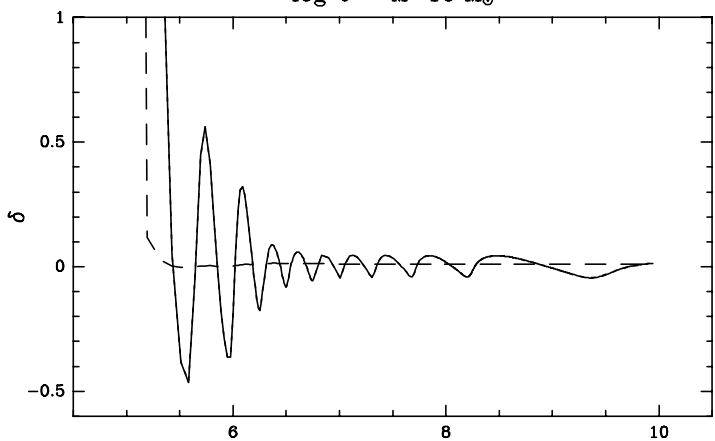
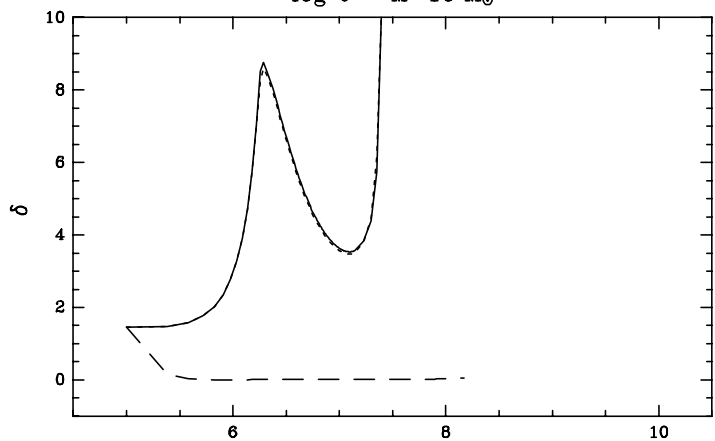
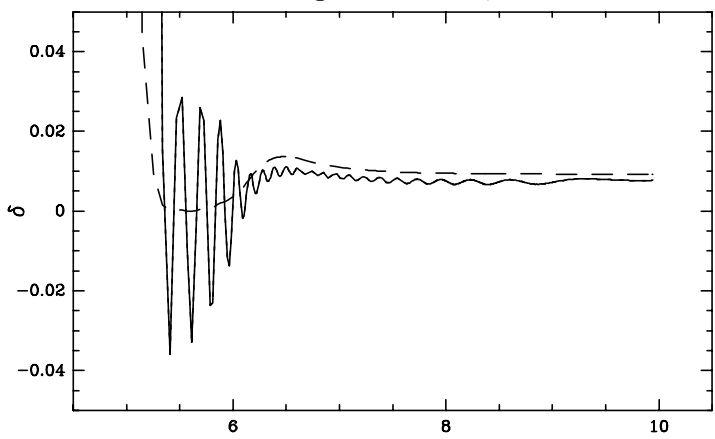
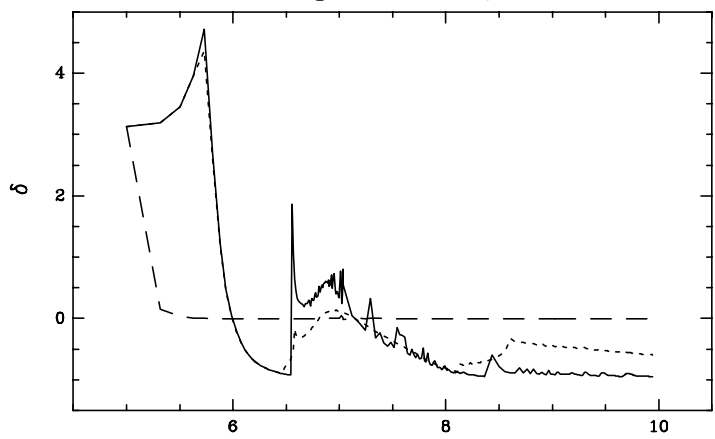
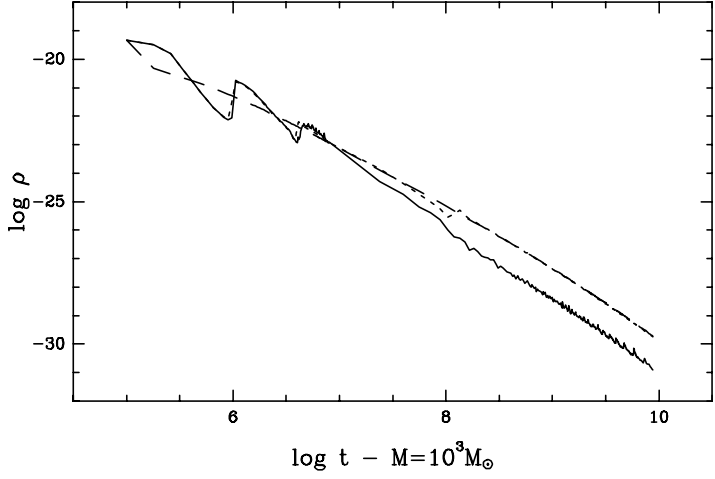
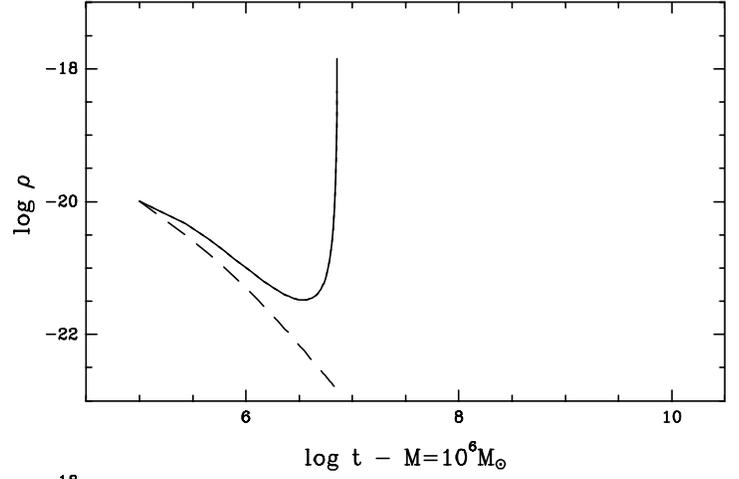
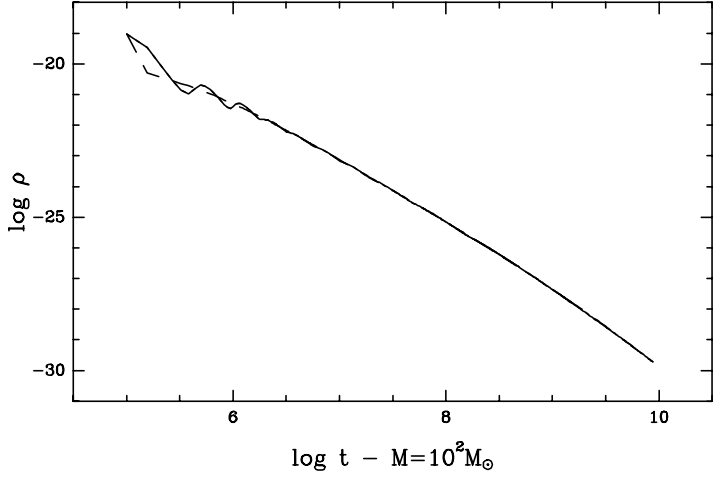
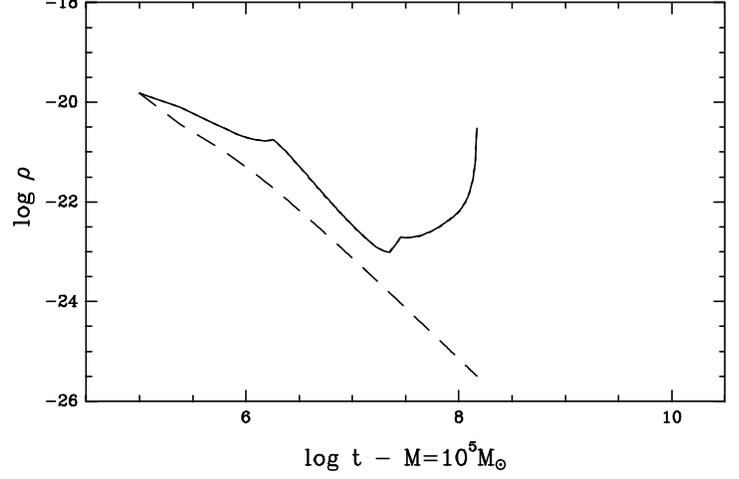
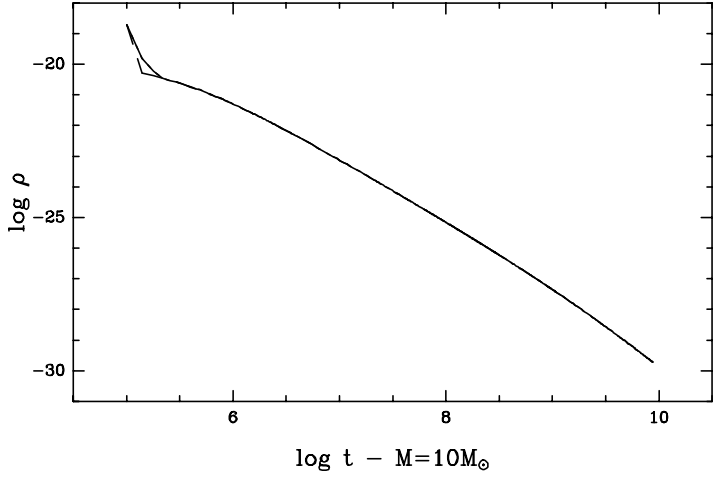
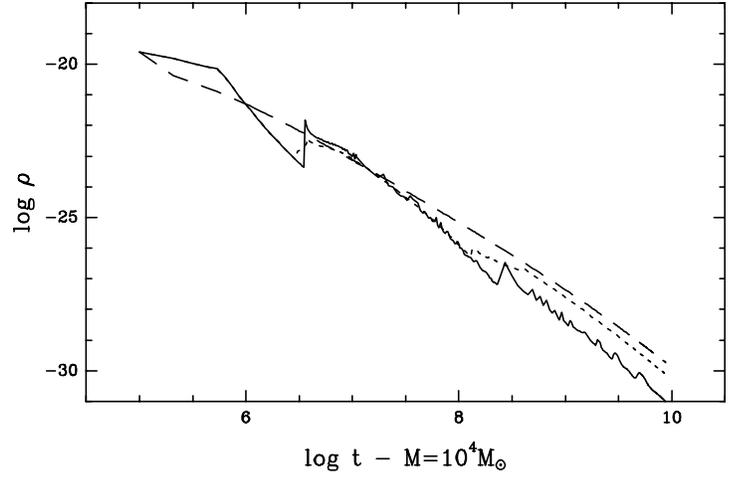
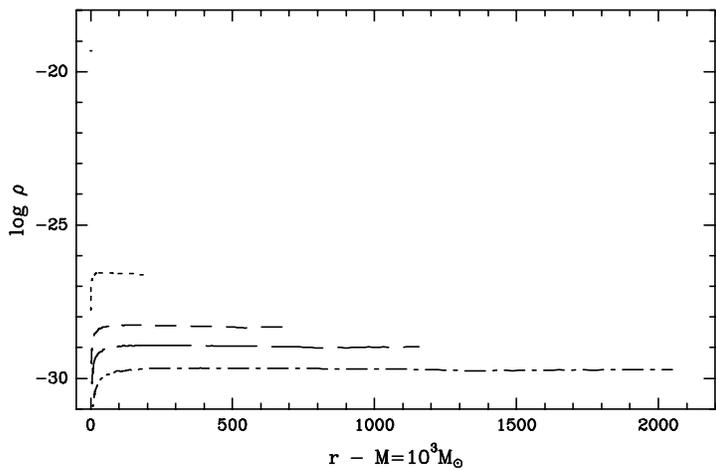
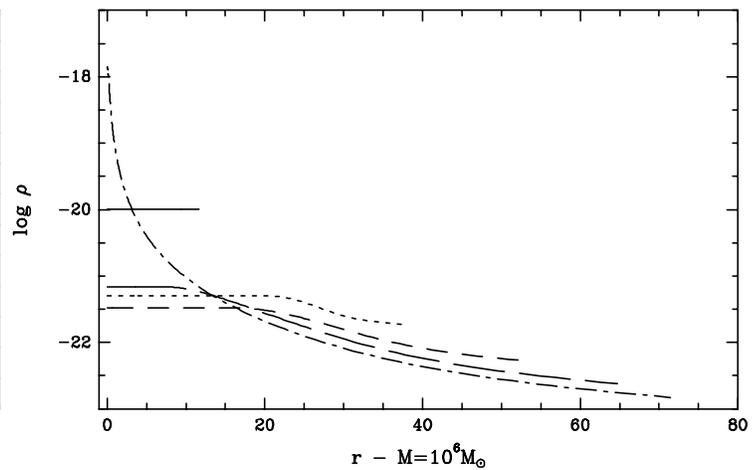
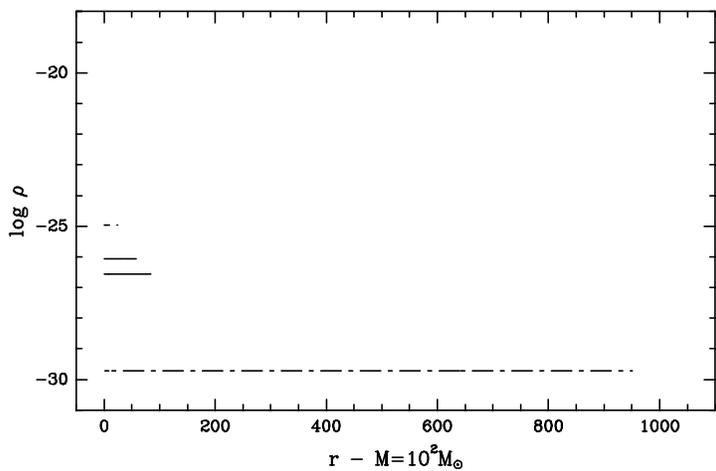
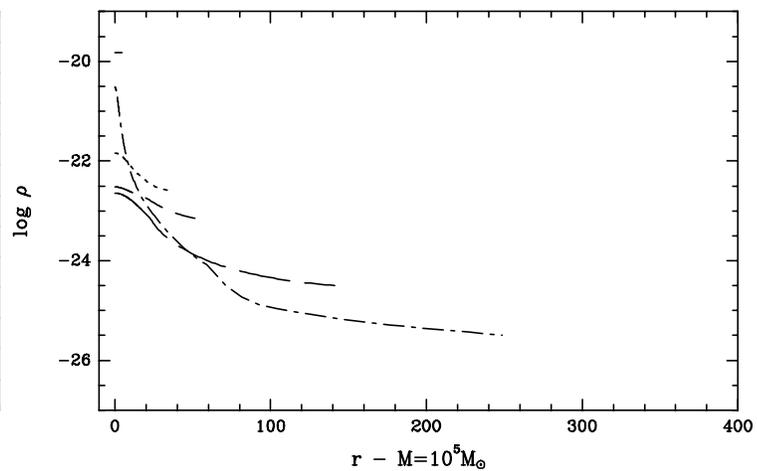
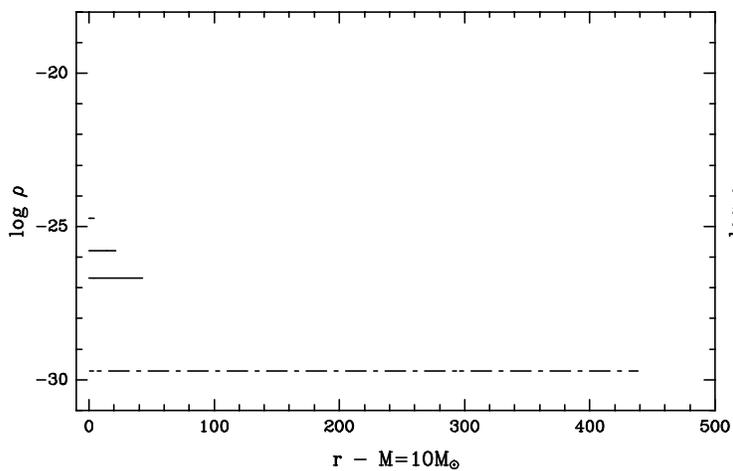
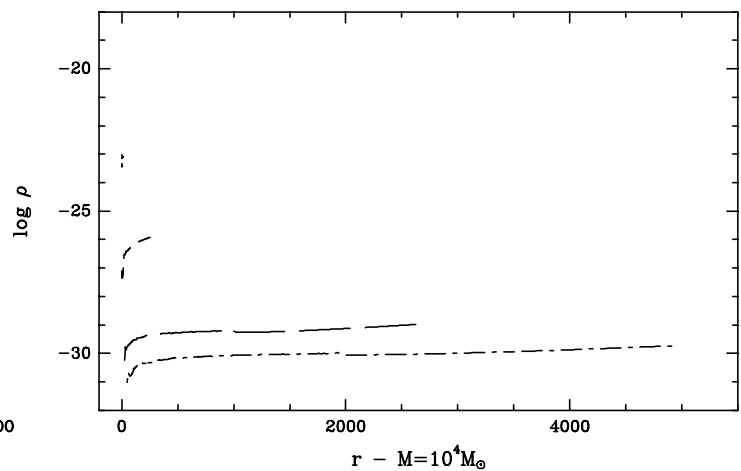
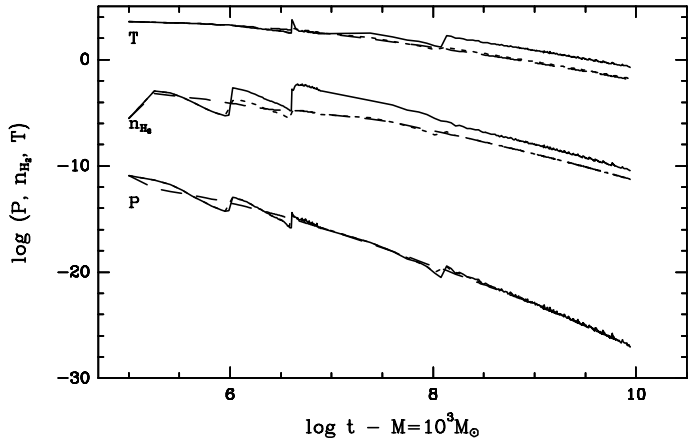
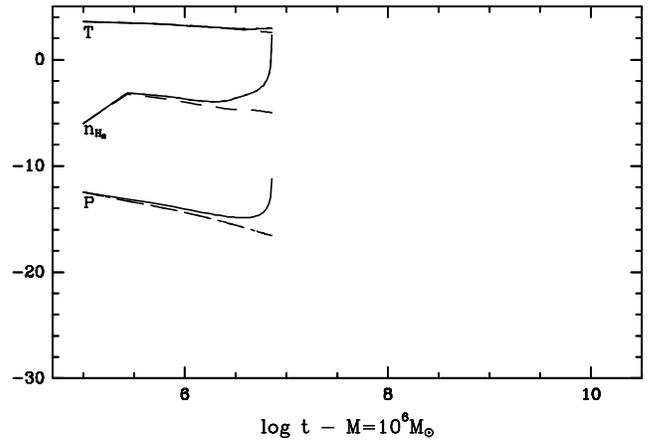
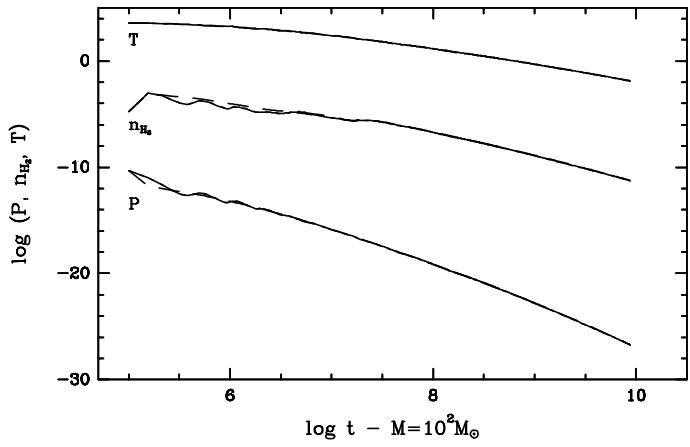
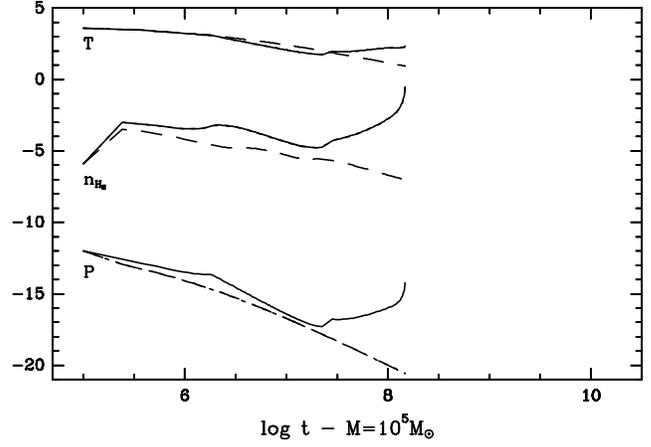
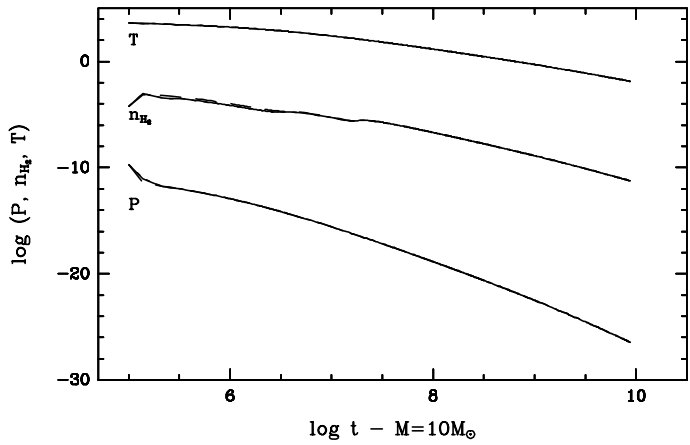
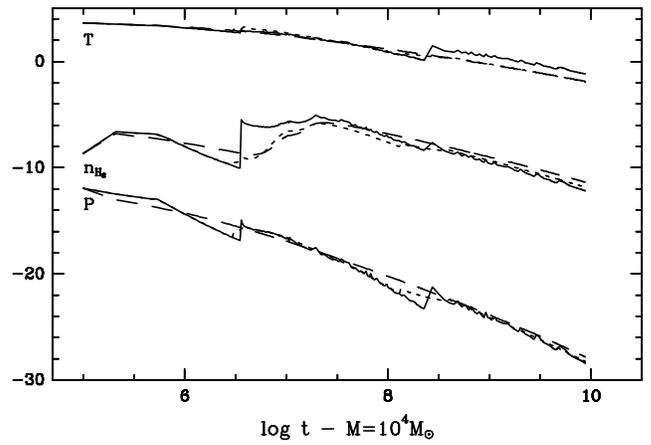
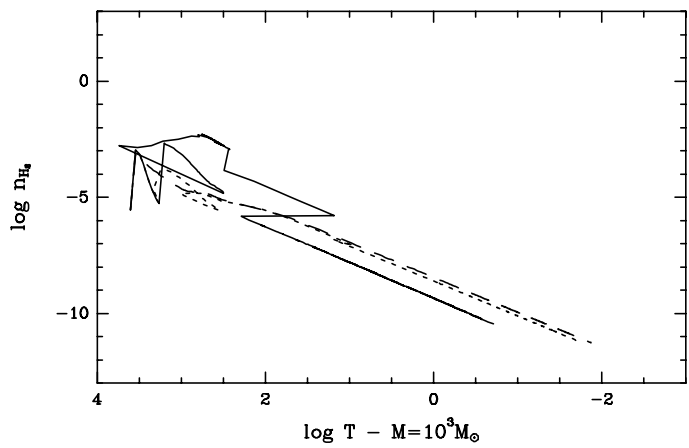
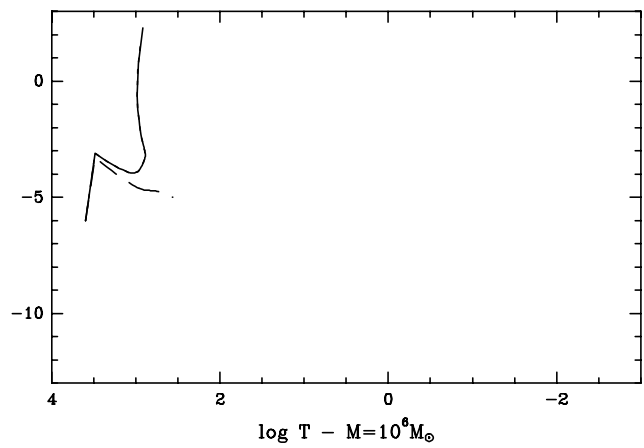
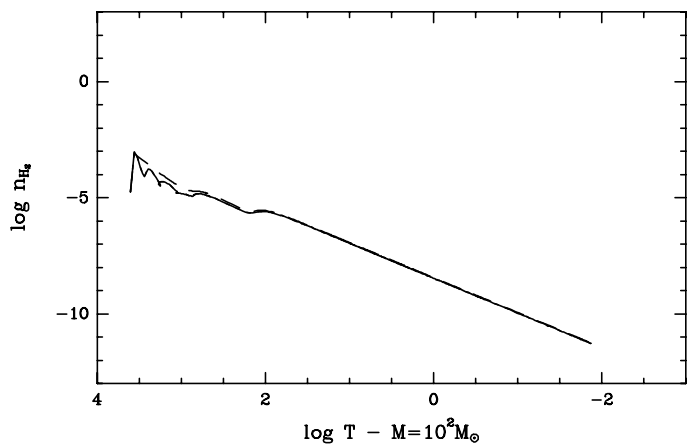
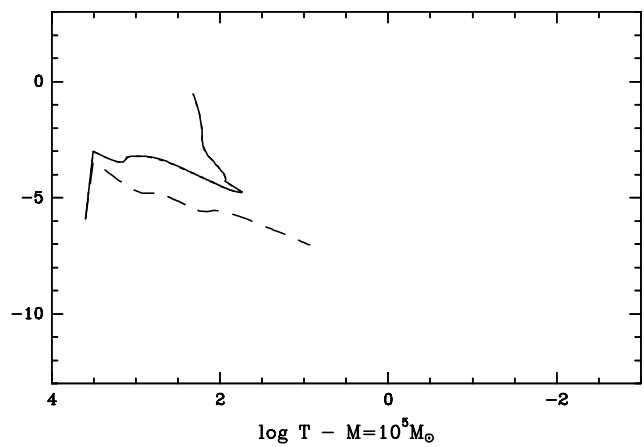
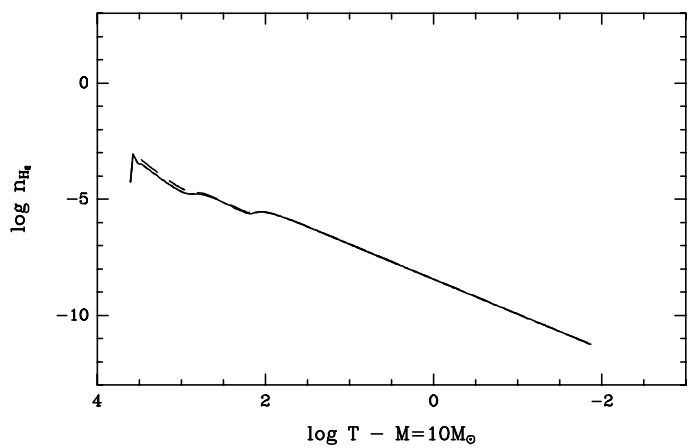
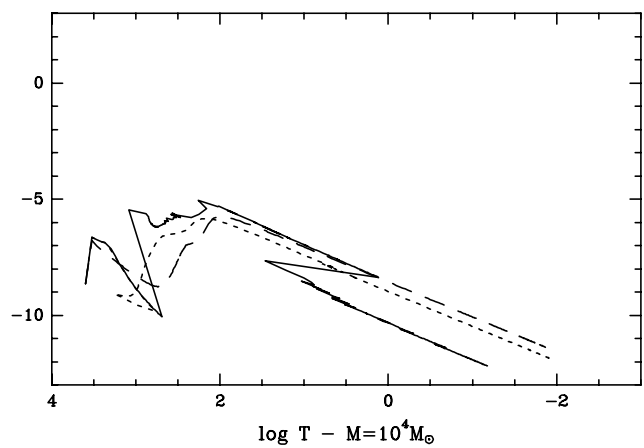
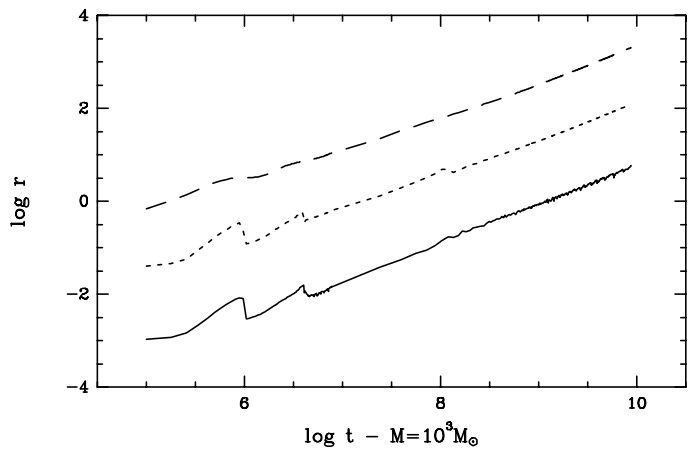
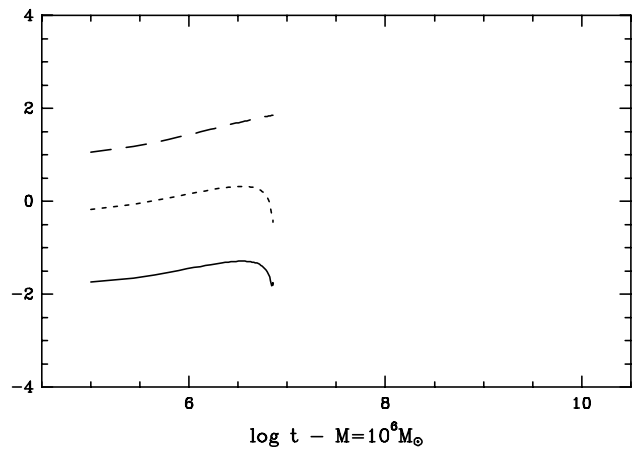
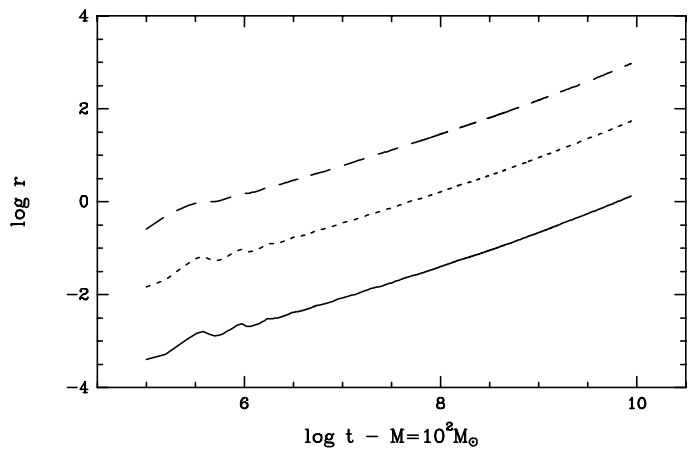
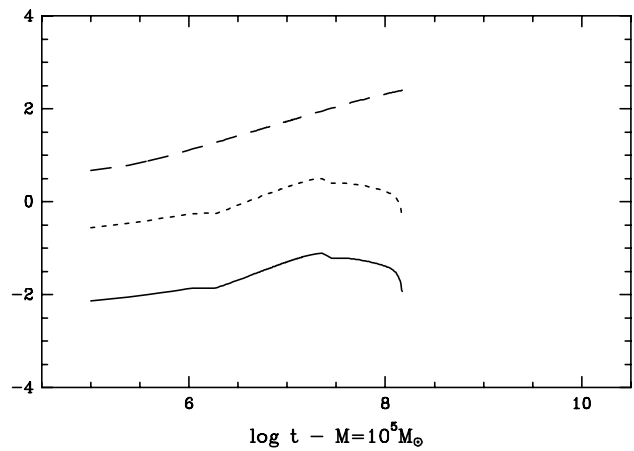
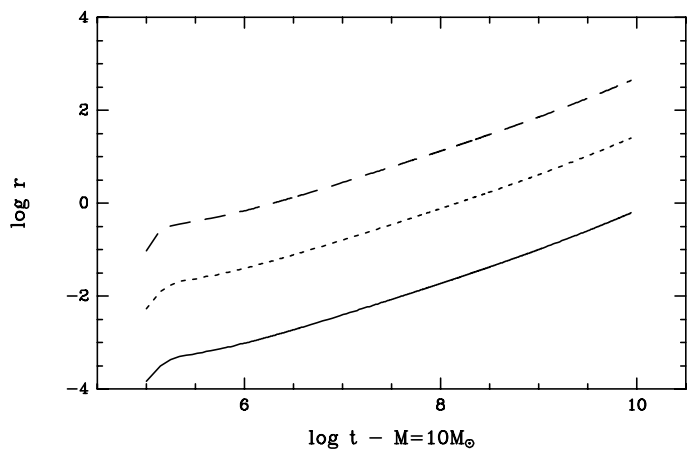
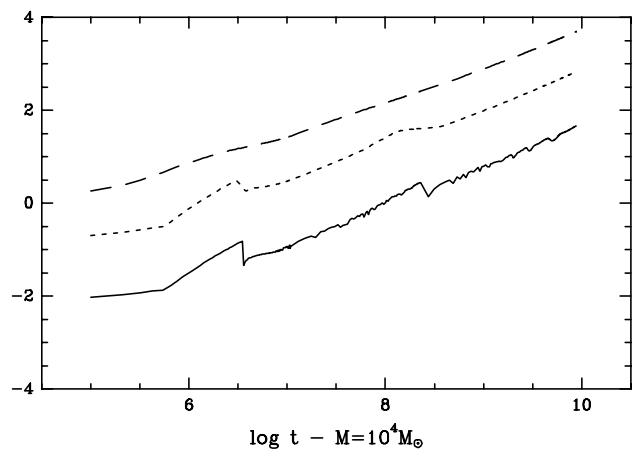
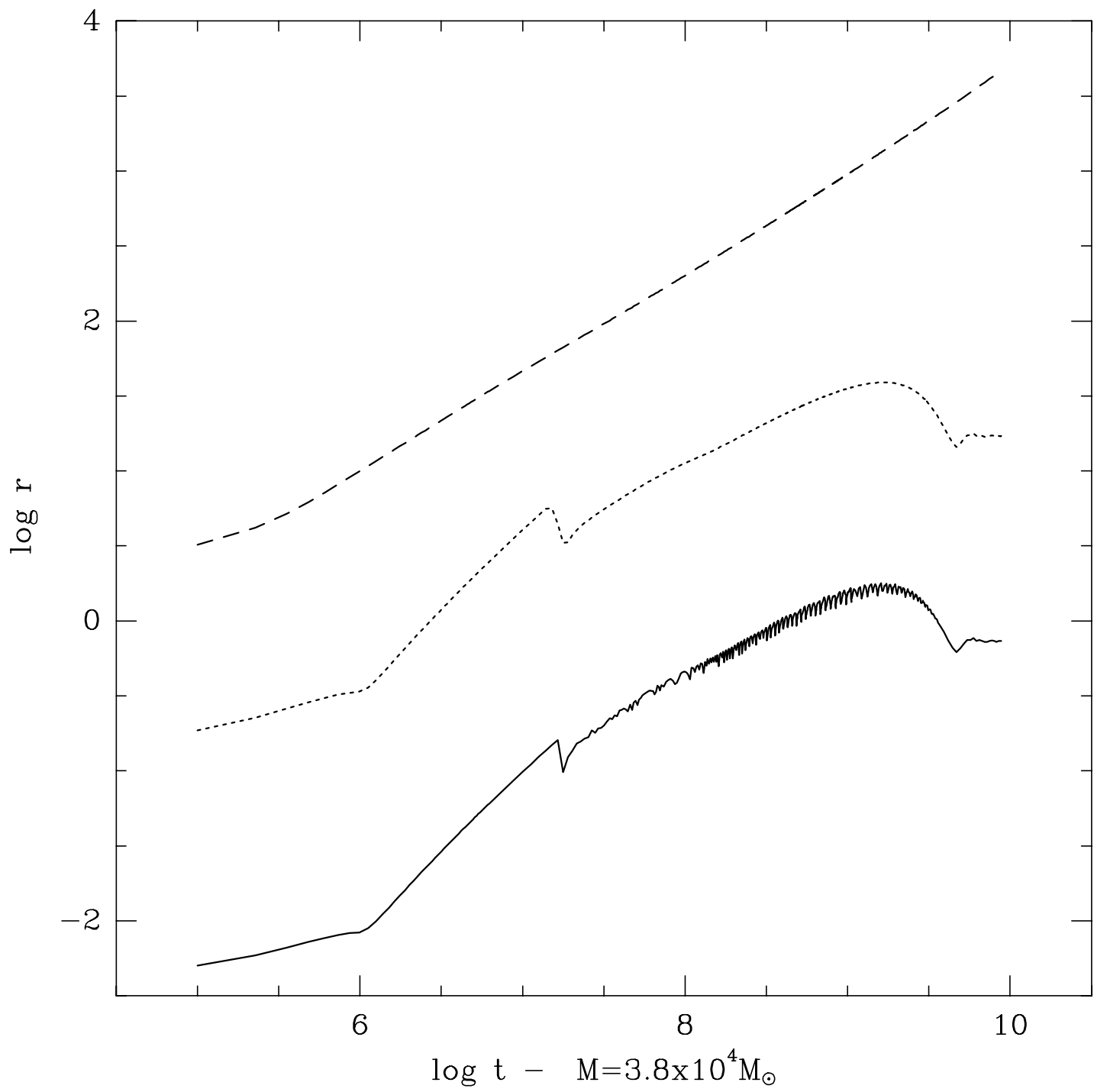
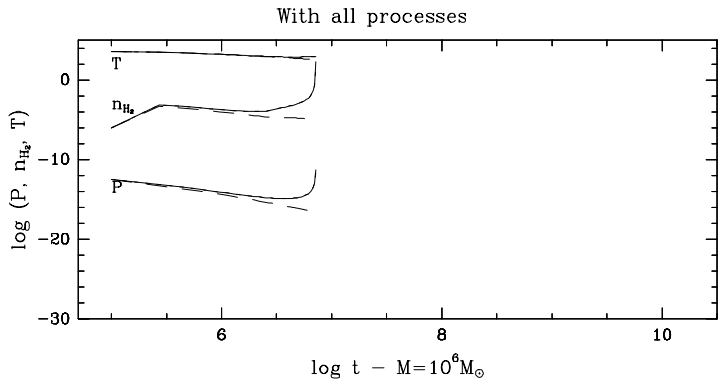
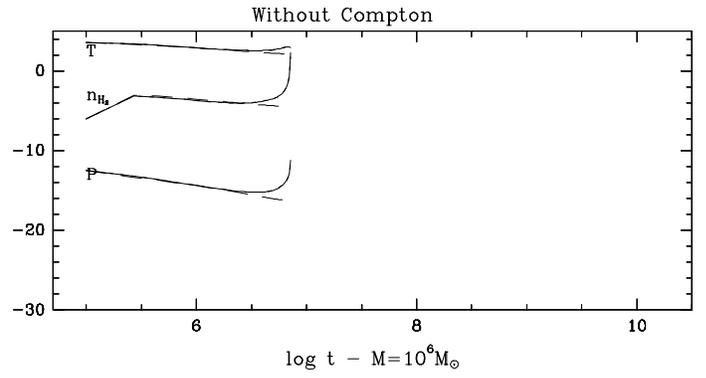
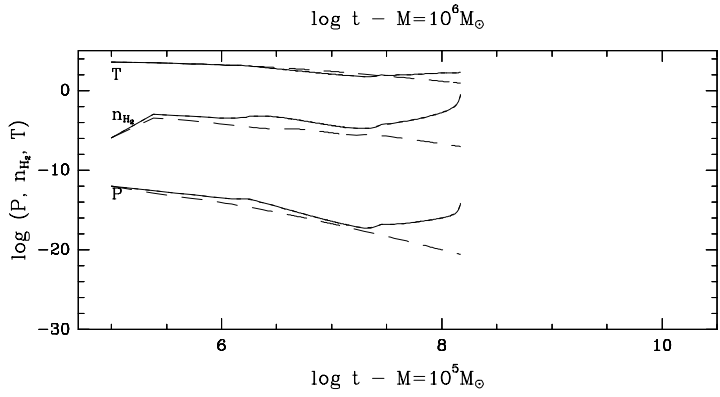
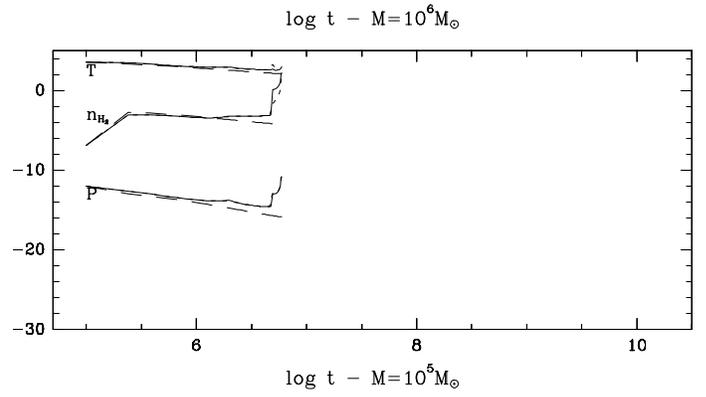
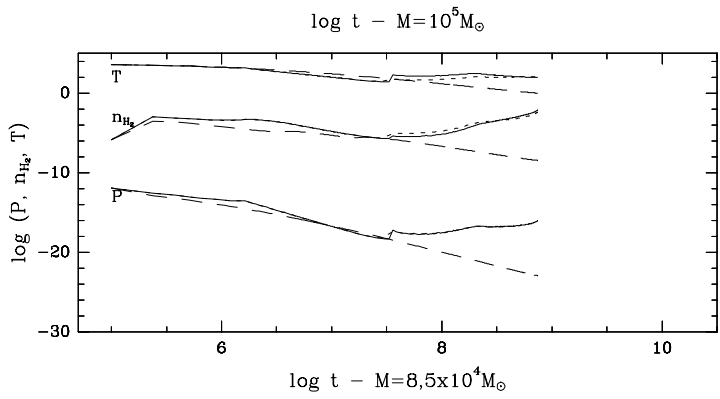
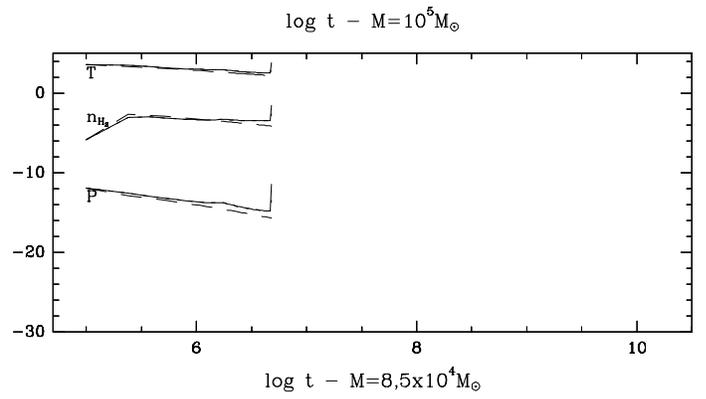
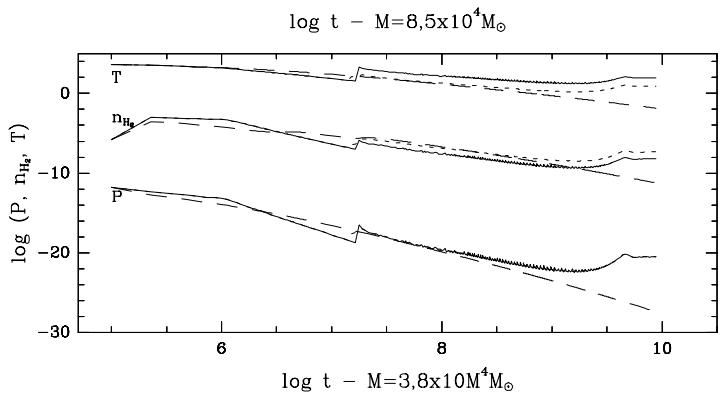
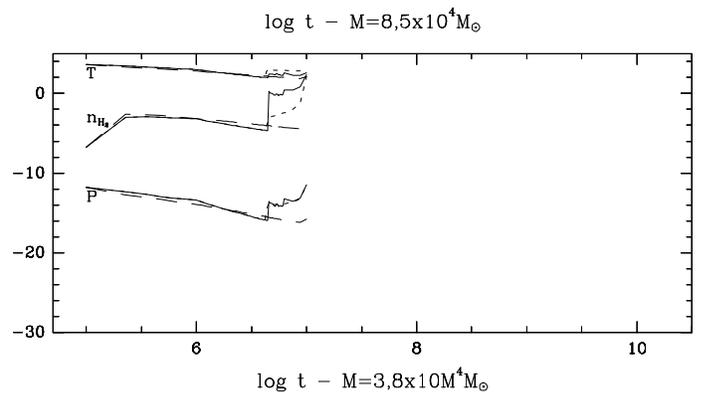
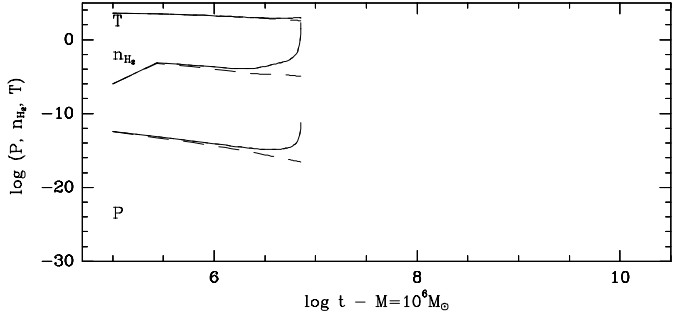
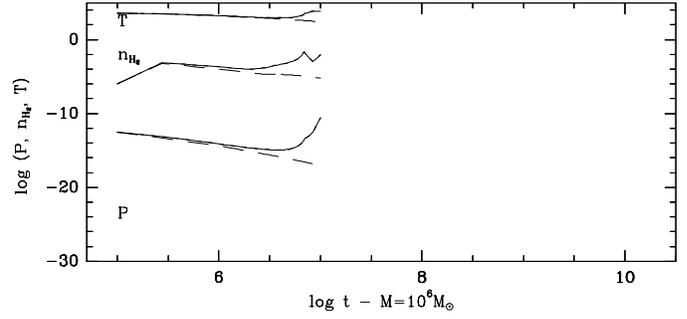
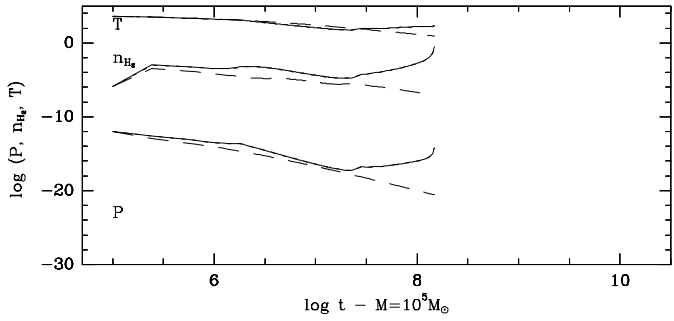
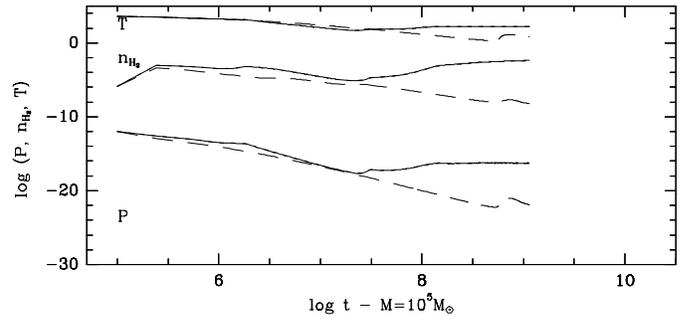
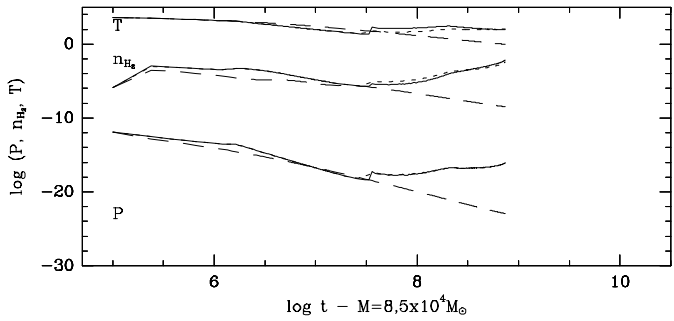
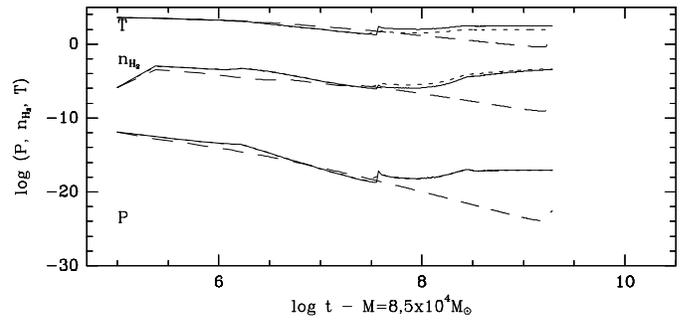
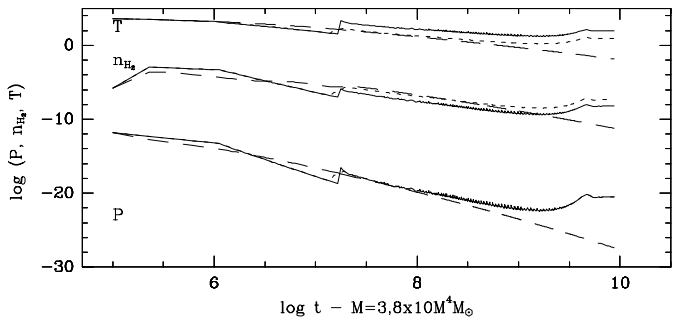
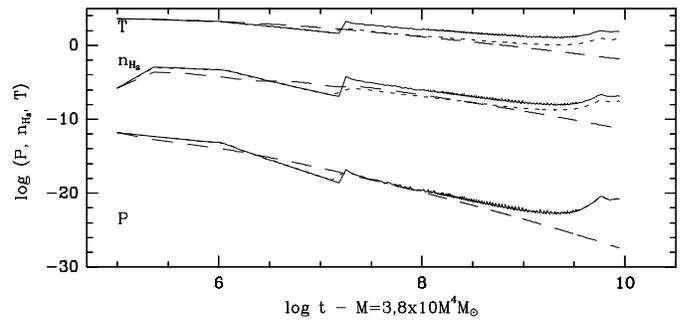
